\documentclass[12pt,preprint]{aastex}
\usepackage{pslatex}
\usepackage{verbatim}
\usepackage{graphicx}
\usepackage{amssymb}
\usepackage{lscape}
\usepackage{amsmath}
\usepackage{wrapfig}
\usepackage{float}
\usepackage{verbatim}
\usepackage{rotating}
\fontfamily{ptm}
\usepackage{hyperref}

\usepackage{pdfpages}

\usepackage{tocloft}
\usepackage{natbib}
\newenvironment{itemize*}%
  {\begin{itemize}%
    \setlength{\itemsep}{3pt}%
    \setlength{\parskip}{0pt}}%
  {\end{itemize}}

\newenvironment{enumerate*}%
  {\begin{enumerate}%
    \setlength{\itemsep}{3pt}%
    \setlength{\parskip}{0pt}}%
  {\end{enumerate}}

\usepackage{color}
\usepackage{ulem}
\newcommand{\ACBc}[1]{\textcolor{black}{ #1}}

\newcommand{\kms}{\rm km~s^{-1}}

\begin{document}

\title{High-Temperature Processing of Solids Through \ACBc{Solar} Nebular Bow Shocks: 3D Radiation Hydrodynamics Simulations with Particles}
\author{A.~C.~Boley\altaffilmark{1,}\altaffilmark{2},
M.~A.~Morris\altaffilmark{3}, and
S.~J.~Desch\altaffilmark{4}}

\altaffiltext{1}{Department of Astronomy, University of Florida, 211 Bryant Space Science Center, Gainesville, FL, 32611, USA}
\altaffiltext{2}{Sagan Fellow}
\altaffiltext{3}{Center for Meteorite Studies, Arizona State University, P.O. Box 876004, Tempe, AZ 88287-6004}
\altaffiltext{4}{School of Earth and Space Exploration, Arizona State University, P.O. Box 871404, Tempe, Arizona 85287-1404, USA}

\begin{abstract}

\ACBc{A fundamental, unsolved problem in Solar System formation is explaining the melting and crystallization of chondrules found in chondritic meteorites.
Theoretical models of chondrule melting in nebular shocks has been shown to be consistent with many aspects of thermal histories inferred for chondrules from laboratory experiments; but, the mechanism driving these shocks is unknown.
Planetesimals and planetary embryos on eccentric orbits can produce bow shocks as they move 
supersonically through the disk gas, and are one possible source of chondrule-melting shocks.
We investigate chondrule formation in bow shocks around planetoids through 3D radiation hydrodynamics simulations.
A new radiation transport algorithm that combines elements of flux-limited diffusion and Monte Carlo methods is used to capture the complexity of radiative transport around bow shocks.
An equation of state that includes the rotational, vibrational, and dissociation modes of H$_2$ is also used.
Solids are followed directly in the simulations and their thermal histories are recorded. 
Adiabatic expansion creates rapid cooling of the gas, and tail shocks behind the embryo can cause secondary heating events.
Radiative transport is efficient, and bow shocks around planetoids can have luminosities 
$\sim$few$\times10^{-8}$ L$_{\odot}$. While barred and radial chondrule textures could be produced in the radiative shocks explored here, porphyritic chondrules may only be possible in the adiabatic limit.
We present a series of predicted cooling curves that merit investigation in laboratory experiments to determine whether the solids produced by bow shocks are represented in the meteoritic record by chondrules or other solids.}

\end{abstract}


\maketitle

\section{Introduction\label{sec:introduction}}

The history of the Solar System's formation and evolution is locked within meteorites. 
Their constituents, especially chondrules and calcium-rich, aluminum-rich inclusions (CAIs; see below), can be radiometrically dated, revealing that these solids formed at the very beginning of the Solar System. 
The chemical, mineralogical, and petrological aspects of chondrules and CAIs can also constrain the energetics and dynamics of the Solar Nebula as planets were forming. 
Thus, these solids provide a chronology of major events during the Solar System's formation.
Meteorites comprise a small amount of mass in the Solar System, but because their components 
have been preserved, they have the potential to tell us more about planet formation than the 
planets themselves. 
 
A principal step toward understanding planet formation comes from chondrules, which are 0.1-1 mm igneous spherules that are found in abundance in nearly all unmelted stony meteorites (i.e., chondrites).  
Roughly 85\% of meteorite falls are ordinary chondrites, which can be up to 80\% chondrules by volume.  
Each chondrule formed from precursors individually melted while floating freely in the solar 
nebula, which then cooled and crystallized.
Estimates based on the flux of chondritic meteorites to Earth suggest that the mass of chondrules in parent bodies is $>10^{24}$ g \citep{grossman_nature_1988}, comparable to the mass of the asteroid belt.  

The thermal histories of chondrules are constrained by their chemical compositions
and their textures, i.e., the number and sizes of crystals within them.
Textures can be reproduced in laboratory experiments for a finite range
of peak temperatures and cooling rates below the liquidus \citep[see][and references therein]{desch_etal_mps_2012}.   
Examples of three common textures are radial, barred, and porphyritic.

Radial textures apparently form from melts that retain no nucleation sites and are 
supercooled.
Upon initiation of crystallization or injection of a nucleating site, the entire droplet
spontaneously crystallizes in a process analogous to freezing rain on Earth.
These textures require peak temperatures roughly 150 - 400 K above the liquidus 
temperature, for most chondrule compositions, to destroy the nucleation sites \citep{hewins_radomsky_1990}.
This constrains the peak temperatures to lie within the broad range $\approx 1820 - 2370$ K.
The cooling rate is believed to be rapid, but is not well constrained.

Barred textures form from melts that retain only a few nucleation sites, also requiring peak temperatures in the broad range $\approx 1820 - 2370$ K.
Laboratory experiments reveal they must cool at rates of a few 
$\times 10^2 \, {\rm K} \, {\rm hr}^{-1}$ to $\approx 3000$ K hr$^{-1}$.
Under these conditions, large parallel laths grow from the melt giving the barred 
structure. 

Finally, porphyritic textures involve hundreds of euhedral crystals growing from
the melt.
This requires milder peak temperatures, only 80 - 120 K above the liquidus
(i.e., peak temperatures $1750 - 2090$ K).
These cooler temperatures allow many nucleation sites to be retained.
Porphyritic textures are reproduced by cooling rates in the approximate range
$2 - 2000 \, {\rm K} \, {\rm hr}^{-1}$.
However, the upper limit of this range is only appropriate to porphyritic pyroxene
chondrules.
Porphyritic olivine chondrules must cool at rates no
greater than 1000 K/hr, with textures more exactly reproduced by cooling rates
under 100 K/hr.
Additional evidence from chemical zoning suggests porphyritic chondrules cooled
at rates toward the slower end of this range \citep[see][]{desch_etal_mps_2012}. 
Other aspects of chondrules, such as their retention of moderate volatiles (e.g., Na and S) and their lack of isotopic fractionation, suggest the following thermal
history: the majority of chondrules started at low ambient temperatures ($< 650 \, {\rm K}$), 
were heated suddenly to $> 2000 \, {\rm K}$ in perhaps tens of minutes or less, cooled from
their peak temperatures to temperatures below the liquidus in a matter of minutes
(i.e., at rates $\sim 10^4 \, {\rm K} \, {\rm hr}^{-1}$), then cooled through their
crystallization range between the liquidus and solidus at rates 
$\sim 10^2 \, {\rm K} \, {\rm hr}^{-1}$ \citep[see][]{desch_etal_mps_2012}.  

A variety of chondrule formation mechanisms have been investigated, using thermal 
histories as a major constraint \ACBc{ \citep{connolly_love_1998,desch_connolly_2002,ciesla_hood_2002}}.
Melting by passage through (1D) shock waves with large lateral extents 
($\gg 10^5 \, {\rm km}$) has been demonstrated to be consistent with chondrule
thermal histories \citep{desch_etal_2005,morris_desch_2010}.
So far, shocks are the only mechanism demonstrated to melt chondrules in a manner 
consistent with their textures and other constraints.
The sources of the shocks have not been conclusively identified, although shocks
driven by gravitational instabilities \citep{boss_durisen_2005} are a promising 
candidate \citep[see also][]{boley_durisen_2008}.
Another promising candidate for chondrule formation is bow shocks around large
planetary bodies on eccentric orbits \citep[][hereafter M2012]{morris_etal_2012}.
So far it is not clear whether these shocks, with their smaller lateral extents,
can be consistent with the thermal histories of chondrules.
Chondrule formation by bow shocks may seem counterintuitive, as chondrules are often thought of as the building blocks of planets.
However, as we discuss next, chronology from meteoritic dating suggests that the majority of chondrules formed late compared with other significant formation events in the Solar Nebula.  

CAIs are the oldest known solids that formed in the Solar System, forming within 
$\sim 10^5$ yr of each other, 4568.2 Myr ago \citep{bouvier_wadhwa_2010}.
This sets a time $t=0$ for planet formation in the Solar Nebula.
Some chondrules may have formed contemporaneously with CAIs \citep{bizzarro_etal_2004,connolly_2012},
but the vast majority of chondrules in chondrites formed over a span of time
$t \approx 1.5 - 3$ Myr after CAIs \citep{kurahashi_etal_2008,villeneuve_etal_2009}.
It is unclear whether older chondrules were quickly incorporated into forming 
planets, where all record of their existence was erased, or whether most chondrules started forming
only  a few million years after CAIs.
Regardless, iron meteorites appear to have formed before most chondrules. 
Based on the requirement that their parent bodies retain enough live ${}^{26}{\rm Al}$
to melt, they must have formed within 1.5 Myr of CAIs  \citep{schersten_etal_2006}.
Recently, the age of Mars's core formation has been inferred to be 
$1.8_{-0.8}^{+1.0}$ Myr after CAIs \citep{dauphas_pourmand_2011}.
Taken at face value, {\it large planetary bodies already existed in the Solar
Nebula before the bulk of surviving chondrules formed.} 
We are forced to entertain the notion that chondrules may have formed not only
contemporaneously with such bodies, but perhaps even as a {\it result} of such 
planetary bodies.

If large bodies are present while the Solar Nebula has substantial gas, which, incidentally, is an underlying assumption of the core accretion paradigm for gas giant planet formation \citep{pollack_etal_1996}, then any body that becomes excited in eccentricity and/or inclination can move supersonically with respect to gas on a nearly Keplerian orbit, creating strong bow shocks.
\cite{hood_1998} and \cite{weidenschilling_etal_1998} recognized this possibility and
suggested that planetesimals on eccentric orbits generated chondrule-producing 
bow shocks.  
In principle, a parcel of gas in the Solar Nebula could encounter such
shocks frequently \citep{hood_etal_2009}, but it is not clear that this could produce the majority of
chondrules. 
\cite{ciesla_etal_2004} combined shock speed estimates from rough 2D adiabatic bow shock calculations with detailed 1D shock calculations to predict chondrule thermal histories  for solids interacting with planetesimal bow shocks. 
The calculations showed that chondrule precursors reached the required peak temperatures, but cooled  at rates
$\sim 10^4 \, {\rm K} \, {\rm hr}^{-1}$. 
The cause of these high cooling rates is very efficient radiative transfer through
the small shocked region (roughly the size of the planetesimal itself).
 As a result, planetesimal bow shocks robustly predict cooling rates too high to be consistent with laboratory reproductions of chondrule textures.
\cite{ciesla_etal_2004}  did note that large bodies might mitigate the extremely fast cooling rates.

Building on previous bow shock models and the results from \cite{dauphas_pourmand_2011}, who found that Mars's core had formed before significant chondrule formation, M2012 numerically investigated bow shocks around a large planetary embryo or planetoid roughly half the mass of Mars. 
They found that the larger size of body leads to two very important differences for
chondrule formation, as compared to planetesimal bow shocks.

In both planetesimal bow shocks and planetary embryo bow shocks, the precursors that form chondrules 
must initially be on trajectories that would intersect with the planetoid.
The strength of a shock drops off rapidly with increasing impact parameter in 3D and 2D  bow shock 
geometries, and the peak heating of chondrules is sensitive to the shock speed.  
Only those chondrules initially on trajectories to strike the planetoid achieve the highest peak 
temperatures consistent with chondrule formation.
To avoid accretion, chondrules must dynamically couple to the gas flowing around the body.
Only for planetary embryos is the standoff distance between the planetoid and the bow shock, typically 
10-20\% of the radius of the shocking body, large enough to allow chondrules time to dynamically couple 
to the gas and be swept around the planetoid.
In smaller bodies, any chondrules produced by the bow shock are likely to be accreted directly onto 
the planetoid.

The critical size for a solid to be diverted around a planetoid that is producing a bow shock can be estimated as follows.
Consider a solid of radius $s$ and internal density $\rho_{\rm s}$ embedded in a gas with density $\rho_{\rm g}$ and adiabatic sound speed $c_{\rm s}$.  
At the shock front, the gas is immediately slowed to a subsonic speed with respect to the planetoid, but the chondrules continue at supersonic speeds $\Delta v$ with respect to  the gas until they are slowed (in the planetoid's frame) by the gas.  
Typically $\Delta v$ is a fraction of the shock speed.  
This occurs over a few multiples of the stopping timescale  $t_{\rm stop} \sim (s \rho_{\rm s}) / (\rho_{\rm g} c_{\rm s})$.
As we will justify later, representative values of these quantities in the post-shock  region are $\rho_{\rm g} = 7 \times 10^{-9} \, {\rm g} \, {\rm cm}^{-3}$ and  $c_{\rm s} \sim 3 \, {\rm km} \, {\rm s}^{-1}$.
A chondrule-like object with $s = 0.03 \, {\rm cm}$ and $\rho_{\rm s} = 3 \, {\rm g} \, {\rm cm}^{-3}$ will have a stopping time of about 1 minute.
The distance the chondrule travels past the shock front before dynamically coupling to
the gas is roughly $(\Delta v) t_{\rm stop} \sim 400 \, {\rm km}$.  
After it has traveled this distance past the shock front the chondrule will be swept up by the gas and carried around the planetoid.
If the standoff distance is less than about 400 km, then chondrules will be accreted.
The need for the shock standoff distances to exceed 400 km or so shows that the critical
planetoid radius for producing chondrules that can escape the body and cool in free space over hours is about 2000 km.

Planetary embryos also have larger cross sections for producing chondrules than smaller bodies.
They also have relatively long orbital eccentricity damping timescales.
A single planetoid can melt a substantial amount (possibly up to $10^{24}$ g) of nebular solids.
M2012 found that for larger bodies, more solids can be melted, and these solids can escape
accretion onto the shocking bodies.

The recent advances by M2012 show significant promise for the bow shock model of 
chondrule formation. 
Nonetheless, there are important limitations to this work. 
For example, although the hydrodynamics bow shocks were run in 2D for calculating shock  structure and fluid flows, the resulting morphologies and shock speeds (as a function of impact radius) are only self-consistent for a constant adiabatic index $\gamma$, a 2D geometry, and no radiative cooling of the shock. 
Multi-dimensional effects, not included in their models, are an essential part of the bow shock model, as we will show. 
Most importantly, chondrule thermal histories were calculated in what amounts to a 1D 
approximation.
The thermal histories of particles were found by taking the shock speed at the point in their trajectory where they crossed the shock, and using that shock speed and the local density as inputs into a 1D detailed shock calculation.
Radiative transfer was only crudely modeled by mixing the locally produced radiation field with the background radiation field, using an approximated optical depth between the unshocked gas and the local gas.
The advantage of this approach is that many physical effects can be included in the calculation of thermal histories, including non-equilibrium ${\rm H}_{2}$ dissociation, and thermal decoupling between solids and gas.
By these approximations, M2012 estimated that chondrule cooling rates might be slowed
below $10^3 \, {\rm K} \, {\rm hr}^{-1}$.
The disadvantage of this approach is that radiative transfer in the complex geometry, as well as the changing gas heat reservoir in bow shock morphologies, was never calculated.

In this work, we numerically investigate the bow shock model for high-temperature 
processing of nebular solids using detailed, 3D radiation hydrodynamics simulations 
with direct particle integration. 
Our principal science objective is to characterize the thermal histories of chondrule
precursors that pass through planetoid bow shocks, using as much self-consistent physics 
as is tractable at this time and to use our results to inform future laboratory experiments 
for investigating solid processing. 

In section 2 we describe the numerical code and list the physical parameters we use to simulate 
3D bow shocks.
The results of these computations are presented in section 3.
Trends in these data are discussed in section 4. 
We summarize our conclusions in section 5. 

\section{Methods\label{sec:methods}}
  
We have run a series of radiation hydrodynamics simulations with particles to understand the self-consistent dynamical and thermal behavior of solids traversing bow shocks around planetary embryos.  
For these simulations, we use a Cartesian grid code BoxzyHydro (see Appendix) that employs an equation of state that accounts for the rotational, vibrational, and dissociation degrees of freedom in molecular hydrogen, a new radiation algorithm that captures the high, transitional, and low optical-depth regimes, and a particle-in-cell method with gas-solid coupling for following particles and determining opacities.
All simulations are run on grids with $N_x$, $N_y$, $N_z$ = 300, 200, 200 cells, except where noted.  
Each cell has a length $\Delta x$, $\Delta y$, $\Delta z$ = 100 km.
  
 The basic  numerical experiment design  is  a wind tunnel.  
 Gas and particles are set to flow onto the grid from the positive $x$ boundary in the negative $x$ direction at 7, 8, or 9 $\kms$.  
  All other grid boundaries are free flow, i.e., the ghost cells along the boundaries inherit the same gas conservative variables as their adjacent real grid cells.  
 Due to the nature of the problem, these ghost cells behave mostly as outflow boundaries, but do allow for some inflow due to gravitational focusing.  
 Grid cells that are part of the planetary embryo, forthwith {\it anchor cells}, are given a no-flow boundary condition.  
 Any anchor cell that is adjacent to a normal grid cell inherits the pressure of the normal grid cell when calculating the pressure gradient in the fluxing routine (see section \ref{sec:hydro}). 
  The gas flow on the grid is initialized to have a velocity $v_x, v_y, v_z = -V_{\rm w}, 0, 0$ where $V_{\rm w}$ is either 7, 8, or 9 $\rm {km}~{s}^{-1}$.  
We do set the initial flow to zero in  a region centered on the embryo with a radius of 6000 km.  
 This is done to help prevent vacuums from forming directly behind the planetary embryo, which can become numerically problematic.  

 The incoming gas density is set to $10^{-9}$ g/cc and a temperature $T=300$ K. 
The gas is assumed to have a chemical composition by mass of $X_{\rm H} = 0.73$, $X_{\rm He}=0.25$, and $X_{\rm Z}=0.02$ for hydrogen, helium, and metals, respectively, where the mean molecular weight for metals is set to $\mu_{\rm Z}=16.78$. 
 This sets the total mean molecular weight of the gas to $\mu=2.33$ when hydrogen is completely molecular. 
 
At the high temperatures that occur in planetary bow shocks, the internal processes of H$_2$ can have an effect on the dynamical and thermal evolution of solids in the shock. Molecular hydrogen dissociation requires 4.48 eV ($\sim 7.2\times10^{-12}$ erg) per molecular, meaning even a small fraction of dissociation acts as an enormous heat sink compared with a fixed $\gamma$ gas.  
To stress the vital role that H$_2$ plays in gas thermodynamics, consider 1 gram of molecular hydrogen gas at a temperature 2000 K. 
The translational thermal energy of the gas is $E_{\rm trans}=1.5 R_g (T/\mu) 1 {\rm g}\sim10^{11}$ erg, for a mean molecular weight $\mu=2$ and gas constant $R_g$.  
Only $\sim 5$\% of the available H$_2$ mass must be dissociated to buffer a comparable amount of energy.
 However, H$_2$ is even a stronger thermal buffer than indicated by dissociation alone. 
Before dissociation temperatures are reached, the vibrational modes of molecular hydrogen are activated \citep[see, e.g.,][]{boley_etal_2007_h2}.  
As with the rotational modes at lower temperatures, energy goes into internal molecular processes instead of translational energy, which means more  thermal energy can go into the gas without raising the gas's temperature (a lower adiabatic index).
The result is strong thermal buffering by H$_2$, allowing gasses to be compressed to high densities. 

To make the large-scale hydrodynamics simulation tractable, the equation of state is derived assuming chemical equilibrium, which is not strictly met for the first $\sim 10$ s of the shock.
During the first $\sim 10$ s, the dissociation of ${\rm H}_{2}$ acts as an important ``cooling''
\footnote{The internal energy density never decreases during this process.  Instead, a larger fraction 
of the internal energy is partitioned into internal degrees of freedom of the gas, causing a drop in 
temperature and pressure.  This should not be confused with true cooling, as energy is not carried away 
from the system.} 
process in the gas, which may be very important for matching detailed chondrule histories, as 
explored by \cite{desch_connolly_2002}, \cite{ciesla_etal_2004}, and \cite{morris_desch_2010}.
We return to this point when putting our results in the context of laboratory experiments. 
 
 A planetary embryo with radius $R_e=3000$ km and mass $M_e\approx 3.4\times 10^{26}$ g ($\approx0.057$ M$_\oplus$) is used for all calculations.
 As this experiment is intended to be the first, significant step toward understanding the three-dimensional behavior of bow shocks and their interactions with Solar Nebular solids, we restrict our calculations to this single size.
 Future simulations will explore  parameter space in $R_e$.
 The embryo's center is at $x$, $y$, $z$ = 6000, 0, 0 km, where the $x$, $y$, $z$ = 0, 0, 0 km is the center of the computational domain. 
 The gravitational potential of the body is included.

Particles are directly integrated with the hydrodynamics using the particle-in-cell algorithm described in section \ref{sec:particles}.
These particles represent nebular solids, and are randomly distributed throughout the grid. 
 The average chondrule precursor to gas mass density fraction  $\eta_c = \rho_c/{\rho}_g=0.04$.   
 We refer to this ratio simply as the chondrule mass fraction, and it reflects the average values for {\it chondrule-forming regions} only, not the Solar Nebula overall (see Desch \& Connolly 2002).
The assumed $\eta_c$ is consistent with a solar abundance of solids $\eta_s \approx 0.004$ times a chondrule concentration factor of $C=10$ ($\eta_c = \eta_s C$). 
We use this notation to be consistent with previous studies of chondrule formation (e.g., M2012). 
The chondrule concentration factor is ultimately a parameter that accounts for uncertainty in the state of solids when the majority of chondrules formed.
However, there are multiple lines of evidence that suggest $C>1$.  
For example, a combination of dust settling and/or turbulence is expected to concentrate solids aerodynamically \citep[e.g.,][]{cuzzi_etal_2003}.
The measured frequency of compound chondrules, i.e., chondrules fused with other chondrules, \citep[$\sim 5$\%, e.g.,][]{gooding_keil_1981} may require very large $C$ \citep[e.g.,][]{desch_etal_mps_2012}.
Furthermore, the retention of volatiles during cooling may also require high $C$ \citep{cuzzi_alexander_2006}.
The chosen value of $C$ in this study is intended to reflect values that have been used in previous 1D shock studies, and is based on the evidence given above.
The $\eta_c$ parameter space will need to be explored in the future.

The bow shock model implicitly assumes that a very large fraction of the total solids has already been incorporated into planet formation, leaving only a potentially small fraction of the initial solid mass to be processed as surviving chondrules.
As a result, the chondrule concentration factor $C$ may need to be much greater than 10 to be consistent with the $\eta_c$ used here, as the average abundance of small solids (loosely defined as not in a planetesimal or bigger object) will be $\ll\eta_s= 0.004$, changing $\eta_c$. 
While we do not explore lower chondrule concentrations, we do explore different opacity regimes, which will give us a very good idea how different chondrule concentrations will alter chondrule cooling.  
Nonetheless, because $C$ can substantially affect the gas-drag coupling at large values, we will need to explore chondrule concentration factors in a subsequent study.

 We only consider a fixed size of chondrule precursors for this work, as a full size distribution of solids is beyond the scope of this paper.
 We also assume that all the mass of solids is in chondrule precursors.  
 This assumption is both practical and motivated from meteoritics. 
 From a practical standpoint, we want to limit the number of variables we tune during this study.
 Observationally, chondrules are well-known to be size-sorted (0.1-1 mm in size), which may be explained through turbulent, aerodynamic size-sorting \citep[e.g.,][but see section \ref{sec:discussion}]{cuzzi_etal_2003}.  
For these reasons, we take all solids in the simulations to be comprised of grains with a radius $s=0.03$ cm and internal density $\rho_s=3$ g/cc. 

The simulations presented here are run with $10^6$ super-particles (see \ref{sec:particles}), where each super-particle is assumed to be a swarm of chondrule precursors with the properties given above. 
Each swarm has its total mass prolongated to the nearest 8 grid cells, as well as its velocity for drag calculations.  
In the radiation hydrodynamics simulations, this mass derived from the particle distribution is used to calculate the opacity due to chondrules, which we take to be $\kappa_c=10$ cm$^2$ per gram of {\it solid} (see \ref{sec:opacity}).  
We also run simulations where we assume $\kappa_c=1$ and 100 cm$^2$ per gram of solid to explore the effects of changing the opacity. 

\ACBc{All particles are also assumed to be thermally coupled to the gas, i.e., the gas and solids maintain the same temperature.  
Previous 1D shock calculations that track separate temperatures for the gas and solids show that periods of gas-solid decoupling tend to be short-lived \citep{morris_desch_2010}.
Nevertheless, this complexity should eventually be included in fully three-dimensional calculations.}

The basic strategy of the numerical experiments is as follows: 
\begin{enumerate}
\item Three adiabatic, 3D wind tunnel simulations are run with wind speeds $V_{\rm w}=7$, 8, and 9 $\kms$ for about 20 hr of simulation time. 
This corresponds to about 16, 19, and 21 crossing times for each $V_{\rm w}$, respectively, allowing the shock to develop fully.  
The term {\it adiabatic} here is strictly used to mean that there is no cooling other than adiabatic expansion.
The gas is not isentropic, as shocks can increase the entropy of the gas.
The full equation of state with  rotational and vibrational modes of H$_2$, as well as dissociation, is included in the adiabatic simulations unless otherwise noted. 
We stress that although dissociation and other internal molecular processes are often described as ``cooling,'' this description is not strictly correct and is only meaningful when compared with a gas that has a fixed adiabatic index (see \ref{sec:eos}).
Particles are included in these simulations.

\item Twenty particles with impact parameters between 0 to 3000 km ($R_e$) are explicitly tracked in the adiabatic simulations, and are used to establish basal shock profiles and thermal histories for solids.
These particles are tracked near the end of each simulation to ensure that the morphology has reached a steady state. 

\item The end states of the adiabatic simulations are used as the initial states for the radiation hydrodynamics runs.
This is done to reduce computing time, as the simulations with radiation are significantly more expensive than the adiabatic cases.
For the radiation hydrodynamics runs, we focus on only wind speeds with $V_{\rm w}=8~\kms$, and explore $\kappa_c=1$, 10, and 100 cm$^2$ per gram of solid.
As the opacity is derived directly from the spatial distribution of solids, particles are necessarily used in these calculations, too.
The simulations are first run for a little over 1 crossing time, and are then run for an additional crossing time while tracking 20 particles in the same way as done for the adiabatic cases.

\item Two adiabatic bow shock simulations are run with a fixed $\gamma=1.46$ to understand the effects of 2D vs.~3D geometries.  
	   One simulation is fully 3D, where the planetoid is spherical, while the other is only 2D, which implies a cylindrical planetoid.  
	   The 3D simulation is run with particles, while the 2D simulation is not, although we ensured that this omission had a negligible effect on the shock structure. 
	   The differences between these two simulations can be attributed to the dimensionality employed, and is used to emphasize the need to account for all three dimensions in these studies.

\end{enumerate}

In the next section, we will discuss the results of these simulations.
   
 \section{Results\label{sec:results}}
  
 \subsection{2D and 3D Adiabatic Bow Shocks with a Fixed Adiabatic Index}
 
\begin{figure}[H]
\includegraphics[width=3.5in]{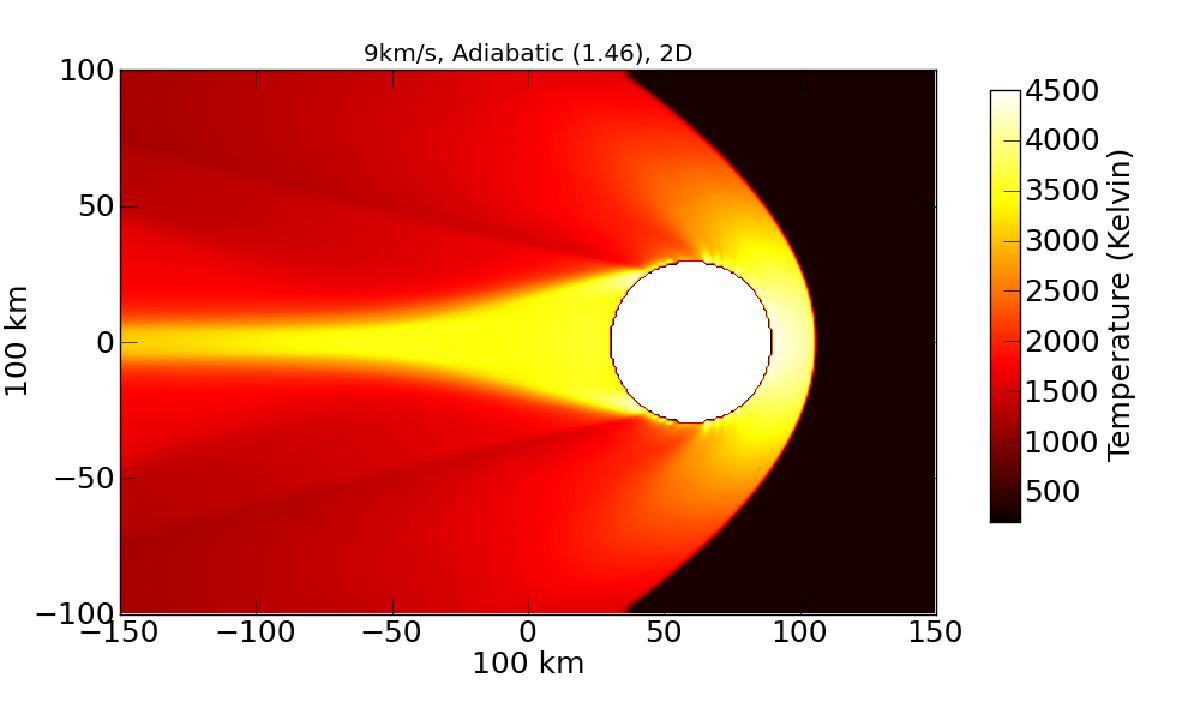}
\includegraphics[width=3.5in]{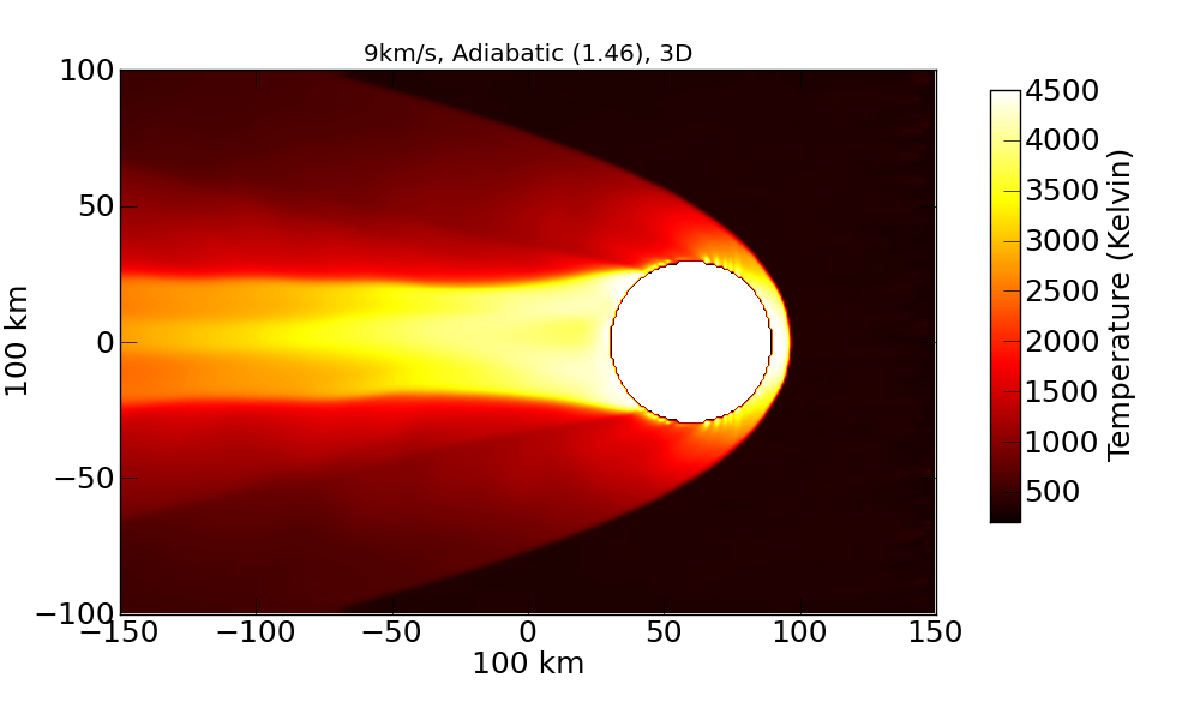}
\caption{Comparison between 2D and 3D simulations with a fixed adiabatic index $\gamma=1.46$.
 \label{fig:2d3d}}
\end{figure}

 In this section, we show the results of two bow shocks, each of which was run with a fixed adiabatic index $\gamma=1.46$ and a wind $V_{\rm w}=9~ \kms$.  
 One simulation explores the shock in 2D ($N_x$, $N_y$ = 300, 200 cells), while the other explores the fully 3D case ($N_x$, $N_y$, $N_z$ = 300, 200, 200 cells). 
 A Cartesian grid is used in both simulations, as has been done in some previous 2D studies.  
The 2D case effectively simulates a bow shock around a cylinder, while the 3D case simulates a bow shock around a sphere.
 The results are shown in Figure \ref{fig:2d3d}.
 
 The overall morphologies of the shocks are similar.  
 Each has a strong bow shock, with very high temperatures directly in front of the planetoid.
 There is also a strong tail shock with a very high temperature (low density) wake. 
 Nonetheless, there are two fundamental differences that will have an impact on characterizing solid processing. (1) The standoff distance in the 2D shock ($\sim 0.5 R_e$) is much larger than in the 3D shock ($\sim 0.2 R_e$).  
Simulations in 2D overestimate the retention of solids, as the available stopping distance is much larger than in 3D simulations. 
Standoff distances in simulations using a proper equation of state will be discussed in the next section.
(2) The opening angle of the bow shock is larger in 2D than in 3D.  
This difference will overestimate the available cross section for strong shocks and high temperatures.  
Thus, the amount of solids that can be processed by a bow shock will be overestimated.  
Moreover, the 2D shock morphology gives a false impression that significant processing can occur at impact parameters larger than $R_e$, which is not seen in 3D simulations. 

Both of these effects seen in 2D shocks (larger standoff distance and larger opening angle) will overestimate the efficiency of solid alteration through bow shocks.  
Unless significant care is taken in accounting for the 3D geometry in a 2D code, a 2D bow shock simulation will give unreliable results. 
For the remaining calculations in this manuscript, we focus on fully 3D simulations.
  
 \subsection{Adiabatic 3D Bow Shocks with a full Equation of State}
 
 The 3D simulation in the previous section used a fixed adiabatic index for the equation of state.  
 While $\gamma=1.46$ takes into account the rotational states of molecular hydrogen, it does not include the vibrational states or dissociation.
These internal processes are a tremendous thermal buffer. 
 Here, we explore the morphology of three adiabatic bow shocks using a full equation of state, with $V_{\rm w}=7$, 8, and 9 $\kms$. 
 Solids are tracked through these simulations and thermal histories of solids that pass through the shock are recorded, assuming perfect thermal coupling between the gas and the solids.
 
\begin{figure}[H]
\includegraphics[width=3.25in]{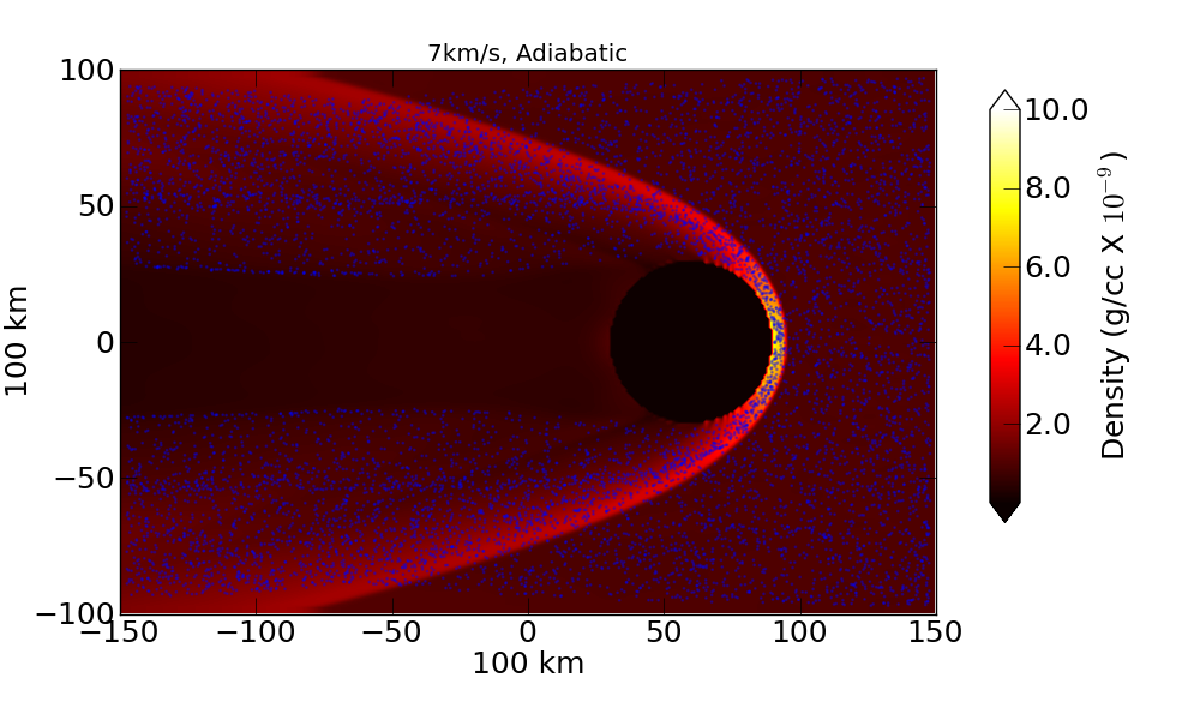}\includegraphics[width=3.25in]{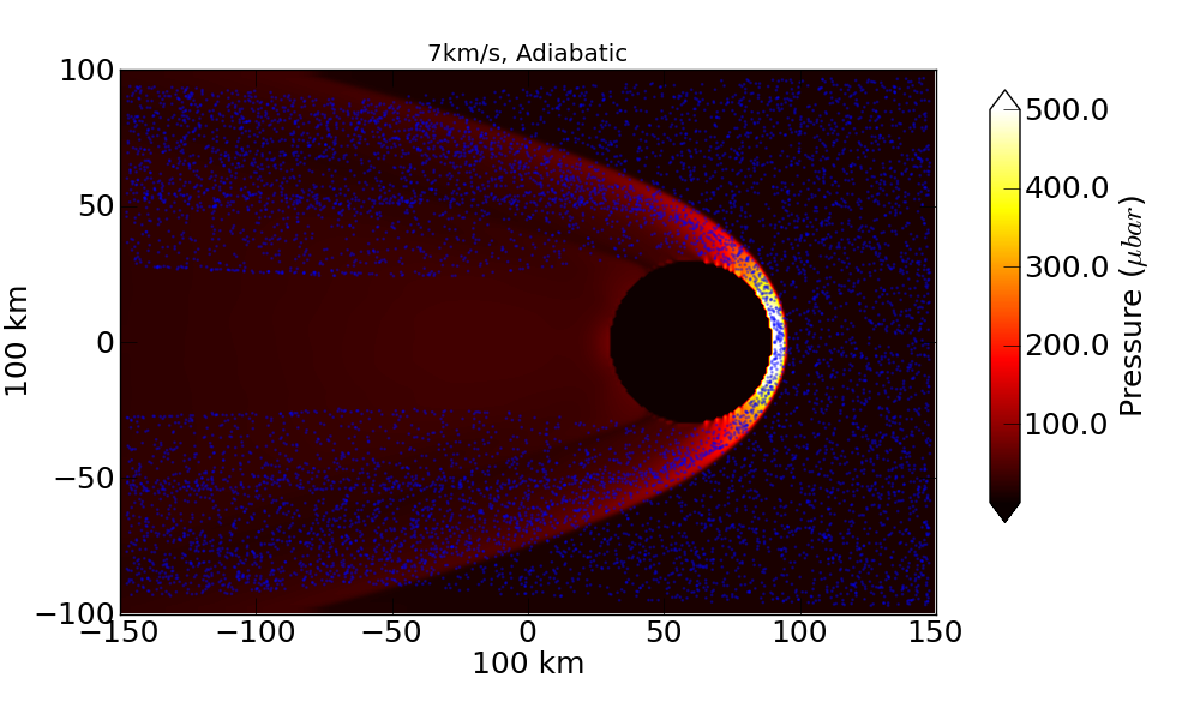}
\includegraphics[width=3.25in]{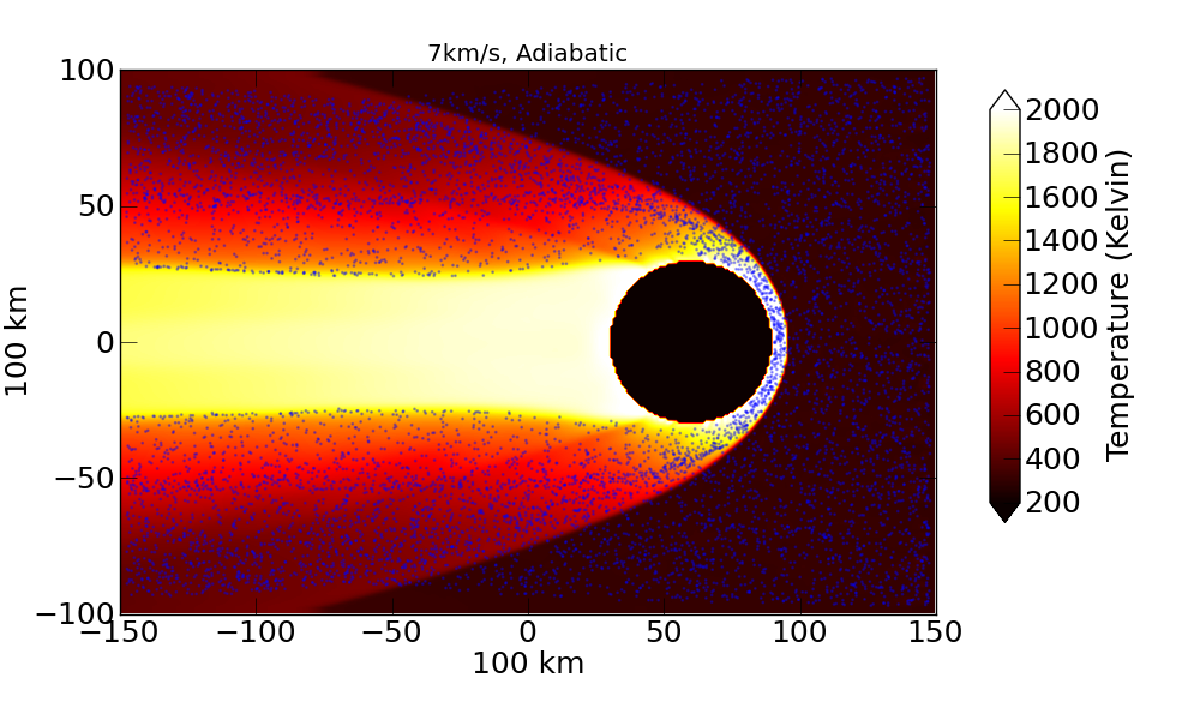}\includegraphics[width=3.25in]{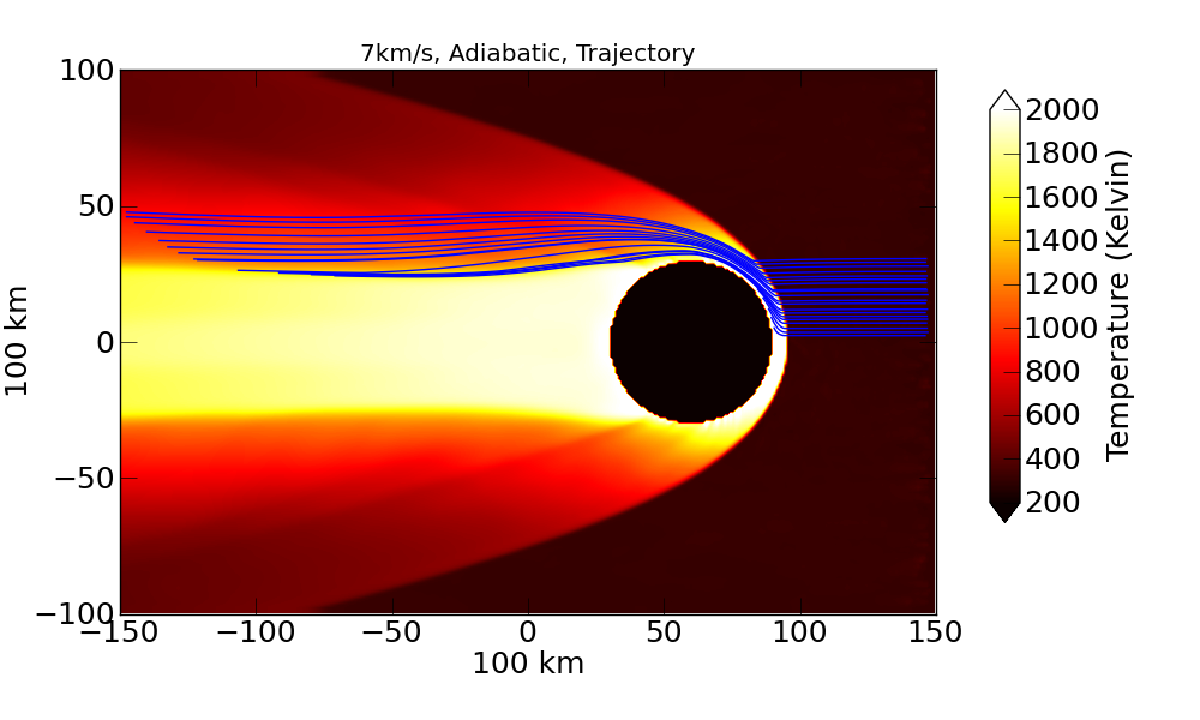}
\caption{A 7 $\kms$ bow shock in the adiabatic limit. 
The images show slices through the simulation along the center of the planetoid.
From top-left, clockwise: Density and particles, pressure and particles,  temperature with 20 particle trajectories (blue curves), and temperature with particles.
The blue dots show particles that intersect the cells depicted in these slices. 
 \label{fig:7adi}}
\end{figure}

\begin{figure}[H]
\includegraphics[width=3.25in]{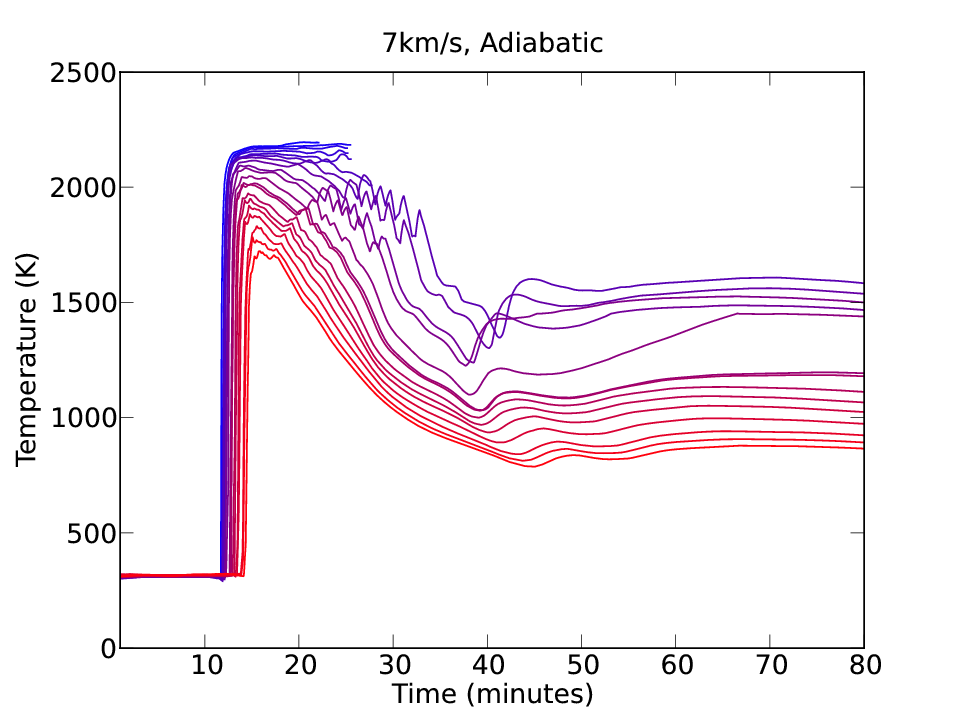}\includegraphics[width=3.25in]{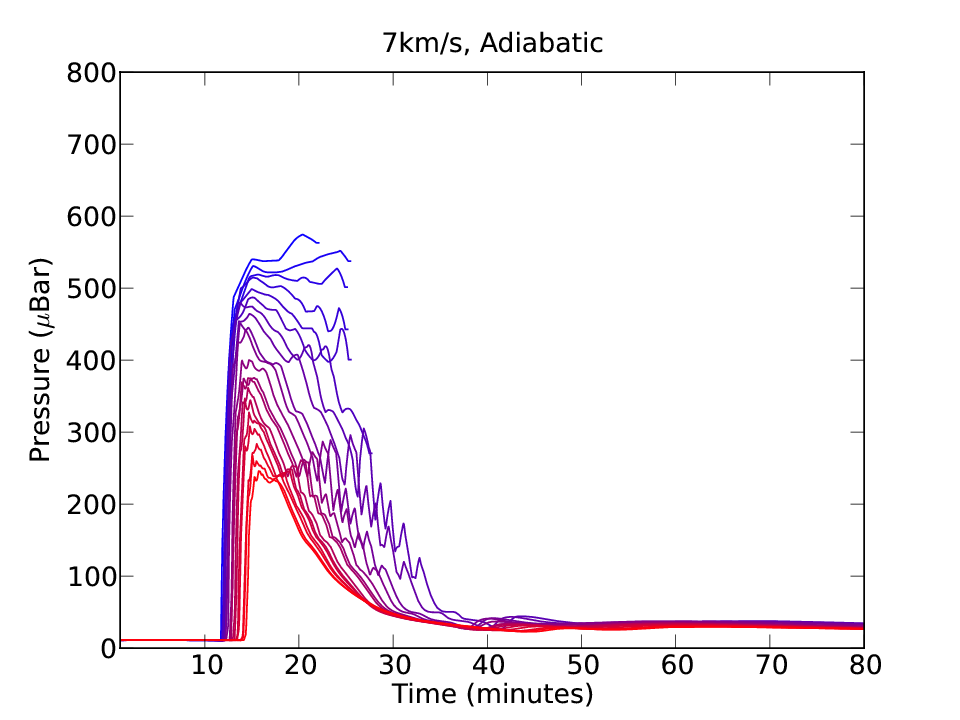}
\includegraphics[width=3.25in]{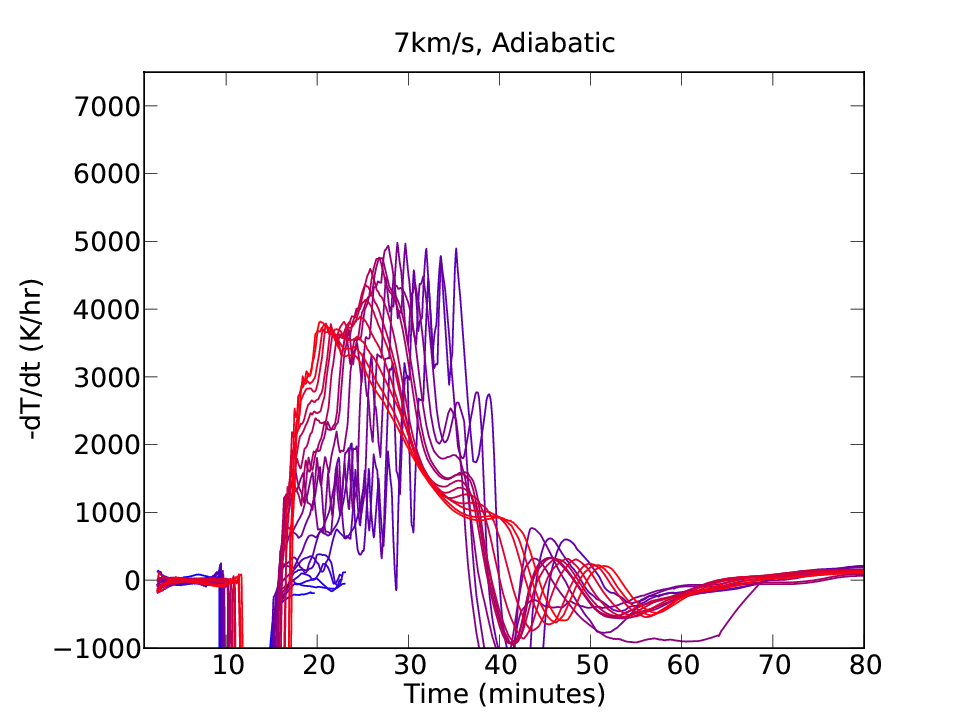}\includegraphics[width=3.25in]{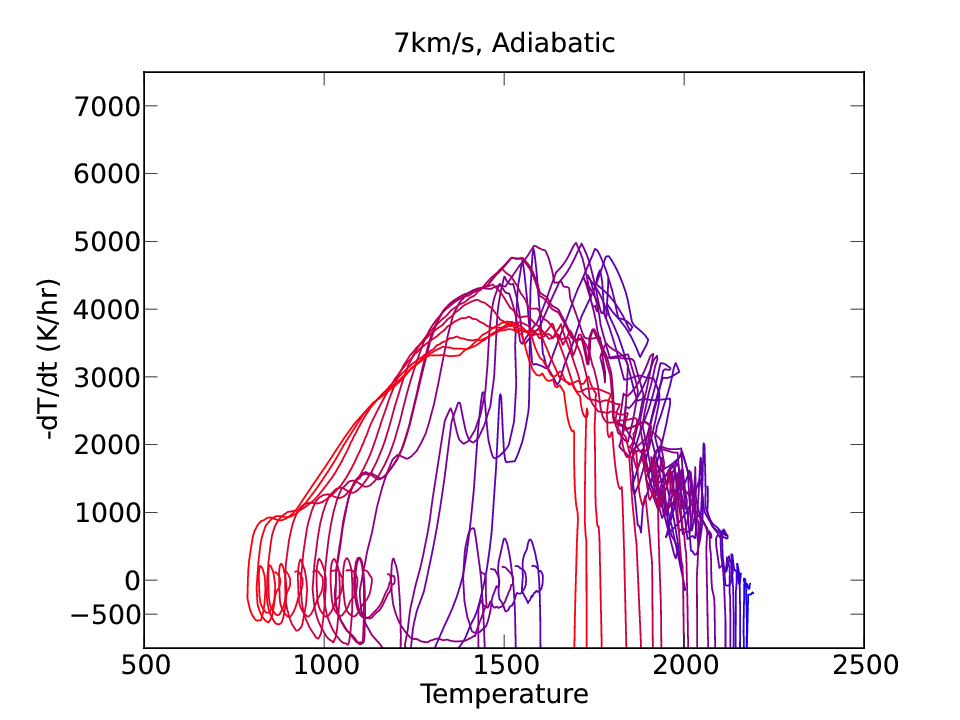}
\caption{The temperature, pressure, and cooling profiles for the solids shown in the bottom-right panel of Figure \ref{fig:7adi}.
 \label{fig:7adi_prof}}
\end{figure}

 Figure \ref{fig:7adi} shows the structure of a 7 $\kms$ bow shock.
 Density, temperature, and pressure are all shown as a slice through the shock and center of the planetoid.
 Particles that are within the cells of the given slice are shown in blue.
 For one of the temperature plots (Fig.~\ref{fig:7adi_prof}), trajectories for 20 particles that have impact parameters between 0 and $R_e$ are shown as blue curves. 
 As in the simulations with a fixed adiabatic index , strong bow and tail shocks form, as well as a high temperature, low density wake. 
 Peak gas temperatures are just over 2000 K, and peak pressures are $\sim 500\mu\rm Bar$. 
The standoff distance is about $0.17 R_e$, enabling most solids to escape accretion onto the body.
 Surviving solids, i.e., solids that are not swept up by the planetoid, are diverted from the wake, creating a dust-free region. 
 We will return to implications of this dust free region in the section \ref{sec:discussion}.

We deconstruct the shock into five regions: the pre-shock, the bow shock, the post-bow shock, the tail shock, and the post-tail shock.  
The dust-free wake could be considered a sixth region, but we exclude it for the following discussion as solids do not enter it.
Particles and gas enter the simulation at the positive $x$ boundary.  
Besides a modest acceleration toward the planetoid due to the planetoid's gravity, the pre-shock region is relatively undisturbed. 
This acceleration will be much more important for planetary-mass objects.  
The temperature and pressure of the gas jumps suddenly at the bow shock.
Near the centroid of the bow shock, the temperature and pressure profiles continue to rise as the solids approach the planetoid's surface.
The curves that end abruptly  result from solids that strike the planetoid.

After a few minutes in the post-shock region, the gas temperature and pressure drop abruptly (see Fig.~\ref{fig:7adi_prof}).
This drop is solely due to adiabatic expansion, and is not captured in 1D models.
The cooling rate after the shock is about 1000 to 5000 K/hr, depending on the impact parameter of the solid and the time after the shock.
These high cooling rates (a few 1000 K/hr) last for about 10 to 20 minutes.
Some curves show oscillations, which are due to the discretization of the planetoid onto the grid, causing a series of minor shocks.
These oscillations, which are artifacts, are most severe for particles that barely miss the surface of the planetoid, but the general trend is indicative of the real behavior. 

  \begin{figure}[H]
\includegraphics[width=3.25in]{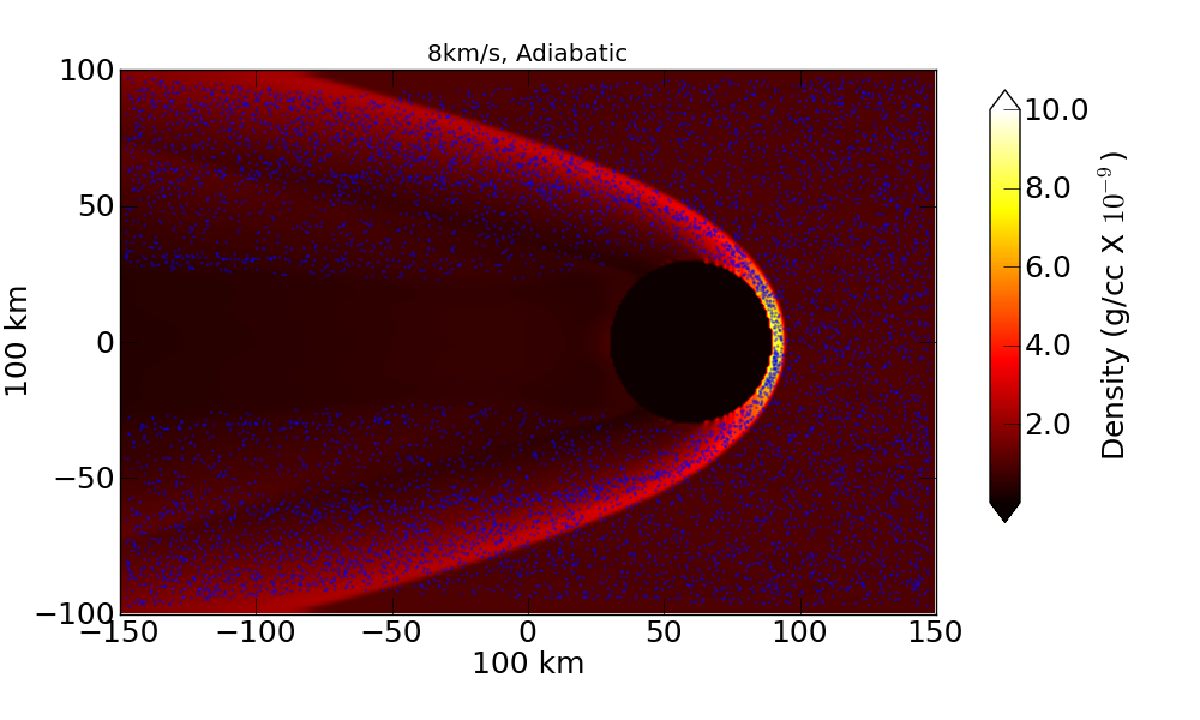}\includegraphics[width=3.25in]{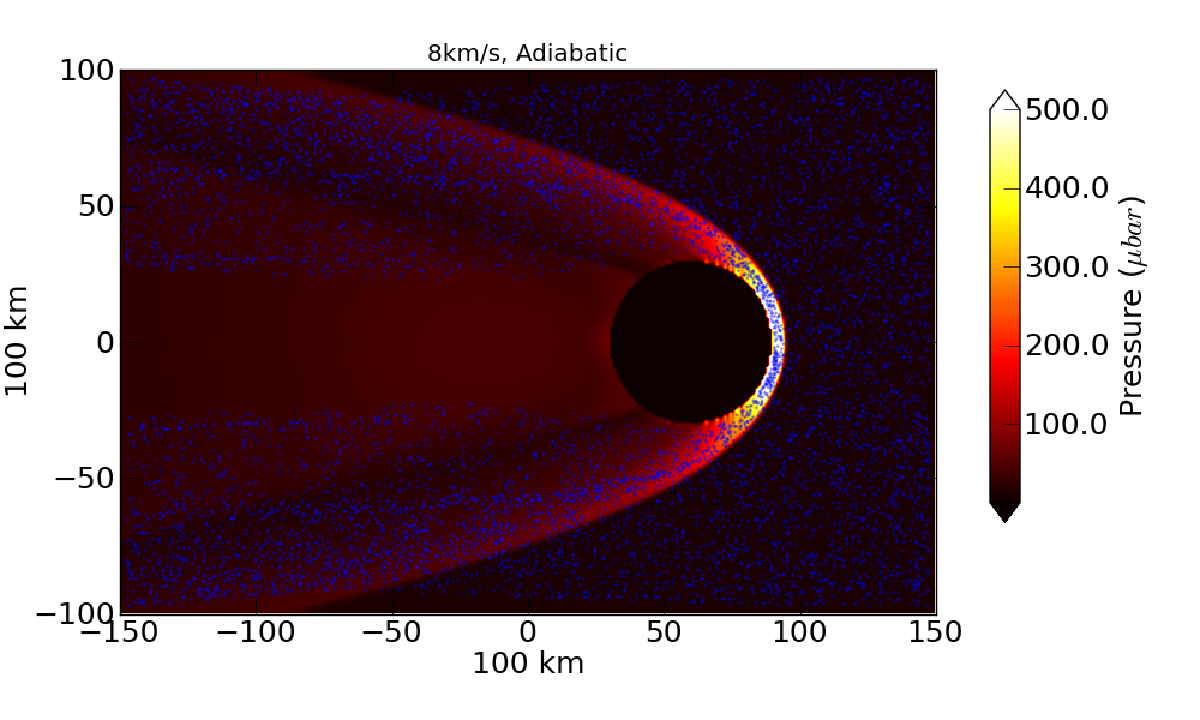}
\includegraphics[width=3.25in]{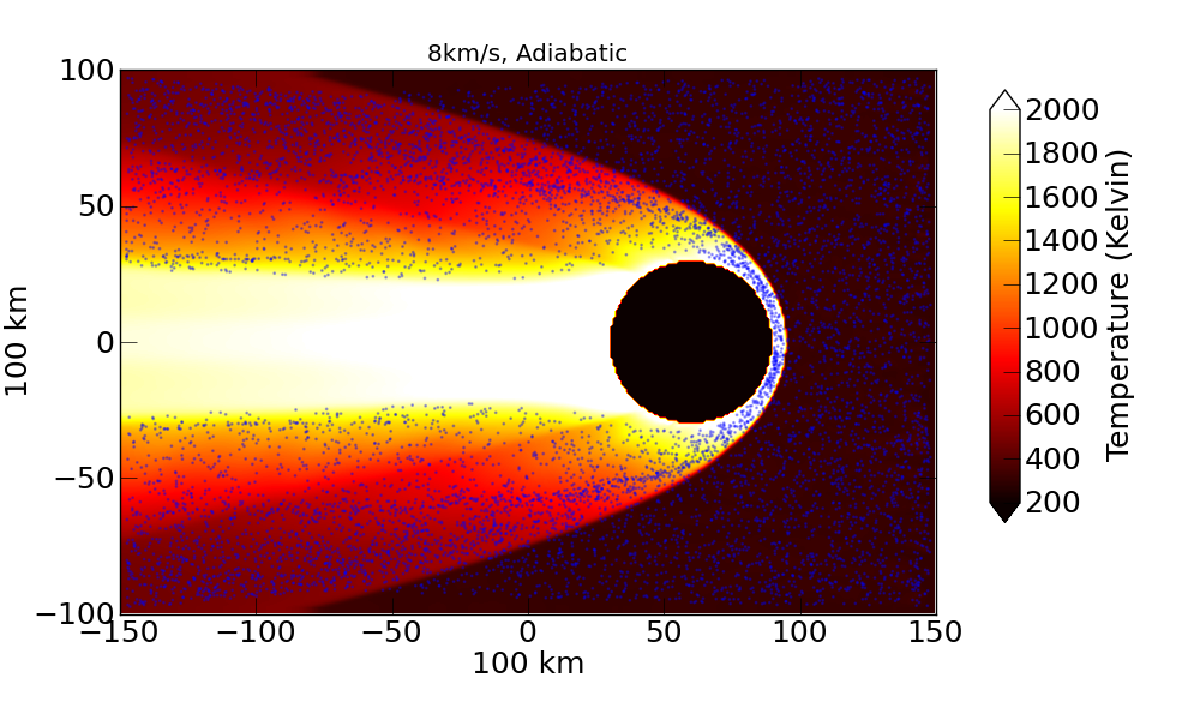}\includegraphics[width=3.25in]{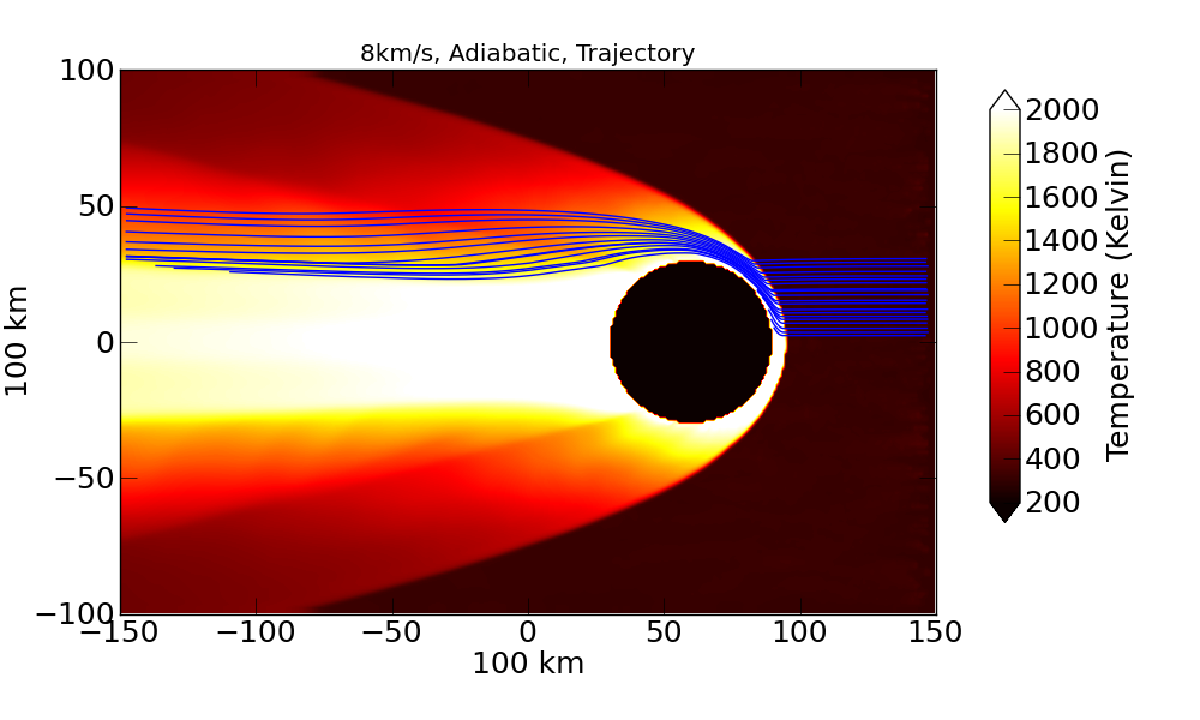}
\caption{Similar to Figure \ref{fig:7adi}, but for an 8 $\kms$ bow shock in the adiabatic limit.
 \label{fig:8adi}}
\end{figure}

   \begin{figure}[H]
\includegraphics[width=3.25in]{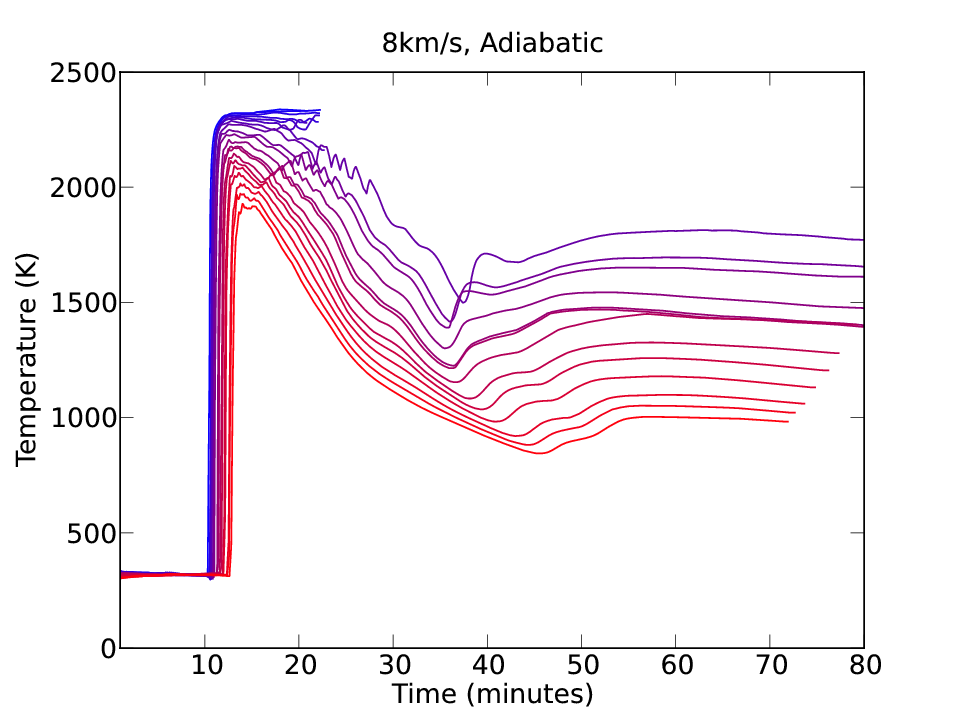}\includegraphics[width=3.25in]{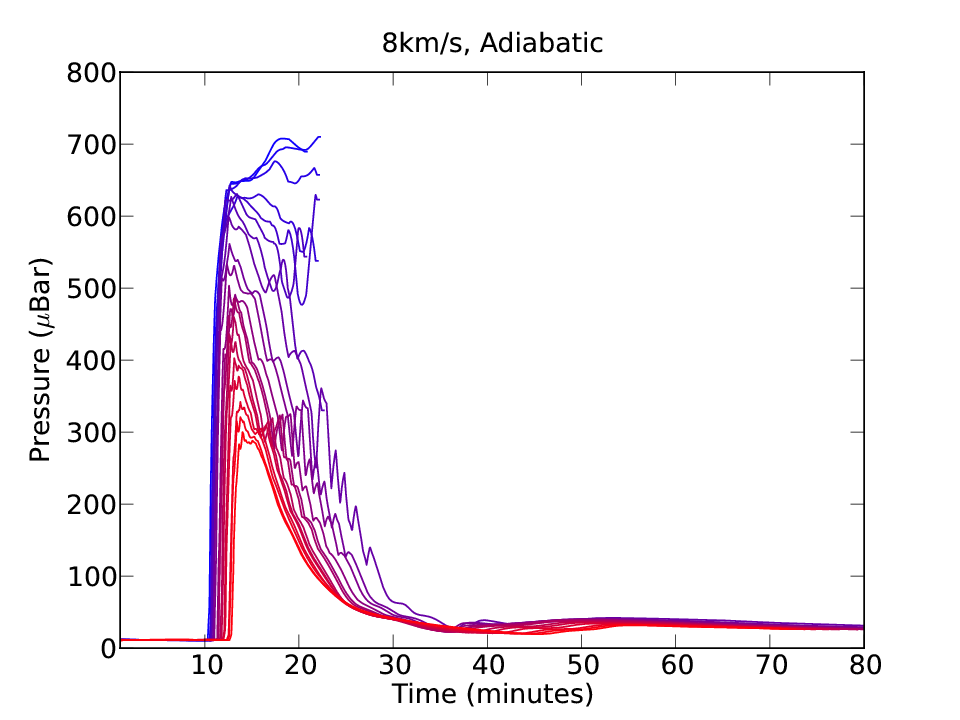}
\includegraphics[width=3.25in]{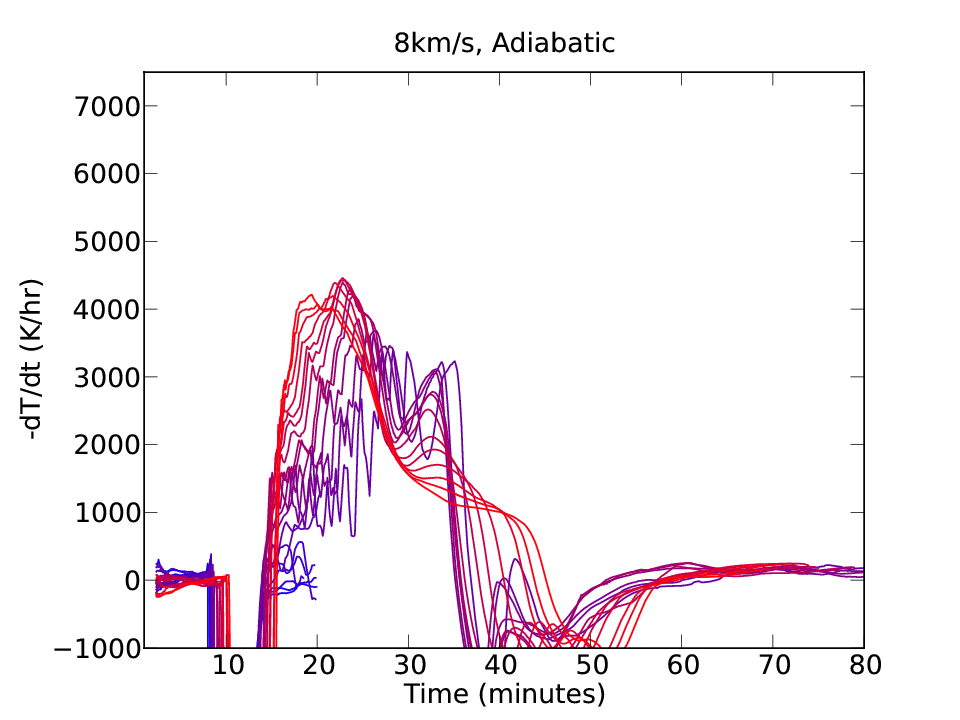}\includegraphics[width=3.25in]{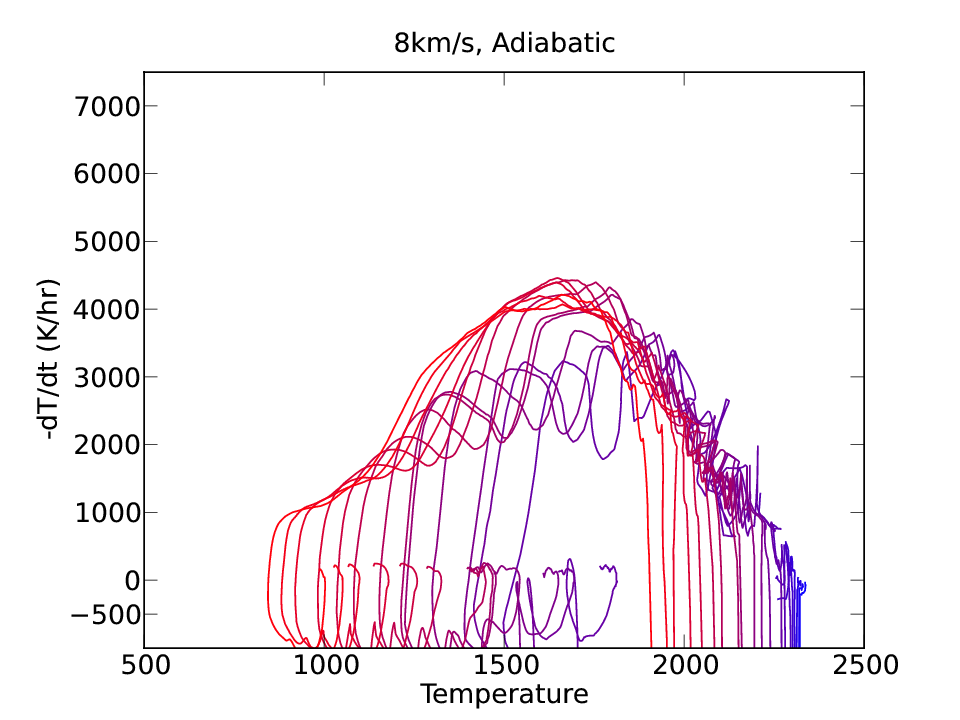}
\caption{Similar to Figure \ref{fig:7adi_prof}, but for the trajectories of the 8 $\kms$ adiabatic shock in Figure \ref{fig:8adi}
 \label{fig:8adi_prof}}
\end{figure}

After the particles have passed through the bow shock and have cooled in the post-bow shock region, they encounter the tail shock, which is produced as gas is diverted around the planetoid toward the wake. 
For this particular case, some particles drop below 1400 K, roughly the lower limit of the crystallization temperature, for a minute or two before they are heated by the tail shock back to or above 1500 K.
The cooling rate in the post-tail shock region then becomes significantly lower  ($\sim 100$ K/hr) than in the  post-bow shock.
While the initial cooling is very fast, the post-tail shock region can potentially set very protracted periods of cooling for the melt, in the adiabatic limit.  

The pressure profile demonstrates that very high pressures are attained  at the bow shock ($\sim 500\mu$Bar), but drop quickly to tens of $\mu$Bar in about 20 minutes. 
For comparison, the pressure in the pre-shock region is about 10 $\mu$Bar.
While the pressure does increase in the tail shock, it never approaches the pressures seen at the bow shock.  

\begin{figure}[H]
\includegraphics[width=3.25in]{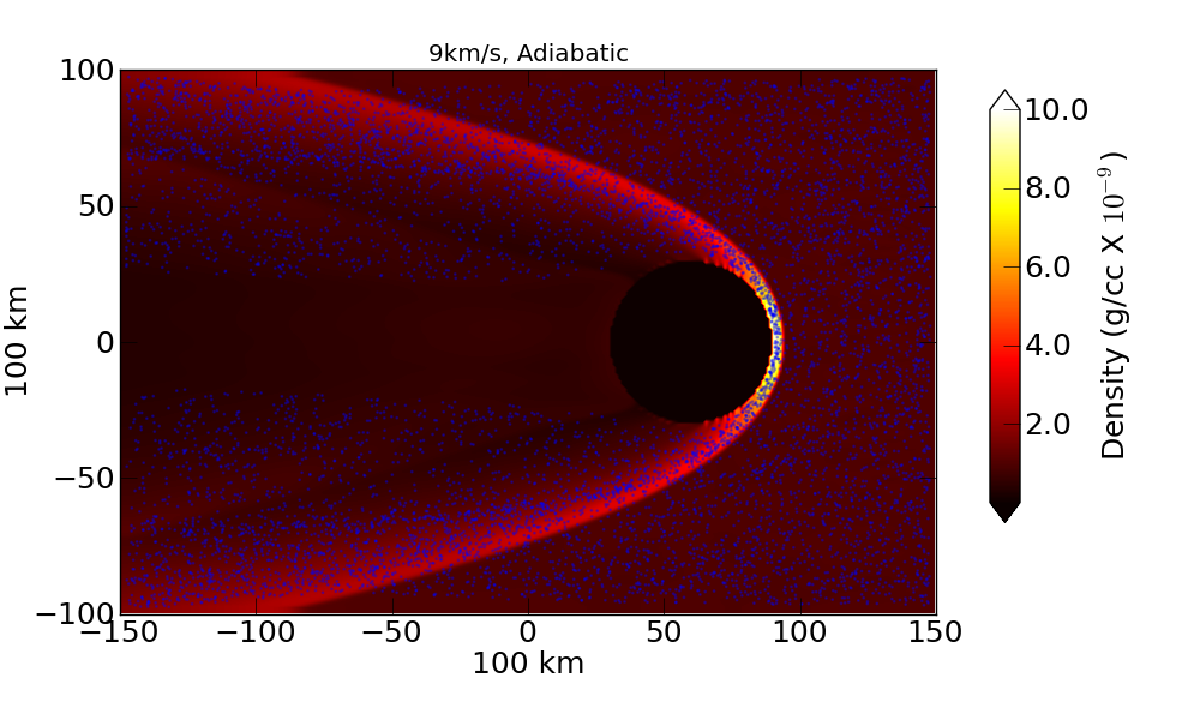}\includegraphics[width=3.25in]{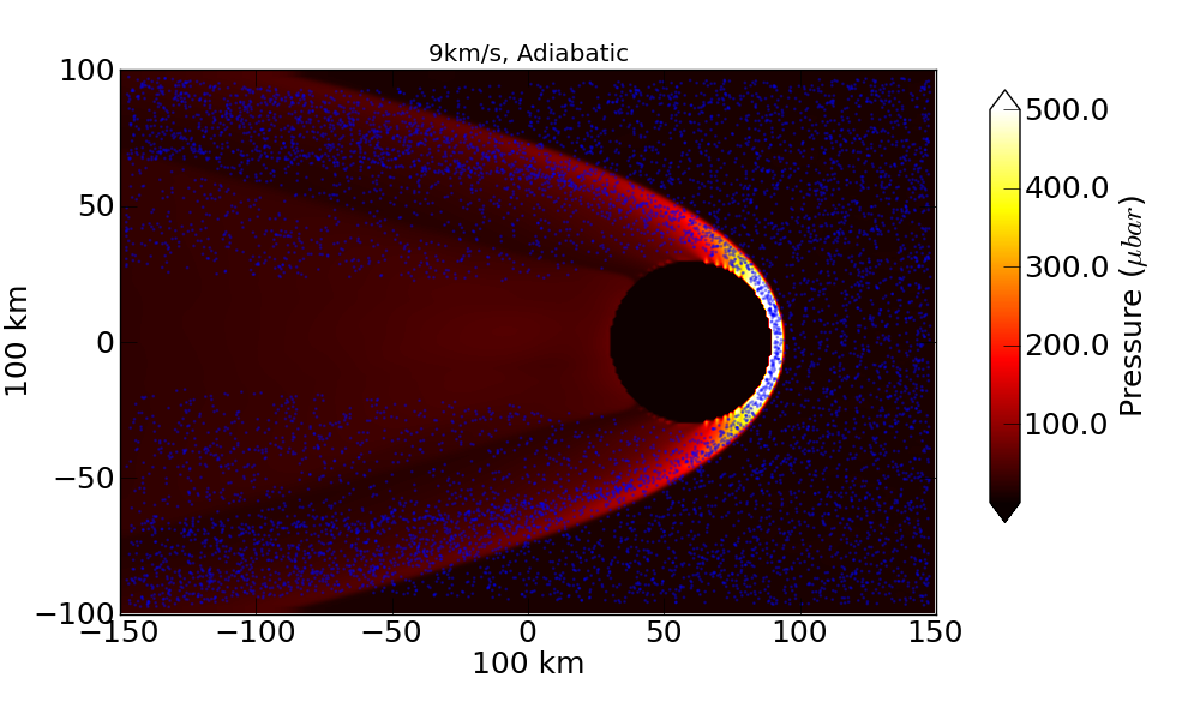}
\includegraphics[width=3.25in]{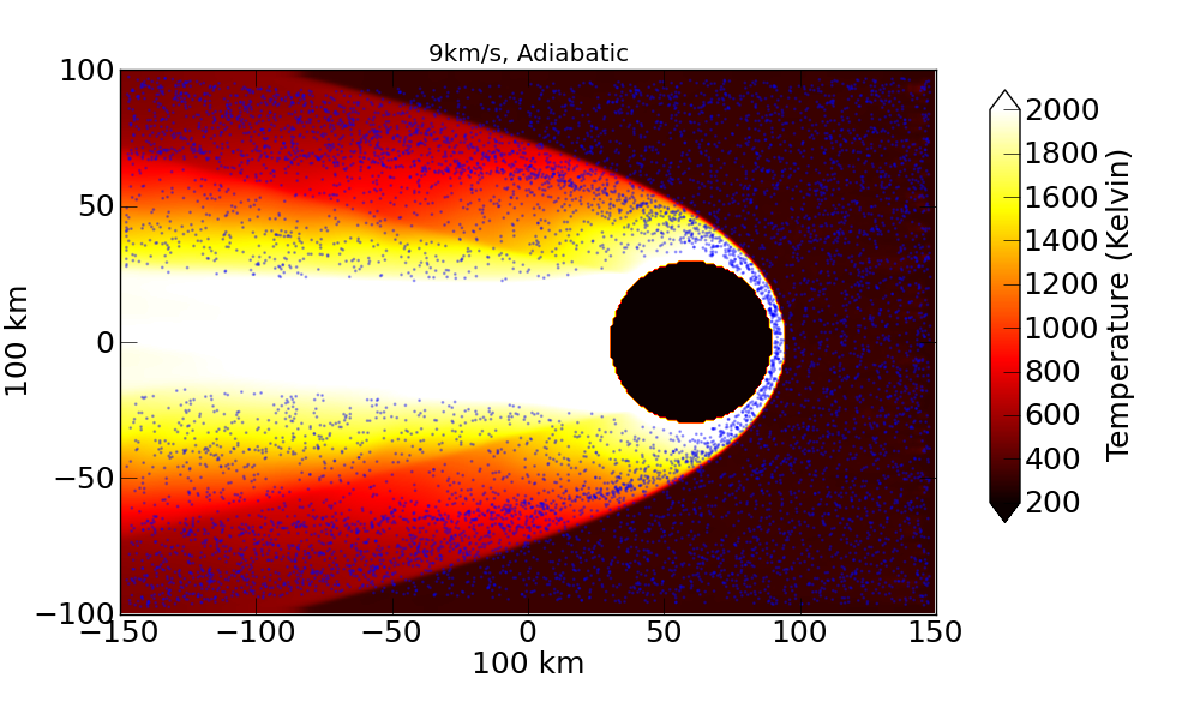}\includegraphics[width=3.25in]{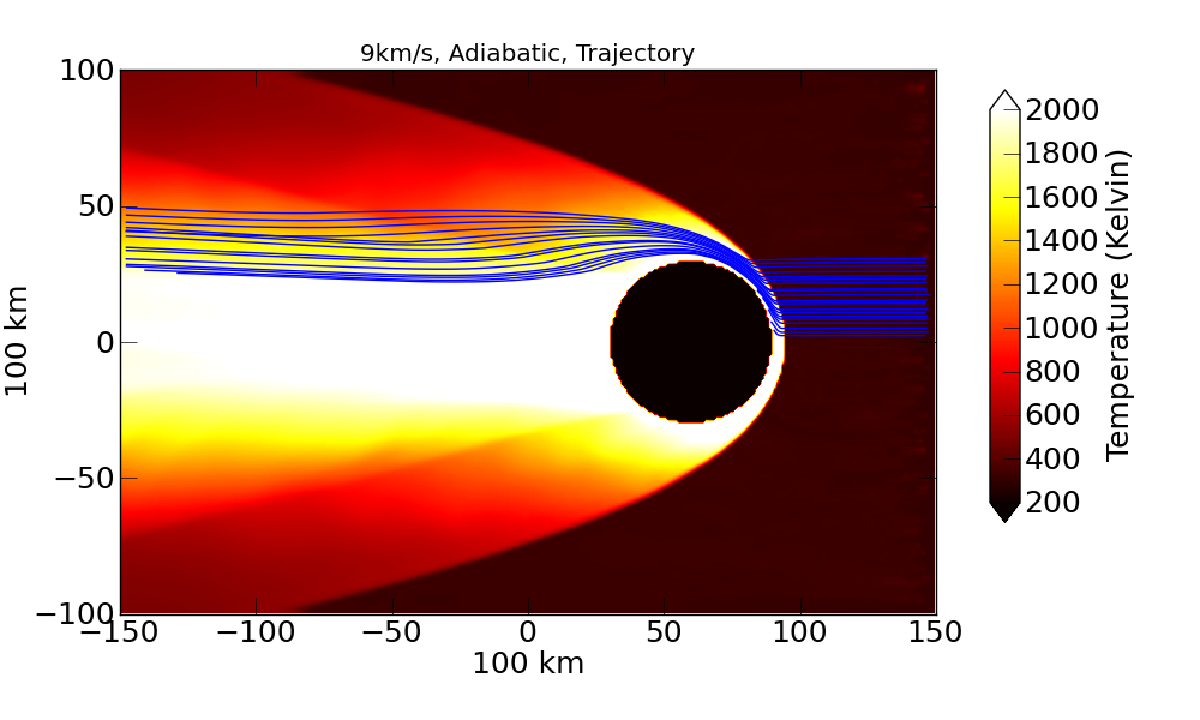}
\caption{Similar to Figure \ref{fig:7adi}, but for a 9 $\kms$ bow shock in the adiabatic limit. \label{fig:9adi}}
\end{figure}

\begin{figure}[H]
\includegraphics[width=3.25in]{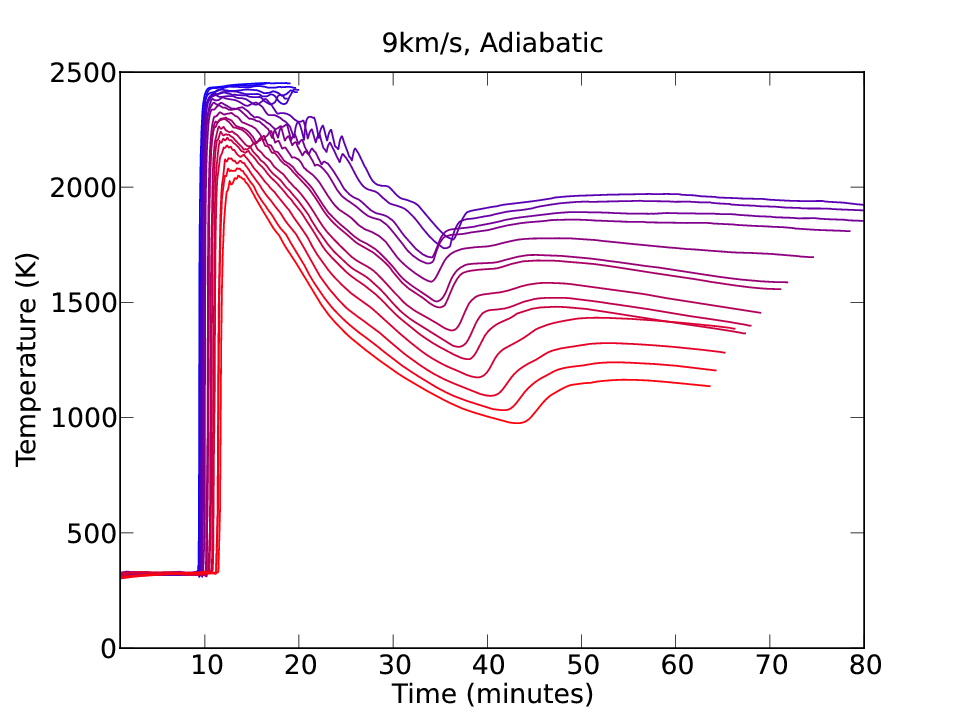}\includegraphics[width=3.25in]{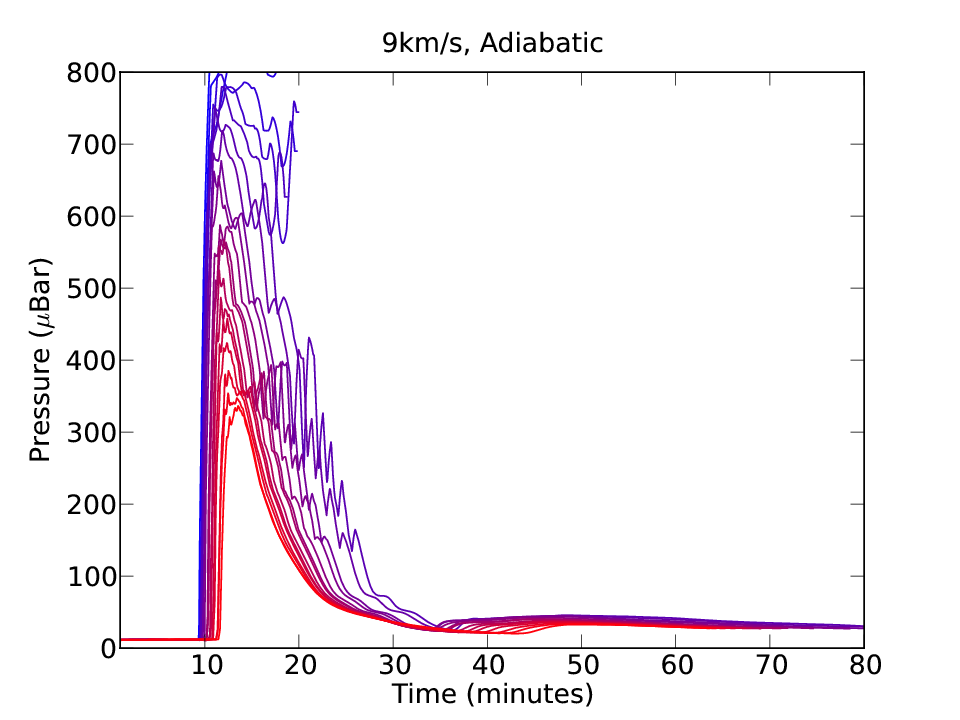}
\includegraphics[width=3.25in]{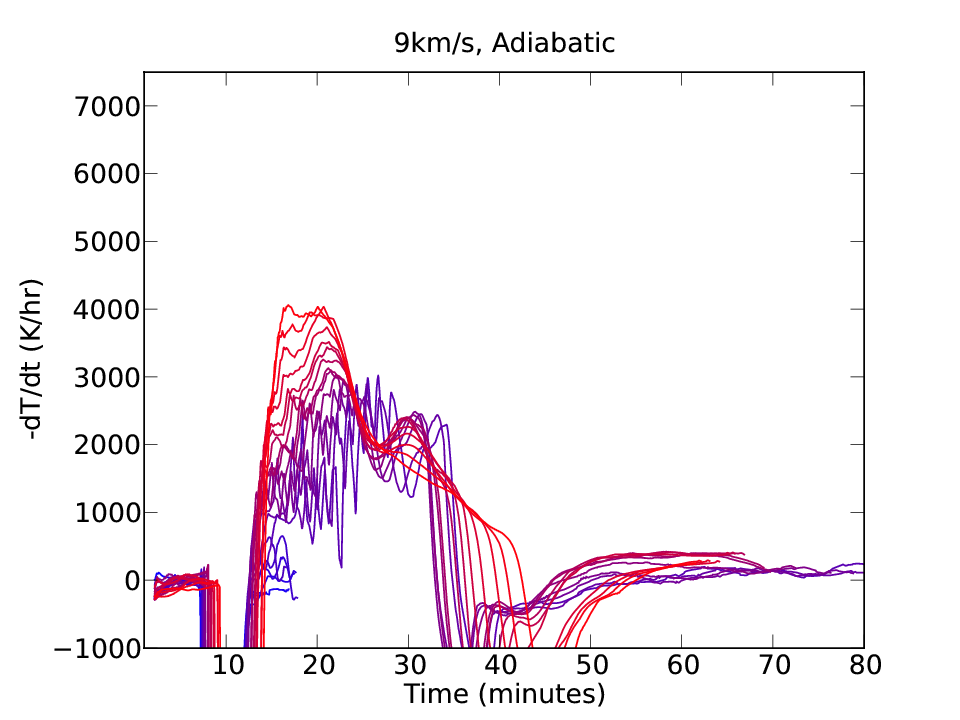}\includegraphics[width=3.25in]{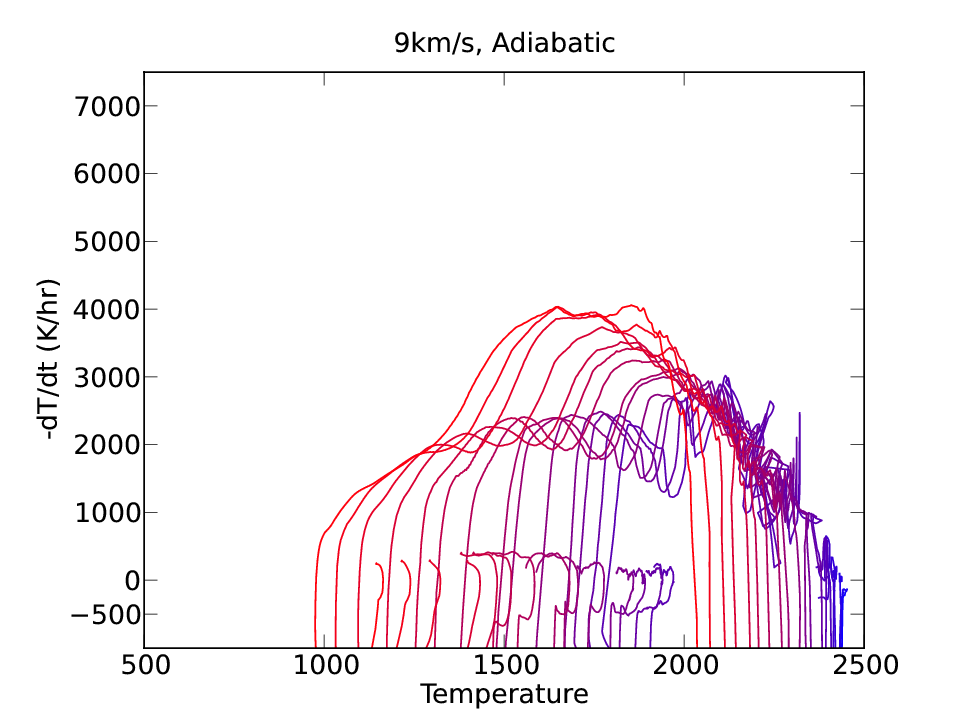}
\caption{Similar to Figure \ref{fig:7adi_prof}, but for the trajectories of the 9 $\kms$ adiabatic shock in Figure \ref{fig:9adi}\label{fig:9adi_prof}}
\end{figure}

The 8 and 9 $\kms$ shocks show very similar morphologies to the 7 $\kms$ shock.  
Figures \ref{fig:8adi} and \ref{fig:8adi_prof} show the results for the 8 $\kms$ shock, and the same is done for the 9 $\kms$ shock in Figures \ref{fig:9adi} and \ref{fig:9adi_prof}.

In each case, we can break the morphology into the same five regions (pre-shock, bow shock, post-bow shock, tail shock, and post-tail shock).
The post-bow shock region again shows very prodigious cooling rates due only to adiabatic expansion. 
At the higher bow shock speeds of 8 and 9 $\kms$, some particles are prevented from dropping below 1400 K, and the tail shocks are stronger than in the 7 $\kms$ case.
Again, the post-tail shock region has very protracted cooling.

These simulations reveal the basic morphology of bow shocks and the heating and cooling that solids will encounter. 
Shock heating and adiabatic cooling make these curves non-trivial, with a striking divergence  from truly 1D shocks, such as those that might occur due to, e.g., large-scale spiral arms. 
We now build on the above complexity by introducing  gray radiative transfer.

 \subsection{3D Radiative Bow Shocks}
  
Having characterized the heating and cooling curves in the 3D, adiabatic limit, we now include radiative transfer using the scheme described in sections \ref{sec:mcrtfld} and \ref{sec:opacity}.  
We assume that any radiation that strikes the surface of the planetoid is reradiated in the direction it came from.
Line cooling is not included at this time.
The only solids that contribute to the opacity are assumed to be chondrule precursors (see sections \ref{sec:methods} and \ref{sec:opacity}).

\begin{figure}[H]
\includegraphics[width=3.25in]{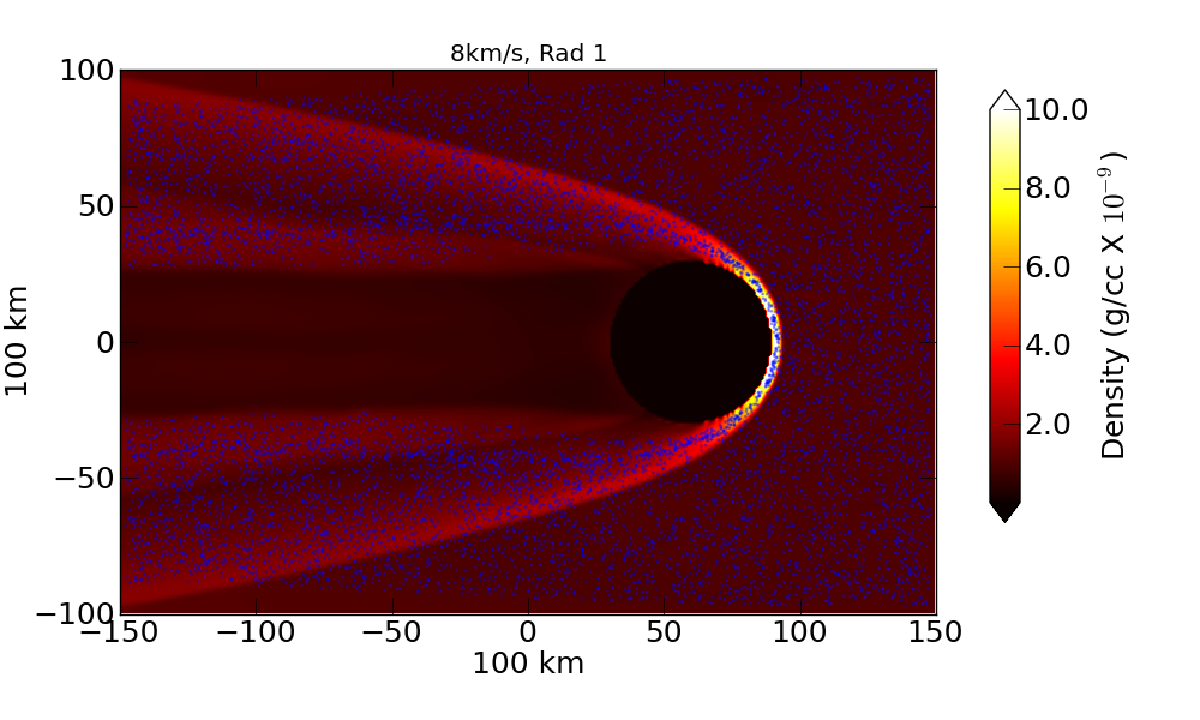}\includegraphics[width=3.25in]{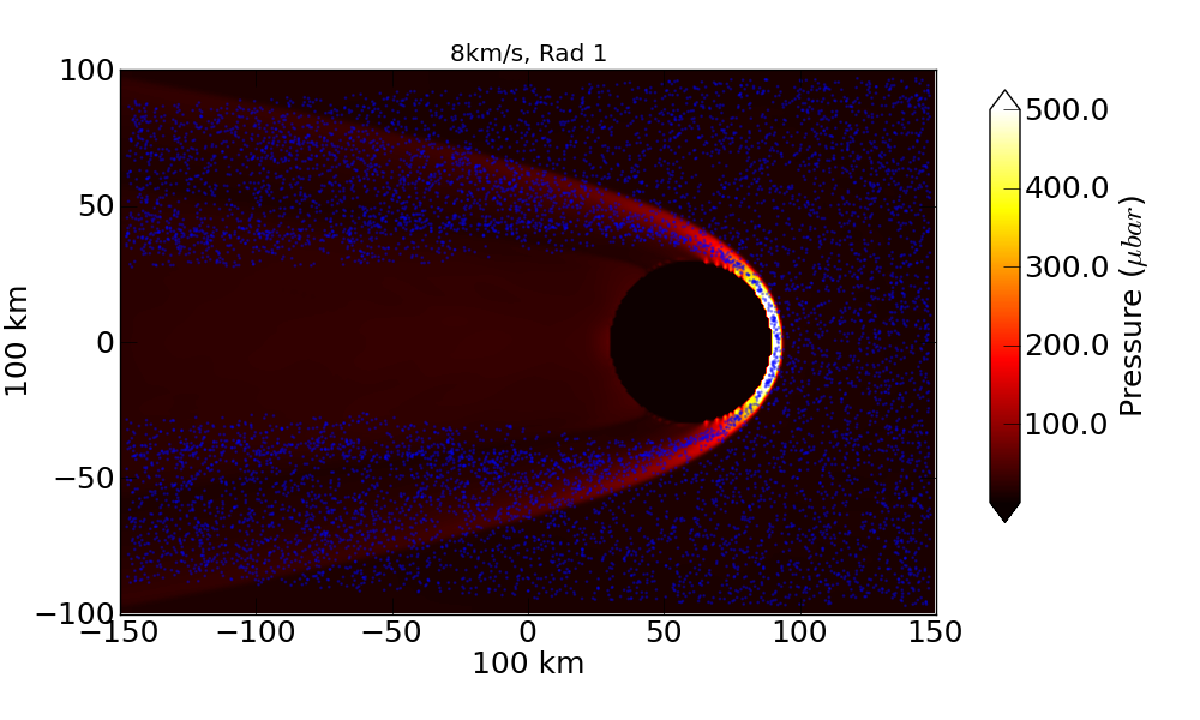}
\includegraphics[width=3.25in]{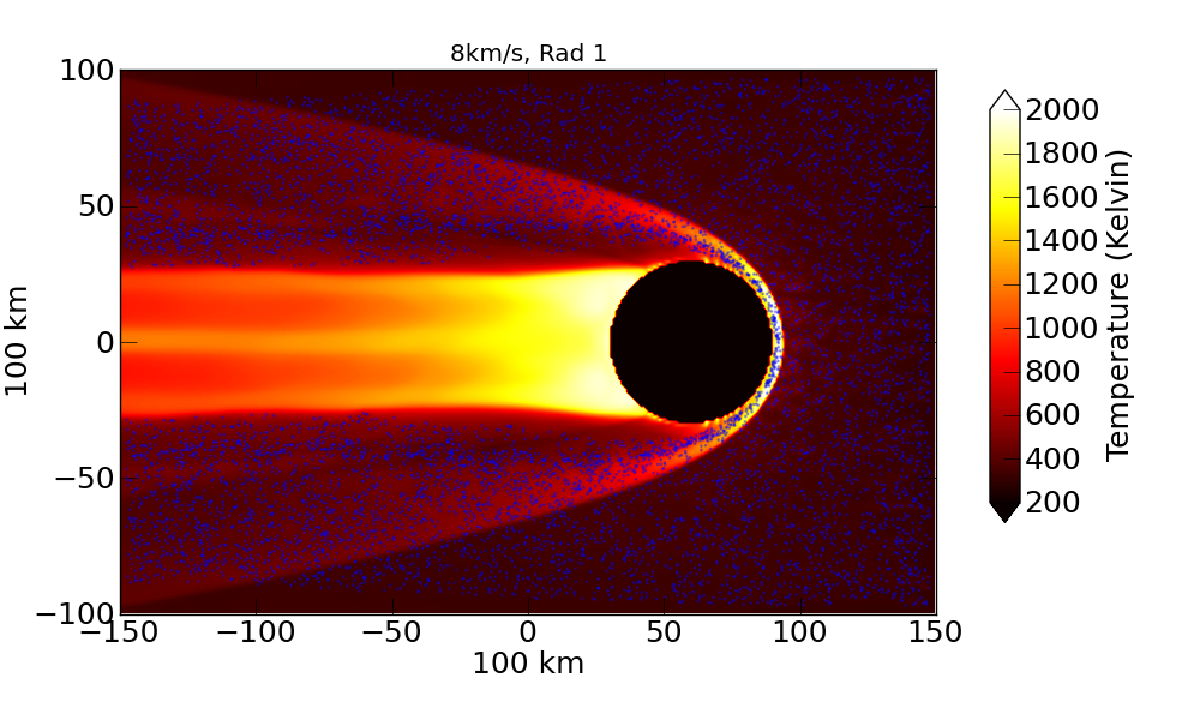}\includegraphics[width=3.25in]{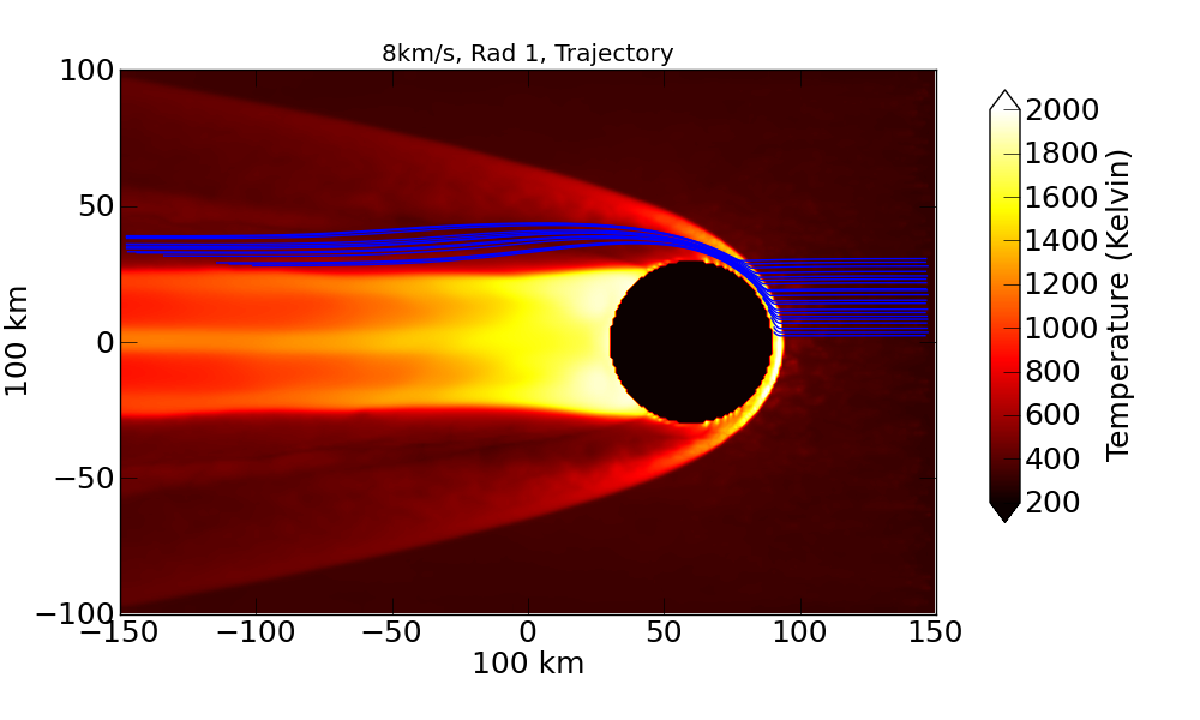}
\caption{Similar to Figure \ref{fig:7adi}, but for a 8 $\kms$ with the standard opacity (Rad 1). \label{fig:8rad}}
\end{figure}

\begin{figure}[H]
\includegraphics[width=3.25in]{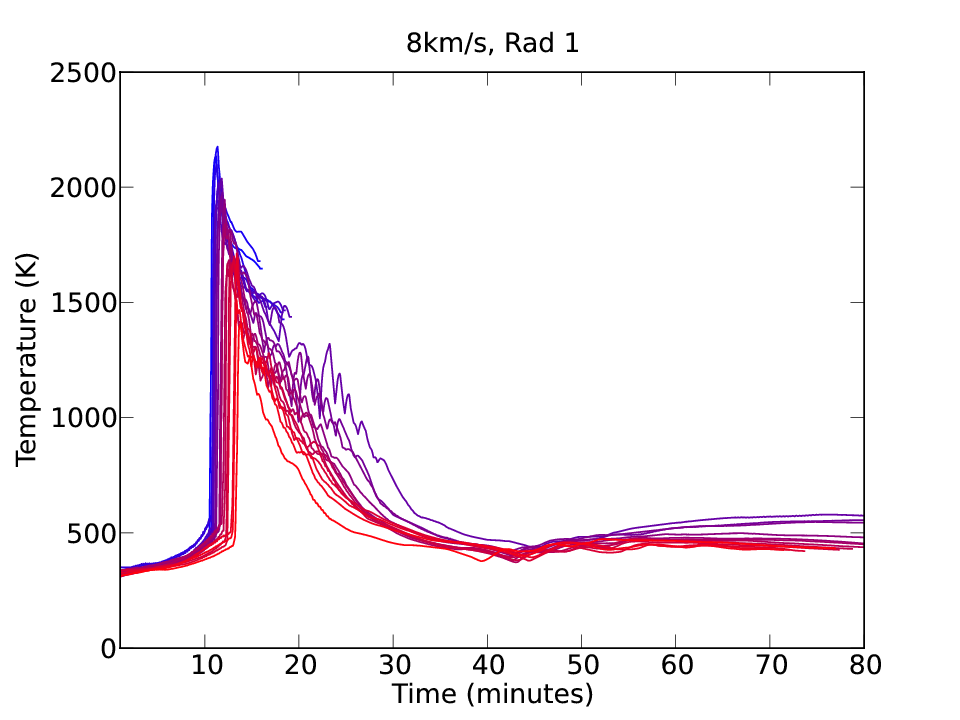}\includegraphics[width=3.25in]{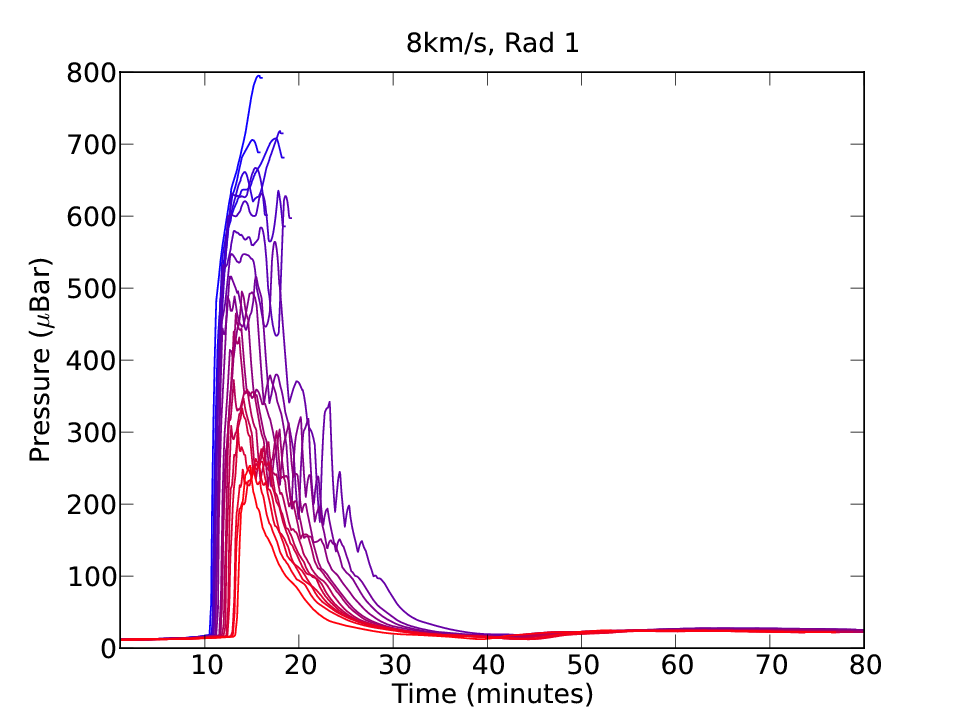}
\includegraphics[width=3.25in]{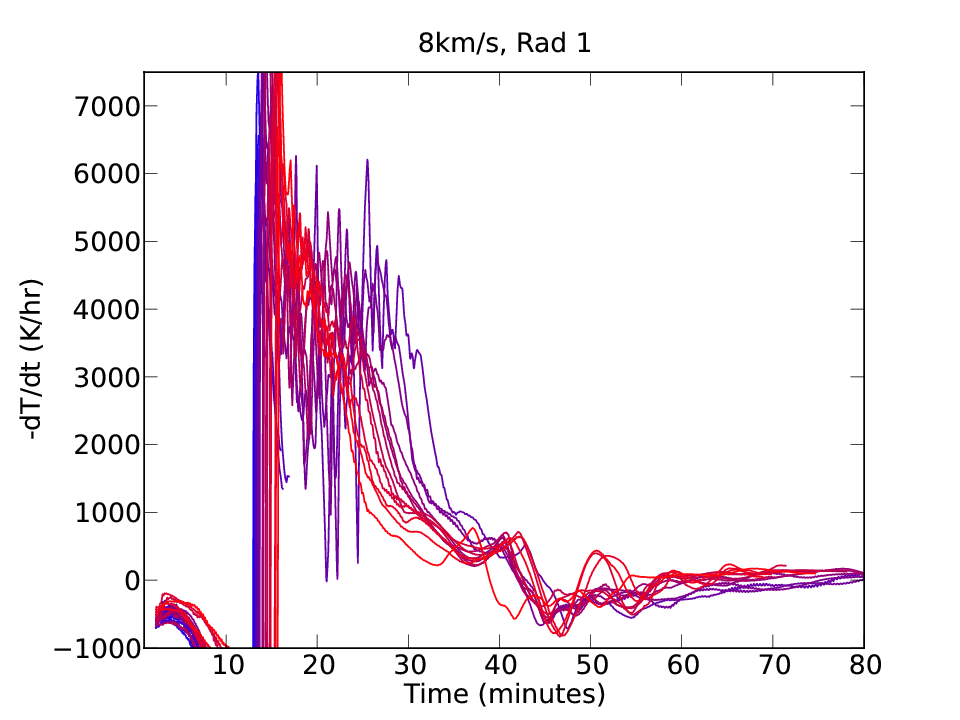}\includegraphics[width=3.25in]{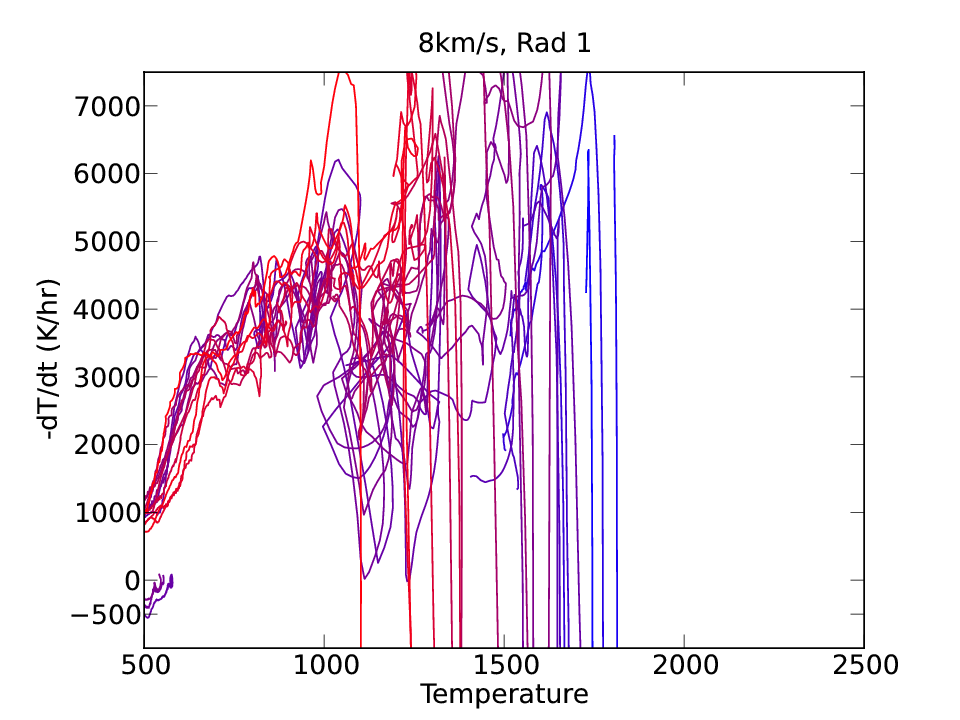}
\caption{Similar to Figure \ref{fig:7adi_prof}, but for the trajectories of the 8 $\kms$ Rad 1 shock in Figure \ref{fig:8rad} \label{fig:8rad_prof}}
\end{figure}

We present three realizations of bow shocks with radiative transfer.  
Each realization is for an 8 $\kms$ shock, but with chondrule precursor mass opacities of $\kappa_c=1$, 10, and 100 cm$^{2}$ per gram of solid, labeled Rad 0.1, Rad 1, and Rad 10, respectively. 
The naming convention reflects the reduction/enhancement of the opacity relative to the ``standard'' opacity $10$ cm$^{2}$ per gram of solid, which is based on our assumed precursor concentration.
Figures \ref{fig:8rad} and \ref{fig:8rad_prof} show the results for Rad 1, where the plots are similar to Figures \ref{fig:7adi} and \ref{fig:7adi_prof}, respectively.

  The standoff distances for Rad 1 are between 0.1 and 0.13 $R_e$ compared with 0.17 $R_e$ for the 8 $\kms$ adiabatic case.
  The highest temperatures are reached at the bow shock, and unlike the adiabatic cases, the temperatures drop immediately, even when approaching the planetoid. 
  Figure \ref{fig:8rad} shows that the opening angle is smaller in Rad 1 than in the adiabatic case due to the radiative case's lower temperatures in all regions behind the bow shock. 
  A tail shock still forms, but it is much weaker than in the adiabatic simulations.  
  The only region behind the bow shock that remains at very high temperatures is the dust-free wake, which cannot radiate due to the lack of dust. 
  
  Figure \ref{fig:8rad_prof} shows the trajectories for 20 solids, as done for Figure \ref{fig:7adi_prof}.
  The curves show that all solids experience a radiative precursor, but this precursor does not increase temperatures to more than 500 K. 
  The radiative bow shock produces a striking spike in temperature, with a sudden rise and drop in the pre- and post-bow shock regions, respectively.
  The pressure also shows rapid evolution, but does not drop as quickly as the temperature due to the increase in density in the post-bow shock region.
  While the tail shock produces some mild heating, the post-tail shock region is not substantially hotter than the pre-shock region, having temperatures between 400 and 600 K. 
  
  At the shock front, the cooling rates approach 10000 K/hr and remain above $\sim 4000$ K/hr through the crystallization temperatures. 
  Radiative transfer is extremely efficient at removing energy from the bow shock that would otherwise keep the chondrules about 1400 K for protracted periods.

\begin{figure}[H]
\includegraphics[width=3.25in]{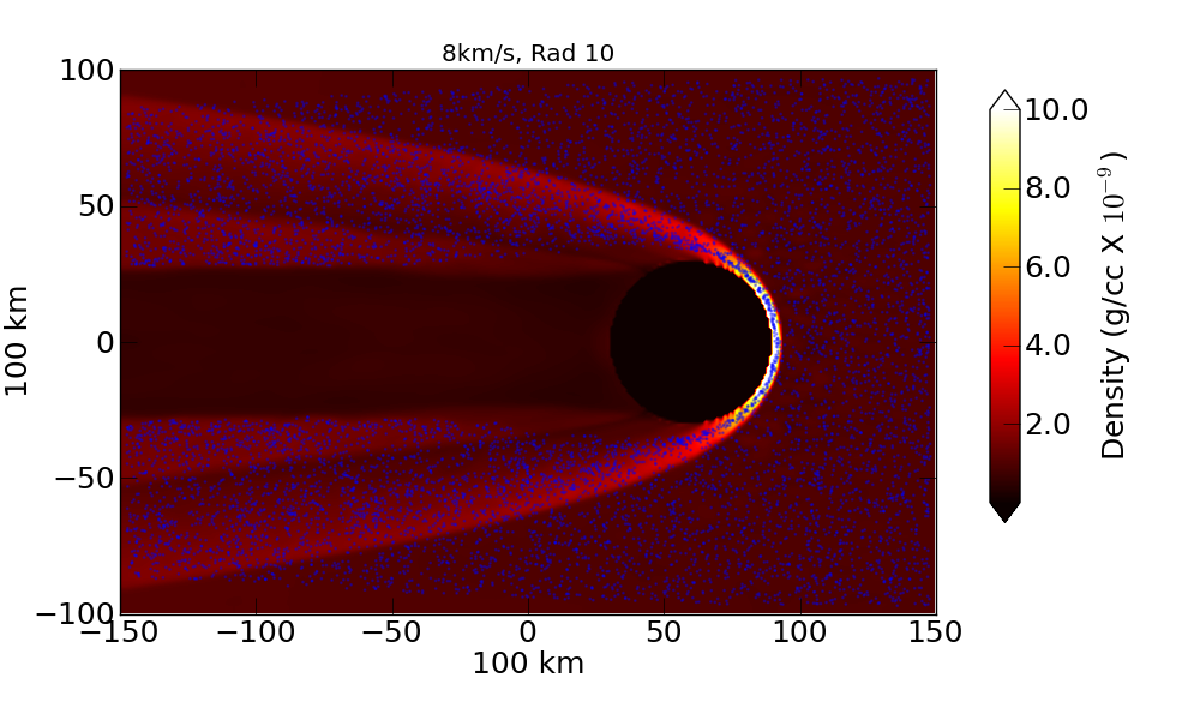}\includegraphics[width=3.25in]{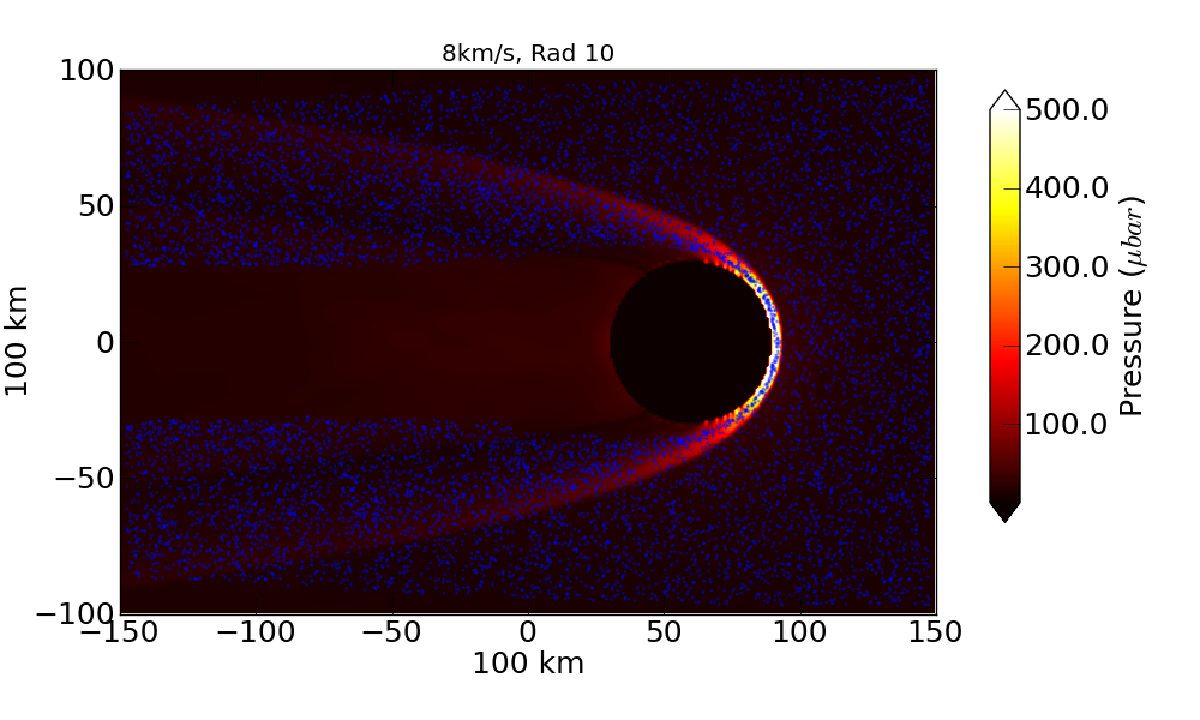}
\includegraphics[width=3.25in]{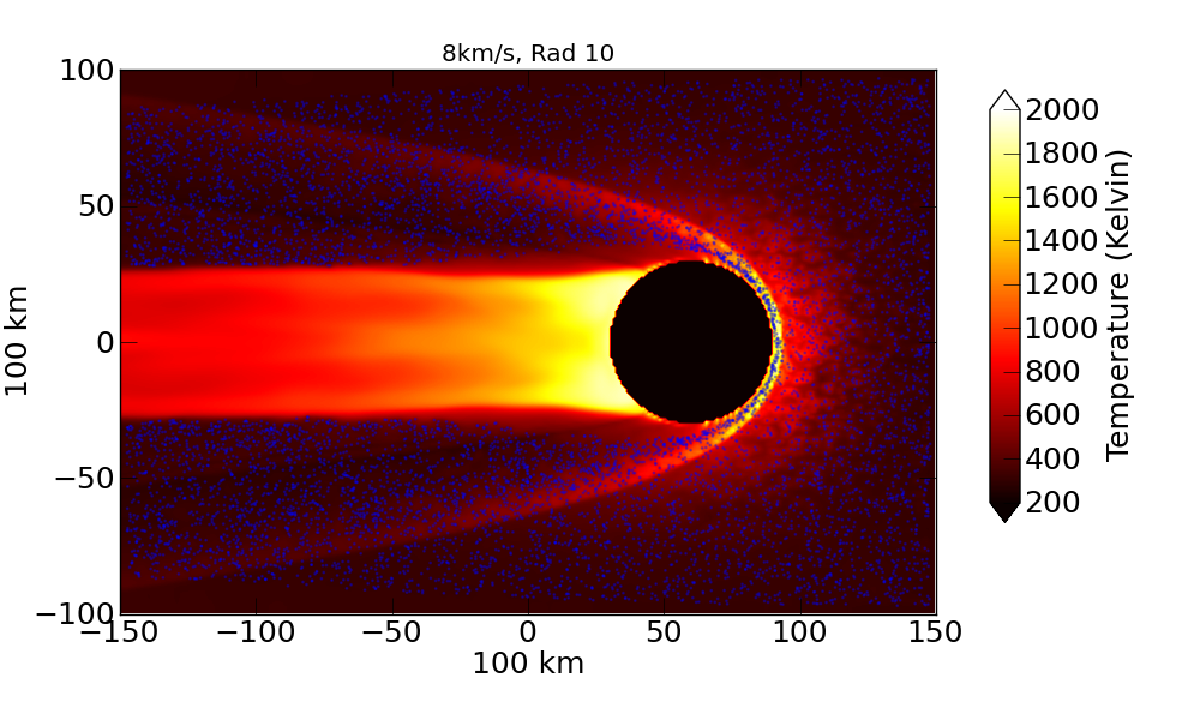}\includegraphics[width=3.25in]{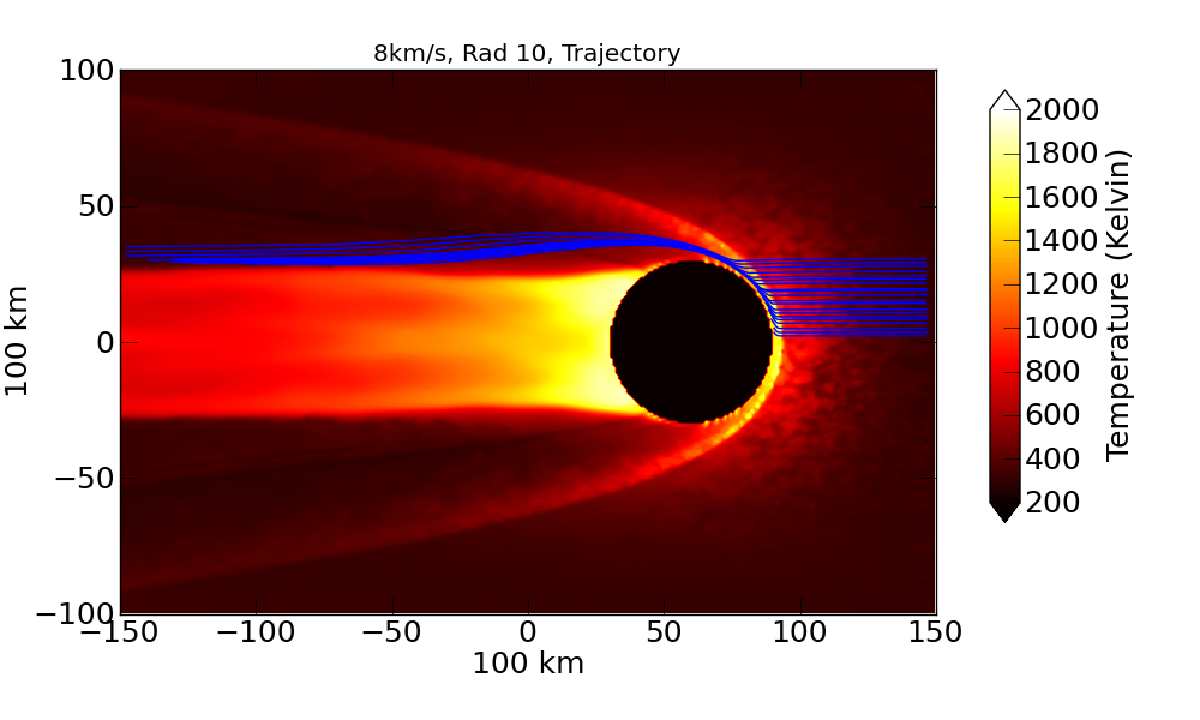}
\caption{Similar to Figure \ref{fig:8rad}, but for ten times the normal opacity (Rad 10).  \label{fig:8rad10}}
\end{figure}

\begin{figure}[H]
\includegraphics[width=3.25in]{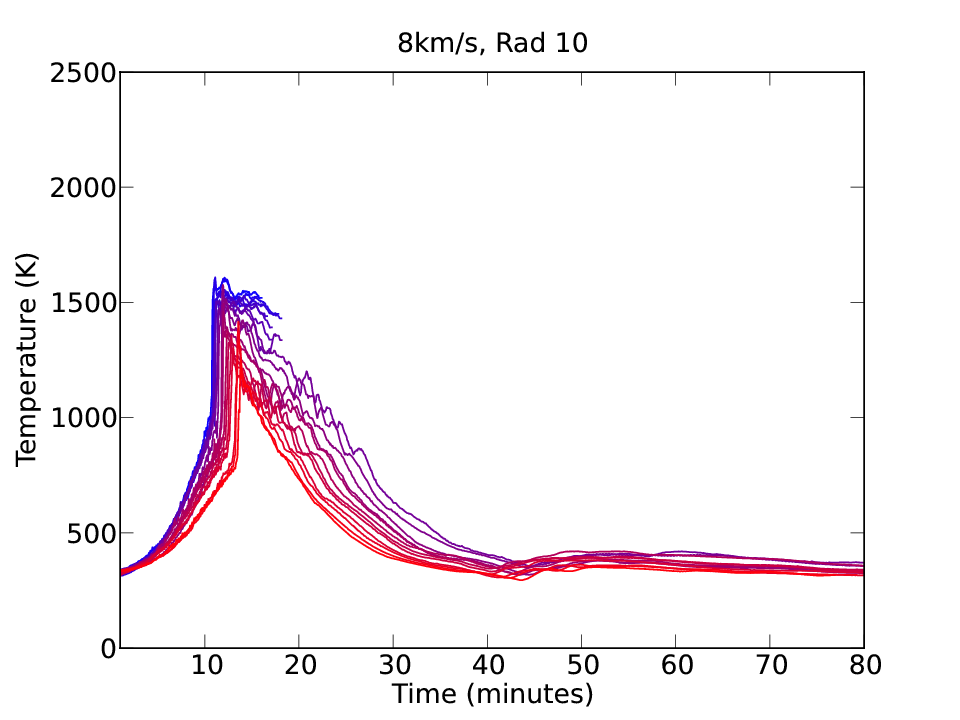}\includegraphics[width=3.25in]{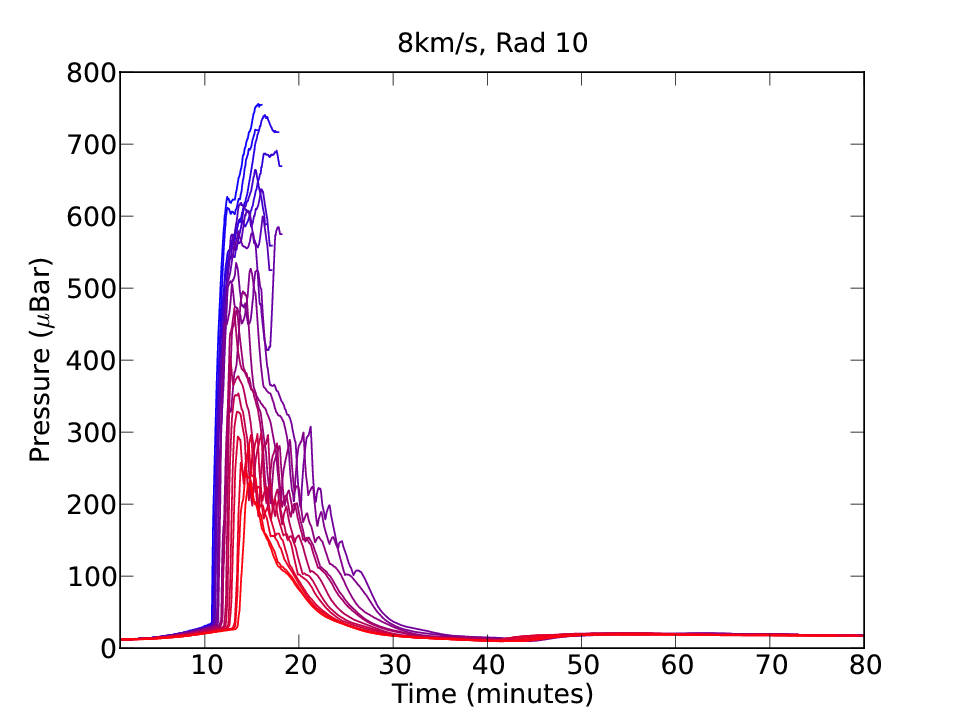}
\includegraphics[width=3.25in]{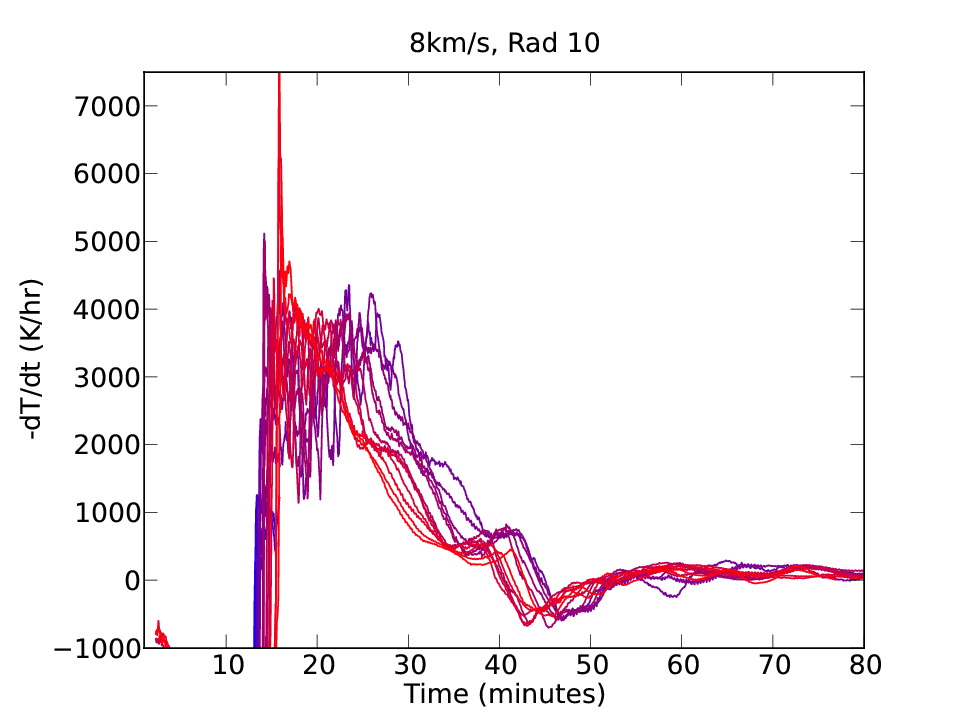}\includegraphics[width=3.25in]{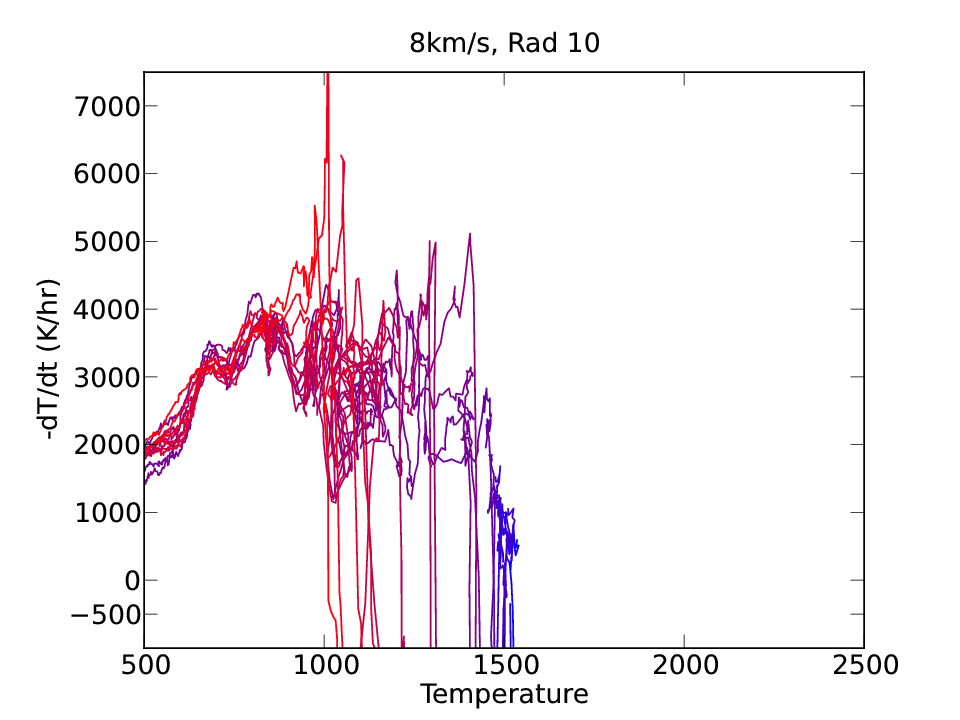}
\caption{Similar to Figure \ref{fig:8rad_prof}, but for the trajectories of the 8 $\kms$ Rad 10 shock in Figure \ref{fig:8rad10} \label{fig:8rad10_prof}}
\end{figure}

Next, we examine the results from Rad 10 in Figures \ref{fig:8rad10} and \ref{fig:8rad10_prof}.
For these simulations, we assumed that the opacity is 100 cm$^2$ per gram of solid, 10 times larger than what we expect for our assumed chondrule concentration.
There are two main changes compared with Rad 1.  First, the radiative precursor is much more significant.  
This can be seen in both the temperature plots in Figure \ref{fig:8rad10} and the temperature profiles for 20 solids in Figure \ref{fig:8rad10_prof}.
However, this radiative precursor also redistributes a significant amount of energy from the shock itself, causing the peak temperatures of the bow shock to reach a value between 1500 and 1600 K, significantly lower than seen in any other run.
The temperature profiles show a radiative precursor that reaches $\sim 1000$ K for some solids just before the shock. 
The pressure spike is very similar to the Rad 1 case, except there is a slight increase in pressure before the bow shock due to the radiative precursor.
The cooling rates are smaller for Rad 10 than Rad 1, but are still between 2000 and 4000 K/hr immediately behind the bow shock.  
However, the cooling rates are from a peak temperature below or barely in the crystallization temperature range, even for an 8 $\kms$ shock.
The post-tail shock region is also cooler than in the Rad 1 case, as this region is able to radiative more efficiently. 

While one might expect a higher opacity to prevent fast cooling, this is only true if the optical depths are already very large.  
The results here suggest that the mass opacity will need to be unrealistically high to reach this limit of long radiative cooling time scales. 
The higher optical depth in the Rad 10 simulation also promotes an efficient propagation of the bow shock energy into the pre-shock region, limiting the peak temperatures of the shock.
However, 
just as very high optical depths can suppress cooling due to long radiation propagation timescales, very low optical depths can also suppress cooling due to a lack of surface area.
The low opacity limit may be more plausible for Solar Nebula bow shocks, and we consider it next.

\begin{figure}[H]
\includegraphics[width=3.25in]{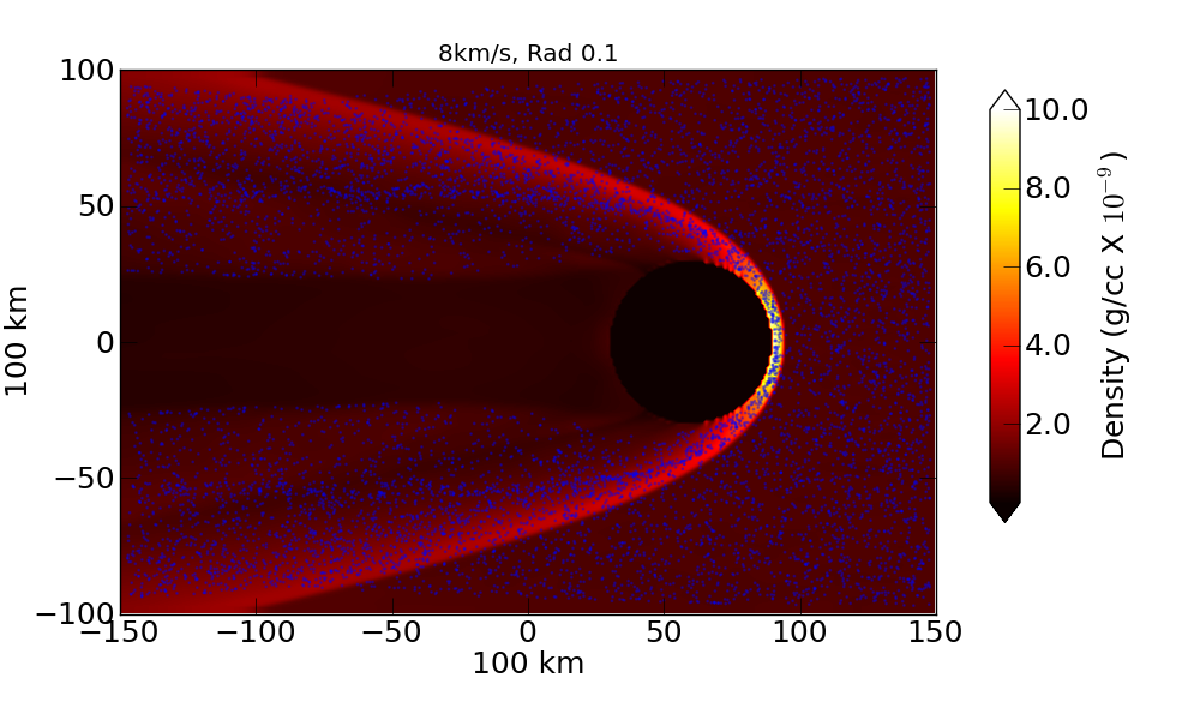}\includegraphics[width=3.25in]{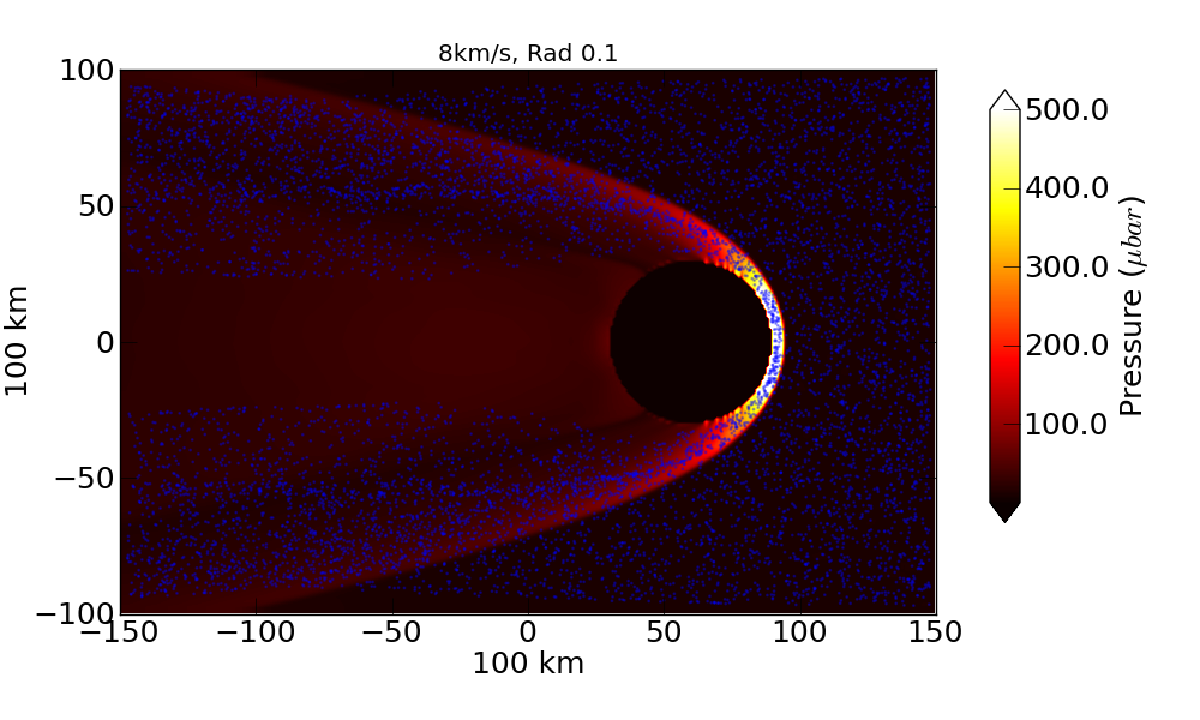}
\includegraphics[width=3.25in]{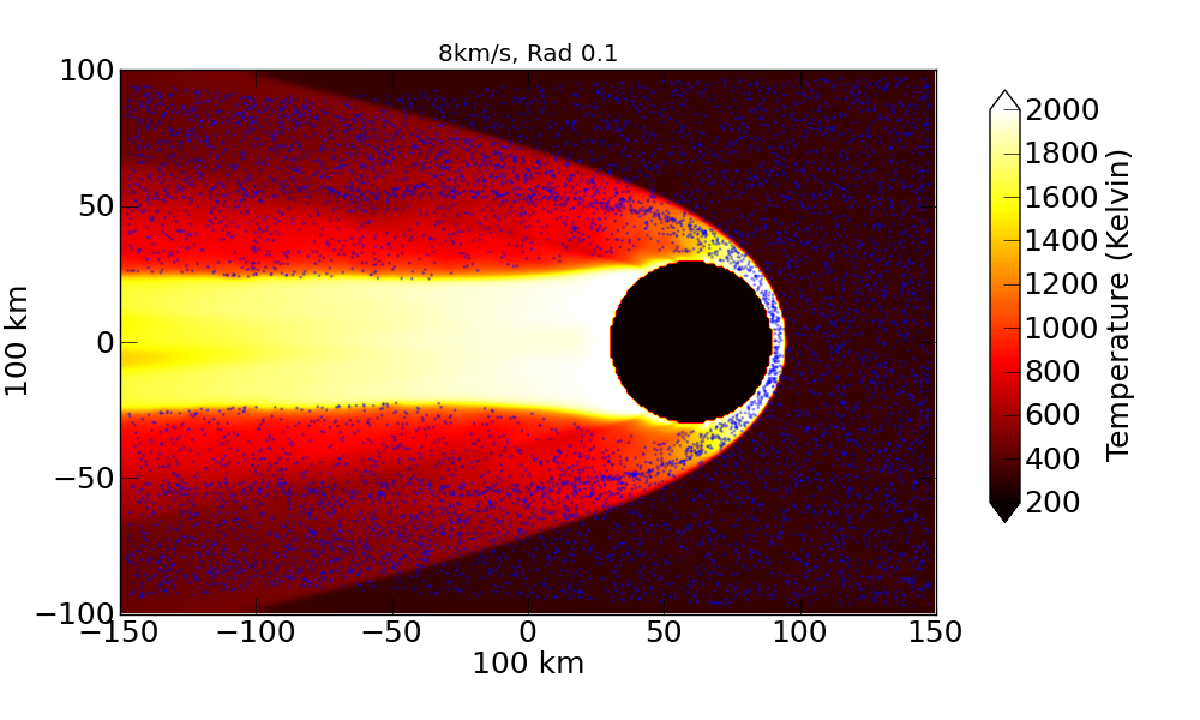}\includegraphics[width=3.25in]{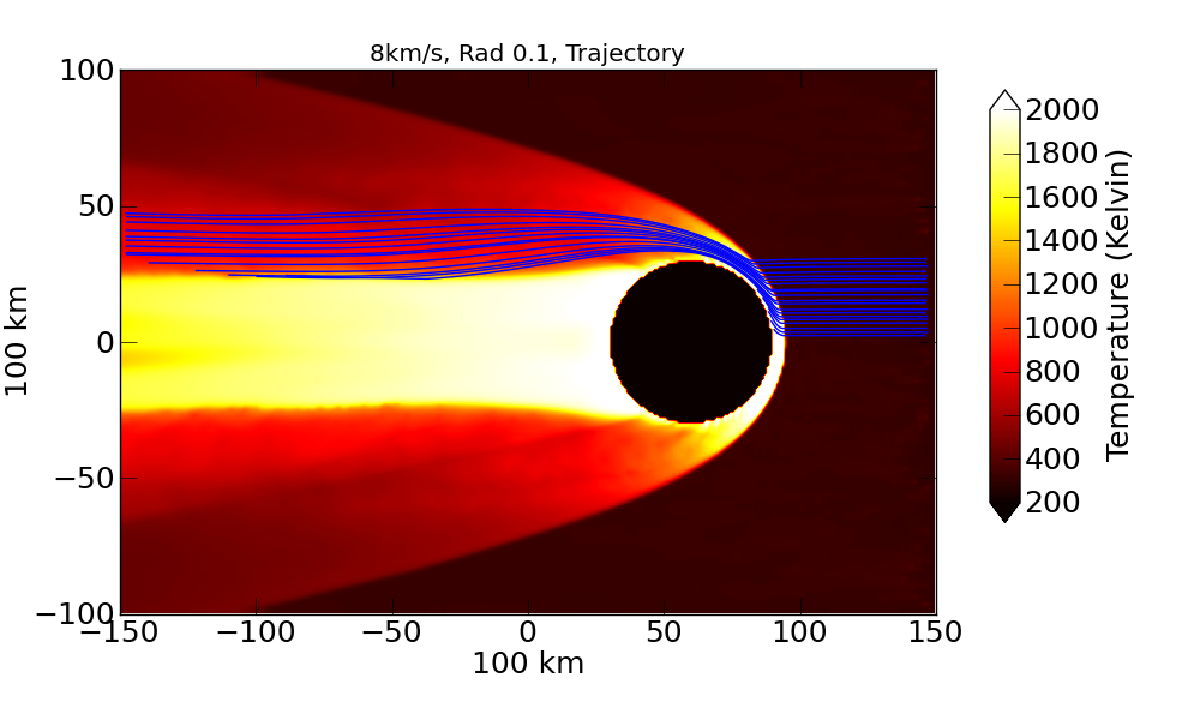}
\caption{Similar to Figure \ref{fig:8rad}, but for one tenth the normal opacity (Rad 0.1).\label{fig:8rad0p1}}
\end{figure}

\begin{figure}[H]
\includegraphics[width=3.25in]{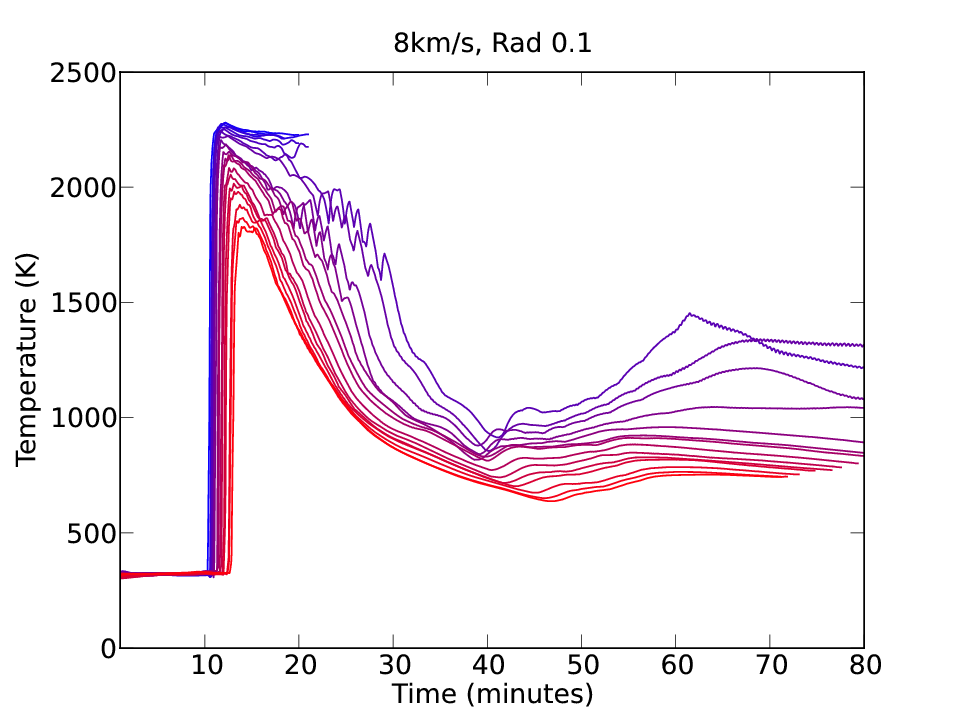}\includegraphics[width=3.25in]{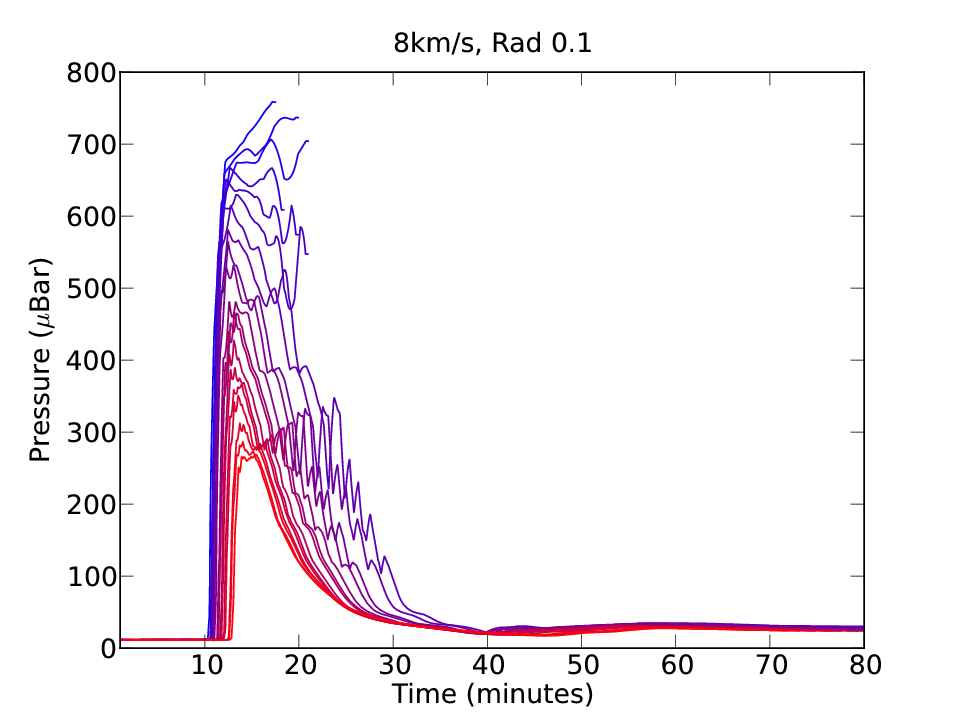}
\includegraphics[width=3.25in]{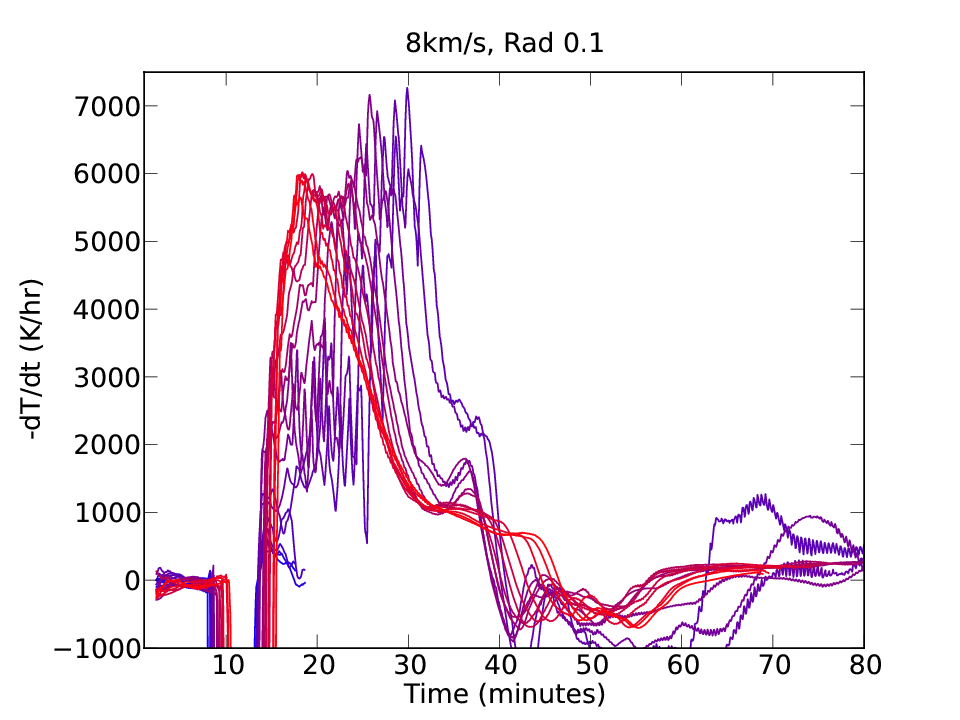}\includegraphics[width=3.25in]{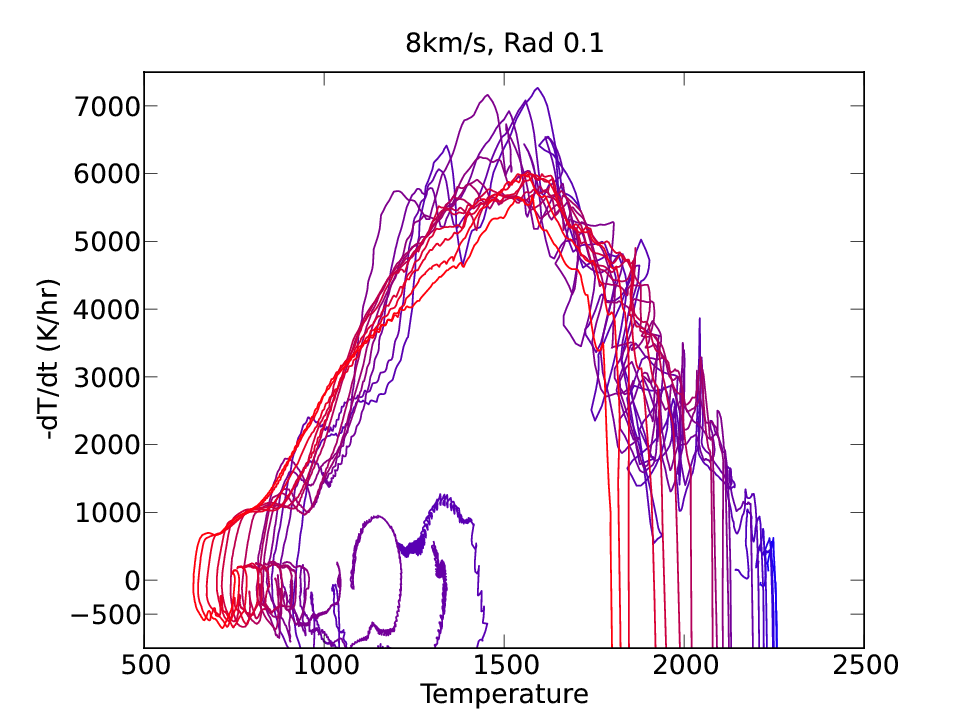}
\caption{Similar to Figure \ref{fig:8rad_prof}, but for the trajectories of the 8 km/s Rad 0.1 shock in Figure \ref{fig:8rad0p1} \label{fig:8rad0p1_prof}}
\end{figure}

For the Rad 0.1 case, we reduce our standard opacity by a factor of ten, resulting in $\kappa_c=1$ cm$^2$ per gram of solid.  
While it makes the gas more optically thin than in the Rad 1 simulation, it also prevents the gas from radiating efficiency, as the standard opacity is already in the low optical depth regime (see section \ref{sec:discussion}).
Figure \ref{fig:8rad0p1} shows the shock morphology, and it is closer to the adiabatic behavior than either the Rad 1 or Rad 10 simulation.  
The opening angle and standoff distances are similar to those for the 8 $\kms$ adiabatic case.
The tail shock once again becomes very apparent.
Nevertheless,  Figure \ref{fig:8rad0p1_prof} shows that radiative cooling causes a consequential deviation from the adiabatic results.
Cooling through the crystallization temperature range is about 6000 K/hr.  
The tail shock does cause strong heating, but the temperatures do not increase above 1500 K. 
Although lower opacity simulations are not presented here,  it is straightforward to infer that a lower opacity will recover the tail shock heating and post-tail shock cooling seen in the adiabatic limit. 

In the next section, we discuss some of the implications of these results.

 \section{Discussion\label{sec:discussion}}

The bow shock model has several appealing aspects regarding chondrule formation.
\begin{enumerate}
\item {\it Bow shocks \ACBc{may} naturally produce  volatile-rich environments.}
 Not all chondrules make it through the bow shock; namely, precursors near the centroid of the planetoid's cross section are likely to be accreted.  
Before their accretion, however, the high temperatures they experience should release volatiles into the gas.
This loss of refractory material to the planetoid and the corresponding enrichment of the post-bow shock gas in volatiles may explain the high partial pressures in volatiles needed to explain chondrule compositions \citep{alexander_etal_2008}.
This would be in addition to any volatile outgassing that should be expected from the planetoid itself, which could be substantial as the planetary embryo should retain large amounts of live ${}^{26}{\rm Al}$ and could possess a magma ocean (M2012). 

\item {\it Bow shocks naturally lead to a maximum chondrule size.}
The sizes of chondrules in chondrites follow an approximate log-normal size distribution, with 
a sharp dropoff in abundance at radii $> 1$ mm
\citep{kuebler_etal_icar_1999}. 
Chondrules as large as 3 mm are very rare in chondrites \citep[e.g.,][]{weyrauch_bischoff_2012}. 

The dearth of large chondrules could be due to size-sorting during chondrite assembly by, e.g., 
turbulent concentration \citep{cuzzi_etal_2003}, but a physical mechanism must impose an upper limit to chondrule sizes.
\cite{susa_nakamoto_2002} considered the ability of molten droplets to remain intact upon entering
a shock and concluded that chondrules with radii $>$ about 2 cm are unstable in an $8~\kms$ shock.
Bow shocks offer a different mechanism imposing an upper size limit. 
As discussed in section {\ref{sec:introduction}}, the standoff distance must be comparable to or 
larger than the stopping distance of a chondrule precursor for the precursor to avoid being accreted 
by the planetary body.  
For chondrule precursors larger than a few millimeters in diameter, the stopping distance will be 
much too large, causing accretion of these solids.
This mechanism may also act to remove large chondrules from Solar Nebula whether or not 
chondrules {\it form} in planetary embryo bow shocks.

\item {\it Bow shocks generate a range of heating profiles.}
Although most of the material in a chondrite has been processed to some degree, not all of the 
material has gone through the same high temperatures as chondrules. 
This range of thermally processed material is naturally reproduced in bow shocks, in which the 
shock strength decreases rapidly with impact parameter.
Even in the $9~\kms$ adiabatic bow shock, the shock temperatures at impact radii larger than about 
2 $R_e$ never exceed those required for melting and re-crystallization.  
This allows some material to be highly processed and other material to escape high-temperature 
alteration, all in a spatially coherent region. 
Whatever process acts to produce parent bodies from these solids, as long as the mechanism is not 
100\% efficient, then some chondrules will survive to be reprocessed during multiple passages of the 
planetoid as its eccentricity is slowly damped over time (M2012).

\item{\it Bow shocks concentrate chondrules.} 
Production of compound chondrules, i.e., chondrules stuck together while molten, requires very
high concentrations of chondrules \citep[see][and references therein]{desch_etal_mps_2012}.
Because chondrules not accreted by the planetary body are diverted around it,
they pass through a relatively smaller area (annulus) around the planetary body.
\ACBc{For example, based on the streamlines shown in Figures 2, 4, 6, 8, 10, and 12, the annulus is roughly the size of the standoff distance of the shock (in some cases even smaller). 
The resulting concentration is thus between 3 and 5 for standoff distances 0.15 and 0.1 $R_e$, respectively.   
This is in addition to the concentration expected from the shock alone (roughly a factor of 10). 
This additional concentration increases the
probability of collisions, which should increase the fraction of compound chondrules, as discussed by, e.g., M2012.}

\end{enumerate}

In addition to these aspects of chondrule melting in bow shocks, there are other attributes that are quite novel and difficult to assess.
Among the main results of this work are that the three-dimensional structure of the bow 
shock gives rise to multiple shocks (one bow shock and at least one tail shock) and that adiabatic 
expansion behind the bow shock can lead to rapid cooling.
 Even adiabatic simulations can produce prodigious initial cooling rates.
Moreover, the temperature history of solids behind the bow shock is not guaranteed to be monotonically decreasing, as multiple heating events are a natural outcome of bow shocks. 
The tail shock can thus give an effectively protracted cooling history.
In the {\it adiabatic limit}, bow shocks remain a possible mechanism for forming  various chondrule types, with a range of heating and cooling histories possible for any one bow shock. 
However, even in this limit, there are multiple curves that would not obviously form chondrules.
 {\it The cooling curves derived from the simulations here should be explored in laboratory experiments} to determine whether any are capable of forming chondrules, and if not, to determine what types of solids are produced by such bow shock passages.

Runs with radiative transfer further highlight the complexity of bow shocks, and place considerable emphasis on the role opacity plays in the cooling profiles. 
Radiative bow shocks reduce the effectiveness of the tail shock in heating the gas, which allows the temperature of the solids to drop below crystallization temperatures shortly after initial shock passage.
The thermal histories of solids again become essentially monotonic, as seen in 1D models, but have very high cooling rates at greater than several 1000 K/hr.  
The standard case has a low enough opacity for a strong shock to develop, but radiative processes quickly cool the entire region behind the bow shock.

\ACBc{{\it Are the cooling rates in the radiative simulations sensible?} To evaluate whether the rapid cooling seen in the radiative simulations is physical or an artifact of the simulation}, consider the chondrule opacity for the standard case of $\kappa_c=10$ cm$^2$ per gram of solid. 
We have assumed a chondrule mass fraction $\eta_c=0.04$. 
The pre-shock gas density is $10^{-9}$ g/cc, and the average chondrule density is $4\times10^{-11}$ g/cc. 
The distance a photon will travel before reaching an optical depth of unity is $1/(\kappa_c \eta_c \rho_g)$, which is 25000 km, or about 8 $R_e$.
The pre-shock region is not very efficient at absorbing photons\ACBc{, consistent with the lack of any sizable radiative precursor in the standard case.}
At an opacity 10 times larger than the standard case (Rad 10 run), \ACBc{an optical depth of $\delta \tau\sim1$ through} the pre-shock region becomes comparable to the $R_e$, which is apparent in the radiative precursor in Figures \ref{fig:8rad10} and \ref{fig:8rad10_prof}.
When the opacity is lowered by a factor of ten (Rad 0.1 run), radiative cooling and heating becomes much less efficient, i.e., although the photons can stream freely from any emitting solid, the reduction of the number of emitting solids prevents the gas from cooling quickly.  
As such, contact with the gas can keep any solid at a high temperature.  

The optical depth through the standoff distance offers another way of characterizing the optical depths of the bow shock.  For a standoff distance of about 0.15 $R_e$ and post-bow shock gas and chondrule densities about 10 times higher than in the pre-shock region, the optical depth through the bow shock is approximately \ACBc{0.02, 0.2, and 2 for Rad 0.1, Rad 1, and Rad 10, respectively.  
Although the bow shock  optical depth in the Rad 10 simulation is approaching the optically thick regime, the pre-shock region is still too optically thin to prevent rapid cooling of the gas. }

\ACBc{The energy that is radiated from each bow shock has a total luminosity between 2 and $3\times10^{-8}$ L$_{\odot}$ for the Rad 0.1 and Rad 10 cases, respectively, making these shocks potentially observable.
This luminosity is comparable to the expected heating by the bow shock ($\sim \rho_g V_w^3 \pi R_e^2$), showing that most of the energy dissipated in the shock is radiated away. }

\ACBc{In the calculations presented here, we do not include additional radiation from the planetoid. 
If a planetoid has a magma ocean, its surface temperature must be at least $\sim 1400$ K. 
In this situation, the radiation of the planetoid itself could alter a solid's temperature history. 
However, the largest solid angle that the planetoid can subtend is 2$\pi$. 
A precursor near the surface of a planetoid will have an  equilibrium temperature of $\sim 1200$ K, assuming a planetoidal surface temperature of 1400 K.
While radiation from a very hot planetoid could keep a chondrule precursor warm for about $\sim 10$ min during the precursor's passage around the planetoid, it is not expected to keep a melt above its crystallization temperature. 
We confirmed this behavior in preliminary simulations, but leave detailed calculations for future work.}

Based on these results presented here, we infer the following: 
Radiative cooling, coupled with adiabatic expansion of the gas, is too efficient at cooling the gas to 
match chondrule cooling rates inferred from current laboratory experiments, at least for 
chondrule mass fractions $\eta_c\sim0.04$.
Increasing the opacity by a factor of ten, which is equivalent to assuming either an $\eta_c\sim0.4$ or significant dust in addition to chondrule precursors, makes the problem worse, as much higher optical depths are necessary to prevent efficient radiative transport of energy at  the bow shock and rapid cooling in the post-bow shock regions. 
Recall that the peak temperatures for the Rad 10 simulation never even rise above the melting temperature.

However, if instead the opacity is reduced by a factor of ten, i.e., assuming a smaller $\eta_c$, then solid heating and cooling profiles begin to approach the adiabatic limit \ACBc{due to inefficient cooling}, which does show promise for forming chondrules.  
At even smaller $\eta_c$ \ACBc{than explored here},
it may be very possible to form chondrules in planetoidal bow shocks, in the manner 
constrained by laboratory cooling experiments.
This direction has \ACBc{potentially significant hurdles}, though.  
At very low $\eta_c$, it may not be possible to obtain partial pressures that suppress isotopic fractionation, and it may not be possible to reach densities that can match the frequency of observed compound chondrules.  
\ACBc{If these potentially serious issues cannot be resolved, then approaching the adiabatic limit by only decreasing the chondrule precursor density  may be incompatible with chondrule formation, even if the cooling curves are consistent. 
Ways to reach the adiabatic limit that are also consistent with chondrule formation  need to be explored.}

While the simulations presented here are run with significant complexity, there are nonetheless improvements that can be made.
In principle the solids and the gas can become thermally decoupled, although
\cite{desch_connolly_2002} and \cite{morris_desch_2010} found that chondrules typically remained, 
at most, a few tens of degrees below the gas temperatures, and had similar cooling rates.
Magnetic fields could potentially affect the structure of the shock, although even for the
higher magnetic field strengths thought to be typical of the Solar Nebula \citep[$B \sim 1 \, {\rm G}$; see][and references therein]{desch_mouschovias_2001}, the magnetic pressures $B^2 / 8\pi$
$\sim 0.04 \, \mu{\rm Bar}$ are significantly smaller than the gas and ram pressures
$\sim 10 \, \mu{\rm Bar}$.
Perhaps most importantly, outgassing from the planetary body may create a persistent atmosphere, even with a strong bow shock.
Finally, while we have run these simulations at high resolution, we are still only achieving 100 km 
across a grid cell.  
A finer grid may allow new structure to develop, such as multiple tail shocks and more substantial 
turbulence.
Future work will be needed to address each of these points.

 \section{Summary and Conclusions\label{sec:conclusions}}
 
 In this paper we have used three-dimensional radiation hydrodynamics simulations with direct solid integration to explore the consequences of bow shocks on the thermal processing of nebular solids.  
 Our simulations also include a full equation of state, taking into account the effects of the internal states of molecular hydrogen for both the bow shock structure and the heating and cooling of solids.
 We explore bow shocks with relative wind velocities of 7, 8, and 9 km/s in the adiabatic limit, and explore an 8 km/s shock using radiative physics with three different assumptions for the gas opacity.  
 In each of the radiative cases, the opacity is derived directly from the integrated particle distribution.
 
\begin{figure}[H]
\includegraphics[width=5in,angle=-90]{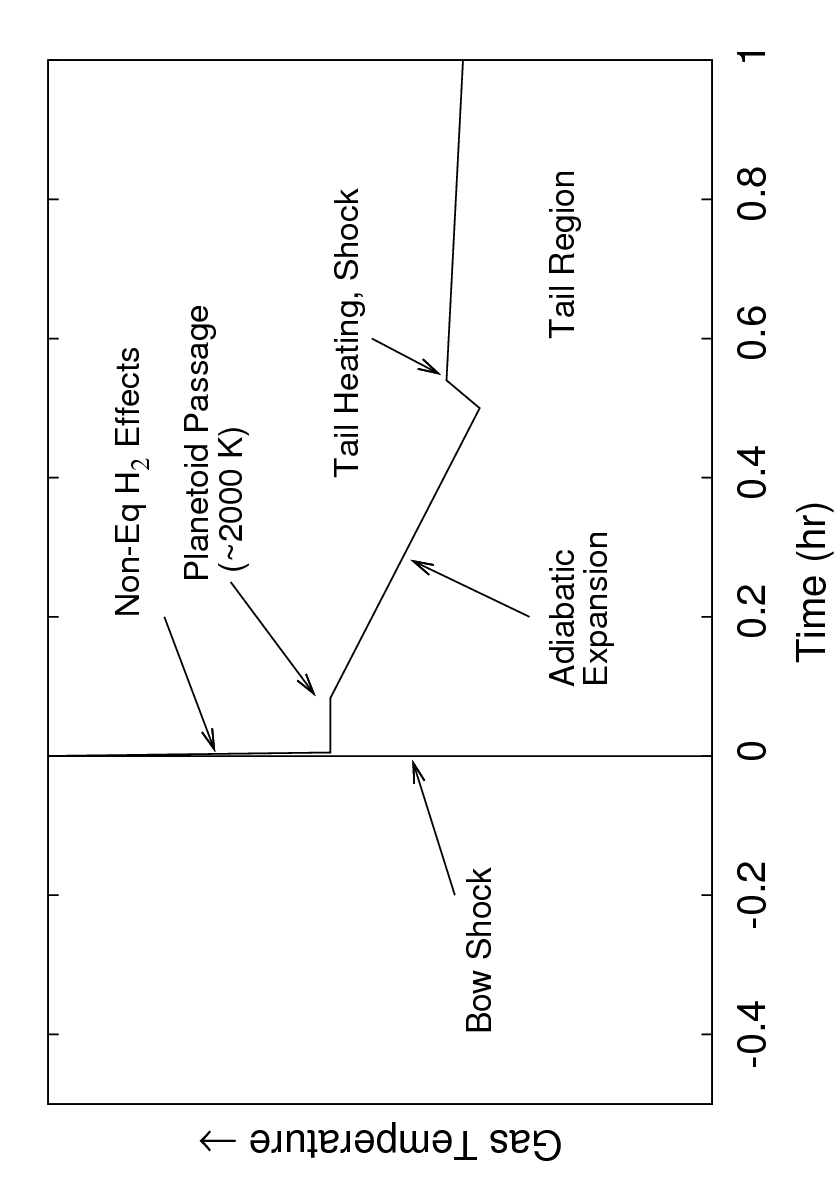}
\caption{ Cartoon depicting the gas temperature history that a solid would experience passing through a bow shock caused by a planetary embryo. 
The temperature profile corresponds to the adiabatic limit.  
Upon reaching the bow shock, the temperature jumps to 4000 K or higher, depending on the pre-shock gas conditions and the strength of the shock.
Within about 10 s, molecular hydrogen dissociation will bring the gas toward chemical equilibrium \citep[][not captured in these simulations]{morris_desch_2010}, and the temperature of the gas will drop to $\sim2000$ K.  
After being diverted around the planetary embryo, a solid will experience gas that cools rapidly with distance from the embryo.
This cooling is due to adiabatic expansion.  
The gas temperature can drop to values below or just above the solid crystallization temperature.  
Tail shocks are expected to form, which will reheat the gas in the tail region.  
Following this reheating, the gas in the tail region continues to cool through adiabatic expansion, but at a slower rate than immediately behind the bow shock. 
\label{fig:cartoon_temp}}
\end{figure}

 The main results are as follows:
 
 \begin{enumerate}

 \item The three-dimensional structure of bow shocks causes significant differences from purely 1D shocks.  
 These deviations include secondary shocks (tail shocks) and rapid cooling due to adiabatic expansion.
As a result, the temperatures of solids behind the bow shock are not monotonically decreasing with time, and cooling rates are substantially different from typical laboratory experiments conducted to characterize chondrule cooling histories. 

\item \ACBc{Fully three-dimensional bow shock structures have smaller shock volumes than previously inferred, leading to a decrease in total shock processing per planetary embryo passage.}

\item In the adiabatic limit, cooling histories may be consistent with radial, barred, and porphyritic chondrule formation.
Adiabatic expansion promotes rapid cooling for over 10 minutes immediately following the bow shock, but high temperatures can be maintained/reached anew due to interactions with tail shocks.
Laboratory experiments with these temperature curves need to be investigated. An idealized temperature history is shown in Figure \ref{fig:cartoon_temp}. \ACBc{Conditions that can realistically reproduce the adiabatic limit and still be consistent with other chondrule formation constraints (e.g., volatile enrichment and compound chondrule frequencies) also need to be explored theoretically, as simply lowering the chondrule precursor density may be insufficient.}

\item Radiative transfer is very efficient.
Tail shock-heating is largely suppressed for opacities $\kappa_c=10$ and 100 cm$^2$ per gram of solid.  
\ACBc{In the simulations with $\kappa_c=1$ and 10 cm$^2$ per gram of solid, barred and radial chondrules could be formed, in principle, but porphyritic chondrule formation may be prevented. 
The gas in the highest opacity simulation does not  reach a peak temperature that could melt chondrule precursors, as radiation transports substantial energy far ahead of the shock.}

\item \ACBc{For each radiative simulation, the bow shock luminosity is $\sim$few$\times10^{-8}$ L$_{\odot}$. Bow shocks around planetoids in protoplanetary disks are potentially observable.}

\item \ACBc{The 3D bow structure gives a finite width to the high-density region of the shock, which is roughly the standoff distance.}

\item The low optical depth simulation may represent the best environment for forming porphyritic chondrules.
However, although a strong tail shock forms, it does not prevent solids from dropping below their crystallization temperature.
Optical depths lower than those explored here may be necessary to approach the adiabatic limit, although if this limit is reached, it is unclear whether the chondrule precursor density will be high enough to form compound chondrules or to produce significant partial pressures of volatiles.

\end{enumerate}

While this work is a step toward building self-consistent model for understanding solid processing through nebular bow shocks, there are additional effects that will need to be included in further studies.
 As described in section \ref{sec:discussion}, separate gas and solid temperatures will need to be tracked to assess fully the impact of bow shocks on the temperature history of the solids.  
 \ACBc{A range of pre-shock densities and temperatures should be explored, and planet surface radiation and line transfer effects should be included in future calculations.}
 Nonetheless, the results here demonstrate that 1D models, while applicable to very large-scale shocks, cannot be used alone to understand how bow shocks process solids, as bow shocks are fundamentally three-dimensional. 
 The geometry of the problem alters the heating and cooling of solids significantly. 
 Even if these shocks do not produce chondrules, the cooling curves found here, even if just in the adiabatic limit, should be used in laboratory experiments to see what types of materials are produced.
 Those results can then be compared with the meteoritic record to constrain planetoidal dynamics during planet formation.
 
 \ACBc{We thank the anonymous referee for comments that improved this manuscript.  
 We thank Fred Ciesla and Conel Alexander for helpful discussions.
 ACB's support was provided under contract with the California
Institute of Technology funded by NASA through the
Sagan Fellowship Program. 
 We gratefully acknowledge support from NASA Origins of Solar Systems program under grant
NNX10AH34G to SJD.}

\appendix 

\section{Appendix}
 
\subsection{Boxzy Hydro\label{sec:hydro}}
 
Boxzy hydro is a Cartesian, second-order accurate hydrodynamics code that has the following capabilities:
\begin{itemize}
\item Hydrodynamics fluxing using either linear or angular momentum.
\item A detailed equation of state that includes the rotational and vibrational states of H$_2$ as well as H$_2$ dissociation.
\item Self-gravity.
\item A radiation algorithm that is valid without directional bias and couples the thick and thin regimes.  
\item Particle integration using particle-in-cell and particle-particle integration. Drag terms with feedback on the gas is included.
\item Inclusion of arbitrary structures in the grid.
\end{itemize}
Each of these items will be discussed in turn.  First,  the hydrodynamics equations are solved in the following form
\begin{equation}
\begin{array}{lllr}

\frac{\partial \rho}{\partial t} + \nabla \cdot \left({\bf v}\rho\right) & = & 0 &  \mbox{Mass continuity}\\\\

\frac{\partial {\bf v}\rho }{\partial t} + \nabla \cdot \left({\bf v} {\bf v} \rho\right) & = & -\nabla P - \rho \nabla \Phi & \mbox{Momentum equation}\\\\

\frac{\partial E}{\partial t} + \nabla \cdot \left({\bf v} \left(E+P\right) \right) & = & -\rho {\bf v}\cdot \nabla \Phi, & \mbox{Energy equation}\\

\end{array}
\end{equation}
where $E= \frac{1}{2} \rho \vert {\bf v}\vert^2 + \epsilon$ for gas density $\rho$, velocity ${\bf v}$, and internal energy density $\epsilon$. 
In the absence of a gravitational potential field $\Phi$, the energy equation is written in conservative form, allowing total energy conservation to machine precision.  
However, this degree of precision is rarely reached in astrophysical studies due to gravitational fields \citep[see however recent developments by][]{jiang_etal_2013}. 
The pressure $P$ is described in more detail below. 
For the moment, focus on the momentum equation. 
Note that while the equation is conservative for the linear momentum, it makes no such guarantees for the angular momentum.  
This can have dire consequences for simulations
of rotating objects on Cartesian grids.
For example, rotating polytropes in a code fluxing linear momentum can either gain or lose significant angular momentum.  
Any user can verify this behavior in their code of choice by running an $n=1$ polytrope with modest initial rotation.
In the Boxzy hydro, angular momentum tends to decay if the equations are solved as given above. 
To address this, the following two methods are available for momentum fluxing.
(1) The linear momentum, as shown above, can be solved in each direction, as is standard in Cartesian grid codes.  
(2) As an alternative, the $z$ direction can be solved
using linear momentum, but the $x$ and $y$ momenta can be transformed into radial and angular momenta and fluxed accordingly. 
 For this second option, although the grid is Cartesian, angular momentum in the $z$ direction can be conserved in the absence of potential field errors. 
As a result of the transformation for fluxing, the force calculation (sourcing) requires additional terms such that the right-hand side of the momentum equation
becomes
\begin{eqnarray}
\rho v_x & += & \left(-\partial_x P -\rho \partial_x \Phi + \rho \left(x v_y - y v_x\right)^2 x/r^4\right) \Delta t ,\\ 
\rho v_y & += & \left(-\partial_y P -\rho \partial_y \Phi + \rho \left(x v_y - y v_x\right)^2 y/r^4\right) \Delta t , \rm and\\
\rho v_z & += & \left(-\partial_z P -\rho \partial_z \Phi \right)\Delta t.
\end{eqnarray}
 The major trade off is that a singularity is now introduced in the grid, so the choice of fluxing is problem-dependent and must be used with caution.
 At this time, the viscous stress tensor is not included, but is in development. 

Fluxing is performed using an approximate Riemann solver (HLLE) with a generalized MINMOD flux limiter \citep[e.g.,]{toro2009riemann}. 
 In practice, the calculation of the pressure force
is incorporated into the fluxing routine.  Consider flux variables $\bf{F}$ and conserved variables $\bf{U}$ in, for example, the $x$ direction, where
\begin{equation}
  U_x = \begin{pmatrix}
  \rho\\
  \rho v_x\\
  \rho v_y\\
  \rho v_z\\
  E\\
  \end{pmatrix},
\end{equation}
and
\begin{equation}
  F_x = \begin{pmatrix}
  \rho v_x\\
  \rho v_x v_x +P \\
  \rho v_y v_x\\
  \rho v_z v_x\\
  (E+P) v_x\\
  \end{pmatrix}.
\end{equation}
All variables are cell-centered on the Cartesian grid.  
The hydrodynamics is solved for each direction ($x$, $y$, and $z$) independently, but the conserved variables are only updated after the contribution from each direction has been calculated.  
The hydrodynamics time step $\Delta t_{\rm CFL}$ is limited according to the Courant-Friedrichs-Lewy condition $\Delta t_{\rm CFL}[\alpha_x + \alpha_y +\alpha_z]<1$, where $\alpha_x = (c_s+\vert v_x\vert)/\Delta x$ for adiabatic sound speed $c_s$.  
We typically set the time step to be $\Delta t_{\rm CFL}=1/(2[\alpha_x + \alpha_y +\alpha_z])$. 

The pure hydrodynamics equations (no gravity or radiation) are evolved by advancing the initial state $U^n$ over a full time step, which gives a predicted state $U^{n*}$.  The solution is then corrected using a second flux calculation based on the predicted state.  The scheme is second order in time and space. The following example shows how this method is applied in the 1D limit:
\begin{eqnarray}
  {U}^{n*}&=&{ U}^n+\lambda [F(U^{n})_{+1/2}-F(U^{n})_{-1/2}]\\
  {U}^{n+1} &=& ( U^{n}+{ U}^{n*} + \lambda  [ F(U^{n*})_{+1/2}-F(U^{n*})_{-1/2}])/2,
\end{eqnarray}
where $\lambda=\Delta t_{\rm CFL}/\Delta x$ for CFL-limited time step $\Delta t_{\rm CFL}$.  
The quantities $F(U)_{+1/2}-F(U)_{-1/2}$ represent the flux differences for any given cell based on the fluxes at the $+$ and $-$ cell faces, respectively.  These flux values are determined using the approximate HLLE Riemann solver, in combination with a second-order upwind interpolation scheme for the hydrodynamics variables. Slopes for the upwind interpolations are limited (flux-limiter) by the generalized MINMOD algorithm.  Unless otherwise stated, the generalized MINMOD limiter is set to the standard MINMOD limiter (diffusion parameter $\theta=1$).  Please see \cite{toro2009riemann} for additional details.  
Gravity and radiation transfer are included by sourcing $U^{n}$ for 1/2 a time step before advancing the hydrodynamics (before equations A7 and A8).  After the hydrodynamics step is completed, the gravitational potential is updated using the $U^{n+1}$ solution for the density.  Using the $U^{n+1}$ state, a final gravity and radiation update over 1/2 of a time step is performed. This strategy for including gravity and radiation is analogous to a kick-drift-kick method.

\begin{figure}[H]
\includegraphics[width=3.25in]{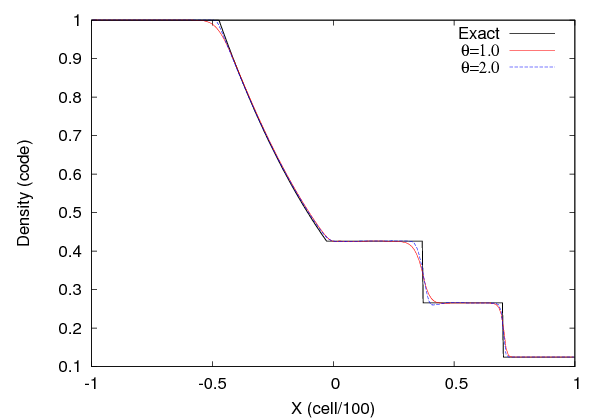}
\includegraphics[width=3.25in]{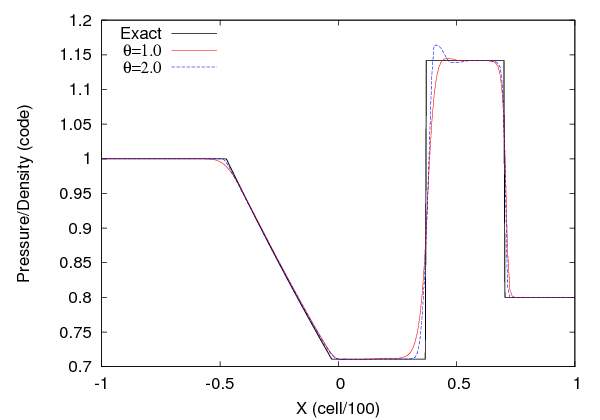}
\caption{Sod shock tub tests. 
The left state ($x<0$) is initialized with $\rho_l=1$ and $p_l=1$ (code units are assumed throughout for this test). 
 The right state ($x>0$) is set to $\rho_r=0.125$ and $p_r=0.1$.
 The grid points are set to 200, with grid spacing $\Delta x=0.01$.
 The solution is shown after time $t=0.4$.
 Two different values for $\theta$ are used, where $1$ gives the classic MINMOD limiter. 
 While $\theta=2$ gives sharper transitions (less diffusive), it also gives rise to a spike in entropy at the contact discontinuity.  
 Unless otherwise stated, $\theta=1$ is used in all other calculations.
 \label{fig:sod}}
\end{figure}

The numerical method is tested using \cite{sod1978} shock tubes.  
For this test, a perfect gas with equation of state
$p=\left(\gamma-1\right)\epsilon$  is placed in a control volume.
The gas is given a left and right state, with one state at a higher pressure than the other. 
When the system is allowed to evolve, a shock, contact discontinuity, and rarefaction wave develop, which have analytic solutions.  
We show a typical example in Figure (\ref{fig:sod}), where we use $\rho_l=1$ and $p_l=1$ for the left states and $\rho_r=0.125$ and $p_r=0.1$ for the right states, each in code units. 
The test uses a 1D grid with 200 grid cells.  
For this test, we explore diffusion parameter values of $\theta=1$ and 2.  
While $\theta=2$ leads to sharper transitions (less diffusive), it also leads a slight dip in density without a compensating dip in pressure, which is a spike in entropy. 
For this reason, we typically set $\theta=1$.
No other form of artificial diffusion is used, and no artificial viscosity is employed.

\subsection{Equation of State\label{sec:eos}}

\begin{figure}[H]
\includegraphics[width=3.25in]{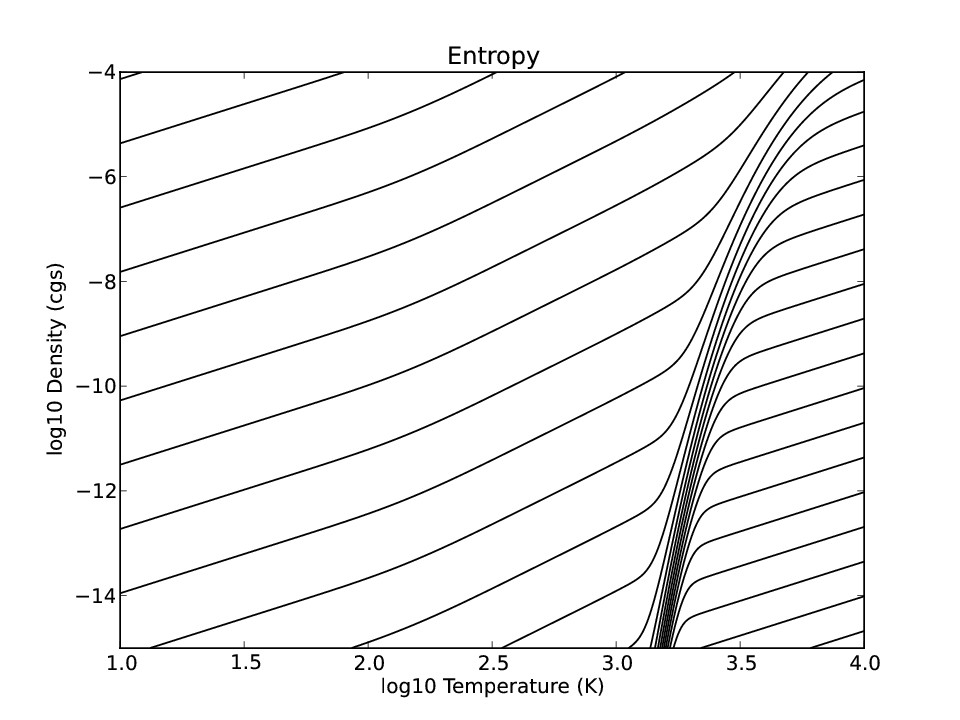}
\includegraphics[width=3.25in]{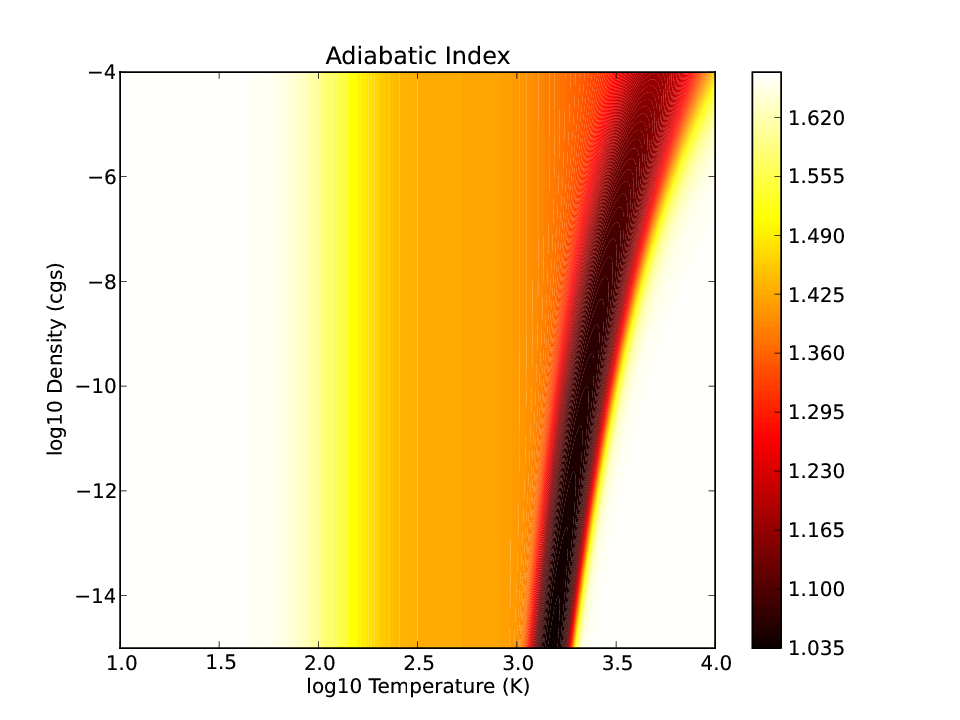}
\caption{Adiabats in the $\rho-T$ plane (left) and the corresponding adiabatic index (right) for gas comprised of 73\% hydrogen, 25\% helium, and 2\% metals by mass.  At low temperatures, the gas behaves as a $\gamma=5/3$ gas.  At $T\sim300$K, rotation adds two degrees of freedom, and the gas behaves as $\gamma\rightarrow 7/5$.  
At even higher temperatures, vibrational modes are activated, which also add two degrees of freedom (partitioned between kinetic and potential ``spring'' energies), giving $\gamma\rightarrow 9/7$. 
When $T\gtrsim 2000$K, depending on gas density, H$_2$ begins dissociating.  
This creates an incredible heat sink, where internal energy goes into breaking apart the molecule instead of providing pressure support for the gas, ultimately driving $\gamma\rightarrow 1$.  
After dissociation is complete, the gas again behaves like a $\gamma=5/3$ gas.  At the point of dissociation, an adiabat can jump several orders of magnitude in density while exhibiting only a relatively small change in temperature.  A slice through the adiabatic index at a constant density is shown in Figure \ref{fig:gamma_spec}.\label{fig:gamma}}
\end{figure}

\begin{figure}[H]
\includegraphics[width=5in]{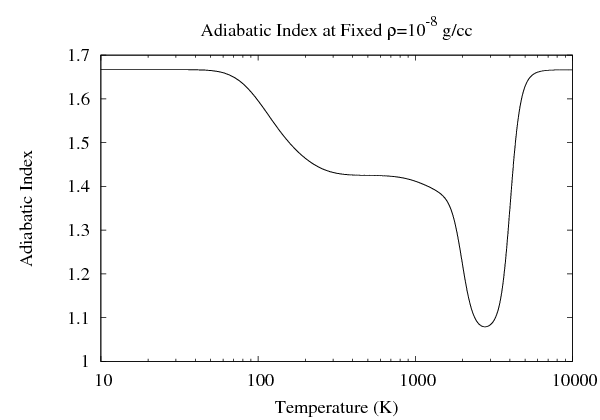}
\caption{Adiabatic index profile for a fixed density. See also Figure \ref{fig:gamma}.\label{fig:gamma_spec}}
\end{figure}

The equation of state is an extension of that used in \cite{boley_etal_2007_h2} and \cite{galvagni_etal_2012}. 
 A mixture of metals, helium, and hydrogen is used to compute tables for the relations between the gas's internal energy density, adiabatic index, and temperature.
With these relations, an ideal gas law can be used that takes into account the partitioning of the gas between translational and internal degrees of freedom, i.e.,  $p= \rho R_g T(\epsilon,\rho)/\mu(\epsilon,\rho)$ for gas constant $R_g$ and mean molecular weight per proton mass $\mu$.
Because ionization is not yet included, only translational degrees of freedom are used for computing helium and metals partial pressures.  
The contribution from H$_2$ is much more complex.
First, the equation of state is derived directly from the partition function, so the effects of heating and cooling due to partitions between translational and internal molecular degrees of freedom are directly included. The internal degrees of freedom include the rotational and vibrational states of molecular hydrogen for different para and orthohydrogen ratios \citep[see][]{boley_etal_2007_h2} and molecular hydrogen dissociation. 
This behavior of H$_2$ for a range of thermodynamic conditions is shown in Figures \ref{fig:gamma} and \ref{fig:gamma_spec}.  
At very low temperatures, rotational states are inaccessible, and a pure molecule hydrogen gas will have an adiabatic index $\gamma=5/3$.  
At temperatures $T\sim300$K, rotation adds two degrees of freedom, and the gas behaves as $\gamma=7/5$.  
At even higher temperatures, vibrational modes are activated, which also add two degrees of freedom (partitioned between kinetic and potential ``spring'' energies), giving $\gamma\rightarrow 9/7$. 
When $T\gtrsim 2000$K, depending on gas density, H$_2$ begins dissociating.  
This creates an  effective heat sink, relative to a fixed-$\gamma$ gas, where internal energy goes into breaking apart molecule H$_2$ instead of providing pressure support for the gas, ultimately driving $\gamma\rightarrow 1$.  
After dissociation is complete, the gas again behaves like a $\gamma=5/3$ gas. 
These tables allow accurate modeling of the compressibility of the gas and naturally take into account the effective heating and cooling rates due to internal molecular processes. 

 \subsection{Self-Gravity}
 
 Self-gravity is included by combining mulit-grid techniques with a tree algorithm.  
 First, the mass on the finest grid is averaged down onto subsequent levels.  
 For each node of the tree, the center of mass and total mass of its 8 children/leaf cells are included.  
As in a standard tree algorithm \citep{barnes_hut_1986}, we use an opening angle to determine whether
a node's contribution to a given local potential is sufficient, or whether the cell needs to be opened, i.e.,  use the finer levels as nodes.
For the algorithm here, we define the opening angle to be $\theta = (dx^2+dy^2+dz^2)^{1/2}/r$, where $r$ is the distance between the point of interest and the center of the parent cell that has sizes $dx$, $dy$, and $dz$.  
The opening angle $\theta$ is adjustable according to the specific problem, but $\theta=0.7$ is typically used in the simulations presented here.  
For any given node's contribution to the potential, we use $\phi_{\rm node} = -M_{\rm node}/\vert{\bf r}-{\bf r_{\rm com}}\vert$, where $M_{\rm node}$ is the node's mass, ${\bf r}$ is the distance to the point where the potential is needed, and $r_{\rm com}$ is the distance to the given node's center of mass (COM). 
Using the COM of the node automatically includes the dipole moment of the cell in the potential calculation. 
At this time, the quadrupole and higher moments are not included, as the tree is only needed to calculate an approximate solution. 

 The tree potential calculation is used to find the grid boundary values, including the value at the finest level, as well as the potential solution everywhere for the penultimate grid level (or coarser if desired).  
 The approximate solution on the penultimate finest level is then successively relaxed until a minimum residual for the grid, set by the user, is obtained. 
 Here, the residual $\mathcal{R}=(\nabla\phi\cdot {\bf dS}-4\pi \rho dx dy dz)/L$, where $L=2 (dy dz/dx+dx dz/dy+dx dy/dz)$. 
 For each iteration $i$ in the relaxation, $\phi_{i+1}=\phi_i+\alpha \mathcal{R}$, where we found $\alpha<1$ provides stability.  
 The relaxed solution on the penultimate finest level is then prolongated to the finest level, except the boundary cells. 
Relaxation is then performed on the finest grid as done on the previous level.
For the simulations shown here, the maximum $\mathcal{R}$ on the grid is typically restricted to be less than $10^{-4}$. 
 This method is reasonably fast and accurate.
 
\begin{figure}[H]
\includegraphics[width=5in]{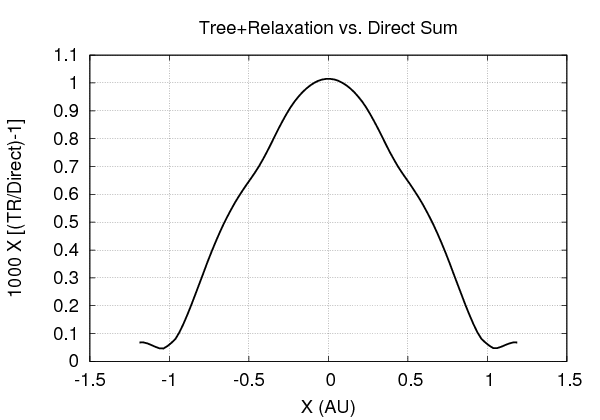}
\caption{Potential test for an $n=1$ polytrope.  The curve represents the code solution  compared with direct summation, assuming each cell is a point mass.  The tree+relaxation (TR), i.e., the code solution, is calculated with an opening angle $\theta=0.7$ and a maximum residual $\mathcal{R}_{\rm max}<10^{-4}$. Note that the actual solution will be slightly different from the solutions for both TR and direct summation. When compared with the exact solution for a constant density sphere, the code solution is comparable to the analytic solution to better than one part per thousand everywhere.  \label{fig:pot_test}}
\end{figure}

 To demonstrate the accuracy of the potential solution, we present two tests.  
 First we place an $n=1$ gas polytrope on the grid, which has the analytic density solution $\rho=\rho_0 \sin(\pi x)/(\pi x)$ for $x=R/R_{\rm surface}$.
 The polytrope's radius is set to 1 AU, the maximum density to $2\times10^{-3}$, and the grid resolution $\Delta x = \Delta y =\Delta z = 0.025$ AU.
 We set the opening angle to $\theta=0.7$ and set the maximum residual $\mathcal{R}_{\rm max}<10^{-4}$.  
The potential is first calculated using the tree+relaxation scheme described above and is then compared the solution based on direct summation, where each cell is treated as a point mass. 
Figure (\ref{fig:pot_test}) shows that the result has a maximum deviation of 0.1\%.  
The actual solution will be slightly different from both the tree+relaxation and direct summation solutions due to discretization effects. 
We did compare the TR solution against an analytic solution for a constant density sphere, and for the same parameters, the solution was better than 0.1\% everywhere. 

The second test uses a hydrodynamics calculation with self-gravity.  An analytic polytrope with $n=1$ is placed on the grid and then allowed to evolve, with the equation of state set to $P=K\rho^{2}$ for polytropic constant $K$.  For this test, we set $R=1 AU$ and the central density $\rho_c=1.19\times10^{-9}$ g/cc ($2\times10^{-3}$ in code units), which gives the clump a mass of about $2.5$ M$_J$. 
The grid is set up to have $96^3$ cells, with grid spacing $\Delta x=\Delta y=\Delta z = 0.025$ AU.  
Figure \label{fig:poly_test} shows the results after about nine dynamical times, where we take the free-fall time ($t_{\rm ff}=(3\pi/[32 G \bar{\rho}])^{1/2}$) to be representative of the dynamical time.

 \begin{figure}[H]
\includegraphics[width=4in,angle=-90]{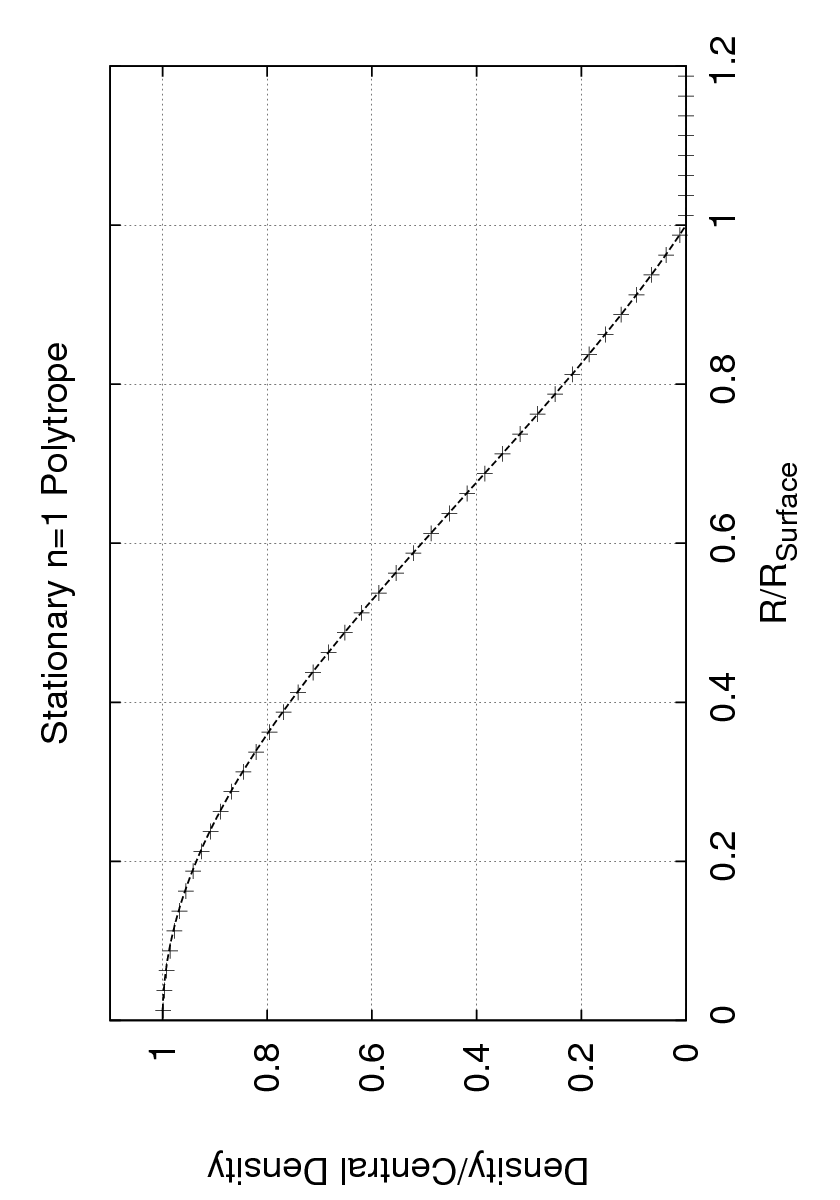}
\caption{A test of the hydrodynamics with self-gravity using a stationary $n=1$ polytrope.  The curve represents the analytic solution ($\rho=\rho_0 \sin[\pi x]/[\pi x]$) for $x=R/R_{\rm surface}$, while the crosses show the simulation results after about nine dynamical times, based on $t_{\rm ff}=(3\pi/[32 G \bar{\rho}])^{1/2}$  for average density $\bar{\rho}$.  \label{fig:poly_test}}
\end{figure}
 
\subsection{Radiative Transfer: mcrtfld\label{sec:mcrtfld}}

Radiative transfer (RT) plays a fundamental role in heating and cooling astrophysical objects, which in turn, has a substantial impact on an object's evolution. 
Unfortunately, integrating radiation transport into simulations can be a technical challenge.  
Diffusion schemes are accurate in regions of high optical depth, but ultimately solve the wrong equation near a system's photosphere, where energy is actually lost.    
Ray tracing schemes do well for the optically thin limit, and will capture energy loss near the photosphere, but can become problematic at high optical depths where temperature gradients, which are used in the diffusion limit, are more informative of the flow of radiation.
In addition, and regardless of the scheme, there is significant potential to have wildly different radiation and hydrodynamics timescales.  
In this case, an explicit scheme can fail severely if Courant timesteps are taken.  
One could choose instead to use an implicit scheme, which will be numerically stable, but will not guarantee accuracy.

Here, we seek to describe a new radiation transport algorithm that combines elements of diffusion, ray tracing, and Monte Carlo methods.  
Our goal is to develop a radiation scheme that works well in the optically thick and thin limits, as well as in the transition regime between these limits.
We also want to attempt to resolve the radiation hydrodynamics time scale.  With these considerations in mind, we will first discuss the diffusion limit, followed by the optically thin limit, and then describe how we have combined these cases to form what we call the  Monte Carlo-Radiative-Transfer-Flux-Limited-Diffusion  (MCRTFLD) method.

Whenever the optical depth through a region of gas is very large ($\Delta \tau\gtrsim 10$),  radiative transfer is well approximated by diffusion. Consider the radiative flux in the direction $x$.  In the diffusion limit,
\begin{equation}
F_x = \frac{16 \sigma \beta T^3}{\rho \kappa}\frac{\Delta T}{\Delta x},\label{eqn:fld_flux}
\end{equation}
where
\begin{equation}
\beta = \frac{2+y}{6+3y+y^2}
\end{equation}
for
\begin{equation}
y=\frac{4}{\rho\kappa T} \bigg\vert \frac{\Delta T}{\Delta x}\bigg\vert.
\end{equation}
The $\beta$ term is the radiative flux limiter, which permits extension of the radiative diffusion treatment into the optically thin regime by ensuring that the divergence of the flux never exceeds the value corresponding to free-streaming. 
 In this way, the flux limiter is an interpolation between the optically thick and thin regimes, where the above $\beta$ is from \cite{bodenheimer_etal_1990}.  
 However, it is this reliance on an interpolation scheme that is a potential problem, as well.  
 Radiative diffusion by itself does not remove energy from the system; it just diffuses it.  
 Radiative losses, which take place near the photosphere, are determined by the flux limiter, so caution needs to be taken when applying a flux-limited diffusion scheme.   
  Moreover, because radiation is allowed to diffuse, some temperature structures can get washed away.  
 The severity of this effect is dependent on the desired accuracy and the problem at hand.

Now consider the optically thin regime ($\Delta \tau << 1$). 
 In this case, a gas can emit as much radiation over $4\pi$ as its emissivity allows.  
 For a source function $S=\sigma T^4/\pi$, the luminosity per volume of gas is $\ell = 4\rho\kappa \sigma T^4$.  
If the contributions of the incident irradiation, $j$, on this same volume are taken into account, then $\nabla \cdot {\bf F} = j-\ell$. 
For this relation, we have already folded in the solid angle components, so $j$ and $l$ are flux-like quantities summed over the area of the volume of gas. 

Many astrophysical objects span both the optically thin and thick regimes. 
Thus, a method for coupling the two limits is highly desired. To reiterate, most of the cooling will occur in the transition region between these limits.  
We build this coupling by readdressing radiative transport in the thin regime. 
 If we have a large volume of gas that represents, for example, a cell in a simulation, then assuming constant density $\rho$, temperature $T$, and opacity $\kappa$ across a cell width $\Delta x$, the intensity at the cell face due to the gas in the cell alone is 
\begin{equation}
I=\frac{\sigma T^4}{\pi} \left(1-\exp{\left[-\Delta \tau\right]}\right),
\end{equation}
where we have assumed black body emission and $\Delta \tau = \rho \kappa \Delta x$.
The amount of energy carried away from the cell by emission  can be calculated by discretizing the cell into six direction, one for each cell face.  
Each cell face then corresponds to a portion of the total emission of the cell over the entire sky, i.e., $2\pi/3$.  
The luminosity per cell volume $V$ that is carried away in any single direction is then $\delta \ell = \frac{2\pi I A}{3 V}$, where $A$ is the area of a given cell face.  
In the limit that $\Delta \tau \ll 1$, then the luminosity per volume $\ell = 6\delta \ell \rightarrow 4 \rho\kappa \sigma T^4$, which is the free-streaming limit.

So far, we have not discussed how the energy is transferred from one cell to other cells using the above assumptions, nor have we discussed how $j$ is ultimately determined.  
Both can be calculated by using particles to represent energy packets.  
At the face of each cell, we launch particles in a random direction, constrained to be within $\delta \theta=60^\circ$ of the normal to the cell face. 
Allowing random angles ensures that directional bias is removed, although it does permit the occurrence of noise.  
Each particle is then allowed to propagate a distance $\Delta s=0.5 \sqrt{\Delta x^2+\Delta y^2 +\Delta z^2}$ along the launch angle.  
During that propagation, the energy is absorbed by $\delta\ell \exp(-\Delta\tau)$, where the optical depth is based on the current cell conditions such that $\Delta \tau  = \rho \kappa \Delta s$.  
The local value of $j$ is likewise determined by summing the energy that is absorbed by a local cell for all particles that enter the cell. 
For simplicity, only the nearest node is used for determining the energy absorbed during any one single propagation.  

A possible disadvantage to this method is that it can easily become time-consuming.  However, this is a tunable feature of the algorithm.  If a poorer treatment of radiative transfer can be tolerated, then combinations of maximum propagation limits, emissivity thresholds, and computational volume limits can be used to increase performance.  This portion of the algorithm also tends to be stable for large time steps, as energy is deposited over many cells unless an energy packet propagates into a very high optical depth region.  This takes us back to the FLD approximation. 

When the optical depth across a cell becomes very high, the variation of the source function along a cell becomes an important component for understanding energy transfer, and the above assumption of a constant temperature for a cell leads to inaccuracies in computing the cell-to-cell coupling.  In the limit $\Delta \tau\gg 1$, radiative diffusion is an accurate description of energy transfer in the gas.  In this spirit, we combine the FLD and the energy packet propagation algorithms by setting
\begin{equation}
\nabla\cdot{\bf F}=j-\left(\ell  \exp\left[-3 (\Delta \tau)^2\right]   + \ell_{\rm FLD}\left(1-\exp\left[-3 (\Delta \tau)^2\right]\right)\right).
\end{equation}
The FLD term is calculated using equation  (\ref{eqn:fld_flux}) using the following method: 
For each cell, the standard FLD flux is calculated for each cell face.  
If the flux is outward from the cell, the value is kept.  If it is inward, the value is set to zero.  The flux is converted into a luminosity per volume using $F_x A/V$.  The energy is then propagated and absorbed in the same way as that used in the low optical depth limit.  This limits exactly to the FLD formalism when the cells are at high optical depth, but allows smooth, directionally debiased propagation of energy in regions where the optical depth becomes low.  
The optical depth-based interpolation ensures both a smooth and sharp transition between the thick and thin regimes and the corresponding algorithm, where the form of the interpolation was ultimately chosen from trial and error with the tests given below. 
We finally note that the above description assumes constant volume between cells. 
For codes where the physical cell size changes, the algorithm needs to be reformulated using luminosity instead of luminosity per volume, which is a straightforward modification.

We build our first test case by considering a spherical distribution of gas with a luminosity point source at $r=0$ of $L=10^{-4}L_{\odot}$. 
The temperature gradient stellar structure equation can be expressed as
\begin{equation}
\frac{dT}{dr} = -\frac{3 \rho \kappa L}{64 \pi \sigma r^2 T^3}.
\end{equation}
If $\rho \kappa$ is parameterized, then a solution for the temperature profile can readily be calculated.  
A function for $\rho \kappa$ that varies rapidly in $r$ will test the algorithm's behavior in both the thick and thin regimes, and will also test the interpolation between the limits. 
For this reason, the first set of tests use $\rho \kappa = \rho_0 \kappa_0 \left(\left(r/r_c\right)^4+1\right)^{-1}$, which
yields the solution
\begin{equation}
T(r)^4=T_0^4+\frac{3 \rho \kappa L}{128 \pi \sigma r_c }\left(f\left(r\right)-f\left(r_0\right)\right),
\end{equation}
where
\begin{equation}
f\left(r\right)=\sqrt{2}\ln \left[ \frac{ r^2+\sqrt{2} r r_c + r_c^2}{ r^2-\sqrt{2} r r_c + r_c^2}\right] + 2\sqrt{2}\left( \arctan\left[1-\sqrt{2}\frac{r}{r_c}\right]-\arctan\left[1+\sqrt{2}\frac{r}{r_c}\right]\right)-8\frac{r}{r_c}
\end{equation}
$T_0$ and $r_0$ correspond to values at the sphere's surface. 

\begin{figure}[H]
\includegraphics[width=3.5in]{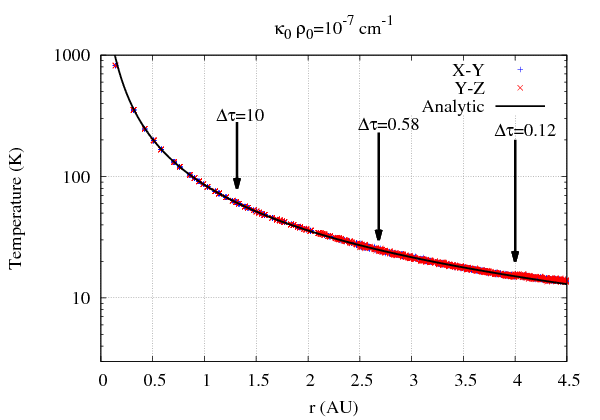}\includegraphics[width=3.5in]{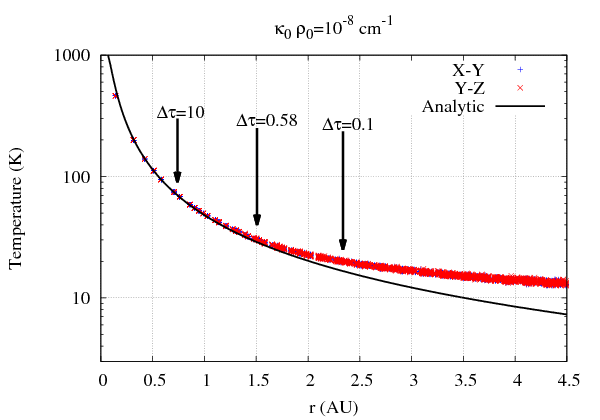}
\caption{Radiative transfer test results comparing the simulation results to the analytic solution for  $\rho\kappa=\rho_0\kappa_0\left(\left(r/r_c\right)^4+1\right)^{-1}$ where $r_c=0.1$ AU.  The $\kappa_0 \rho_0=10^{-7}$  cm$^{-1}$ solution (left) matches very well through out the structure.  Although the sphere is truncated at $r=4$ AU, background density gas, which is much lower than the outer edge of the sphere, is part of the calculation.  The sudden drop in density causes a minor kink in the temperature profile at the surface.  Several values for $\Delta \tau$  throughout the structure of the sphere are also shown with arrows.  The $\kappa_0 \rho_0=10^{-8}$ cm$^{-1}$ solution (right) matches the profile well in the optically thick regime, but then deviates from the analytic solution as the optical depth becomes very low. 
At low optical depths, significant non-local transport of radiative energy occurs, and the actual solution should not be expected to follow the analytic solution, which is derived for regions of large opacity.  \label{fig:radiation_tests}}
\end{figure}

In this test, the fluid is not permitted to move on the grid, isolating the radiation transfer algorithm.  
We use $\rho_0 \kappa_0 = 10^{-7}$ and $10^{-8}$ cm$^{-1}$) for the cored power law profile.  The sphere is truncated at $r=4$ AU and $r_c=0.1$ AU. The former tests how well the algorithm transitions from optically thick to thin, while the latter creates a mostly optically thick sphere. For the analytic solutions, we assume the radiative zero solution, i.e., we take $T=0$ at the surface. To better capture  which is extended to $R=10$ AU.  The solution should not be expected to extend to regions of very low optical depth, as energy transport becomes very non-local.  We should expect the analytic and the simulation solutions to converge rapidly near a radially integrated optical depth $\tau_r\sim 1$, for integrations going from $r=\infty$ to 0.  The simulations are run until the average net energy input on the grid is zero. The results are shown in Figure {\ref{fig:radiation_tests}}, which show very good agreement with the analytic solution for the $\rho_0 \kappa_0=10^{-7}$ cm$^{-1}$ case.  The analytic solution matches the $\rho_0 \kappa_0=10^{-8}$ cm$^{-1}$ solution well until the atmosphere drops to very low optical depths.  At these depths the gas temperature becomes hotter than expected in the analytic solution because the emissivity of the gas becomes small, preventing the gas from cooling efficiently. 

\subsection{Particles\label{sec:particles}}

\begin{figure}[H]
\begin{center}
\includegraphics[width=3.in]{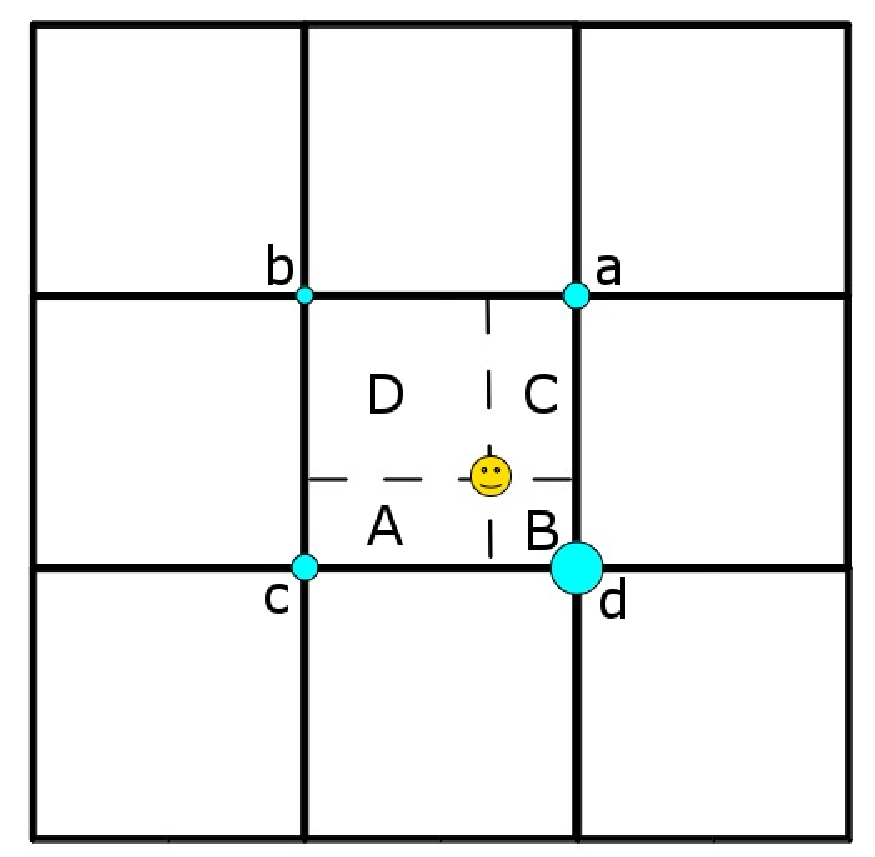}
\caption{Interpolation and prolongation mapping in a 2D example.  
The super-particle is given by the yellow smiley face, while the cyan points represent grid nodes.  
The areas A-D that are formed by the grid nodes and the super-particle set the weights for interpolating and prolongating  values between the super-particle and the grid nodes a-d, respectively. 
In 3D, volumes instead of areas are used to map values between the super-particle and the nearest 8 grid nodes. 
This mapping excludes gravitational self-forces when self-gravity is included. 
  \label{fig:cic_demo}}
  \end{center}
\end{figure}

Particles are included using a particle-in-cell method (PIC), where each particle represents a swarm of smaller particles. 
We refer to the particles that are directly integrated on the grid (the swarm) as {\it super-particles} and any member of that swarm as a {\it sub-particle}. 
All sub-particles have the same mass, density, and size, and are not allowed to evolve in time.  
The particles are connected to the grid using a cloud-in-cell (CIC) interpolation and prolongation, described as follows.  
Consider a two-dimensional Cartesian grid as depicted in Figure (\ref{fig:cic_demo}).  
The intersections of lines represent grid nodes, and the yellow smiley-face represents a super-particle.  
To interpolate values from the nodes a-d to the super-particle, the areas A-D are used for weights, where A is the weight from node a, etc. 
Prolongation is achieved using the same weighting scheme.
In a 3D scheme, the nearest 8 nodes are used for interpolation and prolongation in a volume-weighted scheme analogous to the area-weighted scheme shown Figure (\ref{fig:cic_demo}).

Each super-particle has a position, velocity, and mass.  
For self-gravity calculations, the super-particle's contribution to the total grid mass is included by setting the super-particle's mass density to $\rho_{sp}=m/V$, where $m$ is the super-particle's mass and $V$ is the volume of a single cell on the grid.  
This mass density is then prolongated to the nearest 8 cells, which is used when calculating the total gravitational potential field on the grid..  
The CIC algorithm with volume-weighted prolongation ensures that gravitational self-forces are absent.  
The temperature, density, and pressure of the gaseous fluid are tracked using CIC interpolation.
In addition, the gaseous fluid's velocity is interpolated to the particle's position to calculate drag. 

We use a drag algorithm that is heavily based on \cite{boley_durisen_2010} (hereafter BD10), which includes the back-reaction of the particle onto the gas in addition to gas drag effects.  
As done in BD10, the Knudsen number is used to interpolate between the Epstein and Stokes drag regimes, where piecewise, analytic solutions for the velocity differences  between the gas and the super-particle in the Epstein and Stokes regimes are calculated based on the Reynolds number. 
The velocity difference is then transformed to velocities relative to the grid using conservation of momentum.  
For reference, the Knudsen number ${\rm Kn}=\mu m_p/(2 \rho \sigma_{\rm gas} s)$, where $\mu$ is the mean molecular weight of the gas in proton masses $m_p$, $s$ is the sub-particle size, $\rho$ is the gas density, and $\sigma_{\rm gas}$ is the typical cross section of a gas particle, taken to be $10^{-16}$ cm$^2$.  
It represents the ratio of the mean-free-path of a particle embedded in a gas to the particle's size. 
The Reynolds number ${\rm Re}=3 \sqrt{\pi/8} \mathcal{M}/{\rm Kn}$ for Mach number $\mathcal{M}$, which reflects the ratio of inertial to viscous forces. 

While most of the details of the algorithm are given in BD10, there are several important differences.  
First, BD10 used a nearest node approximation for coupling solid-gas interactions, where we use the CIC algorithm.
After the change in gas momentum has been calculated, the total change is prolongated to the nearest 8 grid nodes. 
Second, we include conservation of energy as well as momentum, while BD10 only included conservation of momentum. 
After the new momenta of the gas and super-particle are calculated, the change in total energy of the gas and super-particle is determined. 
The difference in total energy is then applied to the gas at the nearest 8 cells using prolongation.  
Third, and perhaps most importantly, we include a method for handling drag whenever the drag equations become stiff relative to the hydrodynamics, i.e., the stopping time of the gas becomes comparable to or shorter than the Courant timescale.  

The stopping time of a solid of size $s$ and internal density $\rho_s$ moving through a gas with density $\rho$ and sound speed  $c_a$ is $t_s=\sqrt{\pi/8}(\rho_s s)/(\rho c_a)$. 
The terminal velocity of a solid moving through a gas is ${\bf g} t_s$, where {\bf g} is the local gravitational acceleration.  
Take the velocity difference between the gas and the super-particle after a full time step using the BD10 scheme as $\delta {\bf v}'$.  
If the Courant step is greater than $\rho_s s  \exp(-2)/(\rho c_a)$, then 
\begin{equation}
\delta {\bf v} = \delta {\bf v}' w - {\bf g} t_s (1-w),
\end{equation}
where $w=[\rho_s s\exp(-2)/(\rho c_a)/t_{\rm CFL} ]^2$, which was found to be an acceptable solution based on numerical experiments.

One potential problem with the above method for handling the stiff regime is that a super-particle will always reach the terminal velocity relative to the gas, even if the gas and the particle are in free fall. 
We attempt to circumvent this difficulty by using the pressure gradients $\nabla P$ in lieu of ${\bf g}$. 
When the gas is in hydrostatic equilibrium, the two uses are equivalent, and using the pressure gradient allows the solids and gas to move together when pressure gradients are negligible.

The time evolution of solids' position and velocity is included by first applying a half kick (changing the velocity of the super-particle) during sourcing. 
During a half kick, the velocity is first updated without the drag term, and then the velocity is updated again by incorporating drag.  
After the grid variables are updated from full fluxing, a full drift term is applied (changing the position of the super-particle).  
A second half kick is then calculated using these updates, including the updated potential. 
This creates a second-order leap-frog scheme with time-step averaging, i.e., for a given time step, half a kick is applied using the previous potential field and the other half with the most recent potential field.  

Tests including settling times in gaseous spheres demonstrate that particles reach terminal velocities as expected.
Additional tests show that particles follow the flow of the gas when they are well-coupled and recover the correct stopping times when they are not perfectly coupled.

\subsection{Particles and Opacity\label{sec:opacity}}

The radiation algorithm discussed in section \ref{sec:mcrtfld} requires some estimate of the local opacity of the gas. 
The actual values used are independent of the MCRTFLD algorithm itself, and can be altered based on the problem at hand.  
One possibility is to use a \cite{pollack_etal_1994} opacity table, which assumes some fixed composition of gas and solids, with a set distribution of solids sizes.  
With such tables, only the gas mass needs to be known, and the opacity can then be set according to the local gas density.

However, integrating super-particles directly into the fluid evolution gives rise to another possibility: using the super-particles themselves a the source of the opacity. 
As with a lookup table, such a strategy is not applicable to each problem, as many complexities can potentially arise.  
Nonetheless, there are multiple problems for which such a particle-based opacity method will be advantageous.  
One such example is the limit were there is no size evolution among subparticles.  

Consider  for simplicity a region of some astrophysical object that is populated by dust grains that have the same composition and are of the same size.  
Further assume that each grain has a size $s=0.03$ cm and has an internal density $\rho_s=3$ g/cc.  
A straightforward estimate for the mass opacity of such a population of dust is 
$\kappa\approx\xi \pi s^2/(4\pi/3 s \rho_s) = \xi3/(4 s \rho_s)$.
For our system of dust grains, $\kappa\approx 8.3$ cm$^2$ per gram of dust.
We have assumed that the radiative efficiency of the grain $\xi\sim1$ due to the large size of the grain.
A separate array can be used to track the gas density of the dust only, using the same prolongation technique needed for spatially evolving super-particles. 
With this information, the dust optical depth across a cell of width $\Delta x$ is then $\Delta \tau = \rho_{\rm dust} \kappa \Delta x$, assuming gray transfer.

\bibliographystyle{apj}

\begin{thebibliography}{}

\bibitem[\protect\citeauthoryear{{Alexander} et~al.}{{Alexander}
  et~al.}{2008}]{alexander_etal_2008}
{Alexander}, C.~M.~O.~., {Grossman}, J.~N., {Ebel}, D.~S.,  \& {Ciesla}, F.~J.
  2008, Science, 320, 1617

\bibitem[\protect\citeauthoryear{{Barnes} \& {Hut}}{{Barnes} \&
  {Hut}}{1986}]{barnes_hut_1986}
{Barnes}, J.,  \& {Hut}, P. 1986, Nature, 324, 446

\bibitem[\protect\citeauthoryear{{Bizzarro}, {Baker}, \& {Haack}}{{Bizzarro}
  et~al.}{2004}]{bizzarro_etal_2004}
{Bizzarro}, M., {Baker}, J.~A.,  \& {Haack}, H. 2004, Nature, 431, 275

\bibitem[\protect\citeauthoryear{{Bodenheimer} et~al.}{{Bodenheimer}
  et~al.}{1990}]{bodenheimer_etal_1990}
{Bodenheimer}, P., {Yorke}, H.~W., {Rozyczka}, M.,  \& {Tohline}, J.~E. 1990,
  ApJ, 355, 651

\bibitem[\protect\citeauthoryear{{Boley} \& {Durisen}}{{Boley} \&
  {Durisen}}{2008}]{boley_durisen_2008}
{Boley}, A.~C.,  \& {Durisen}, R.~H. 2008, ApJ, 685, 1193

\bibitem[\protect\citeauthoryear{{Boley} \& {Durisen}}{{Boley} \&
  {Durisen}}{2010}]{boley_durisen_2010}
{Boley}, A.~C.,  \& {Durisen}, R.~H. 2010, ApJ, 724, 618

\bibitem[\protect\citeauthoryear{{Boley} et~al.}{{Boley}
  et~al.}{2007}]{boley_etal_2007_h2}
{Boley}, A.~C., {Hartquist}, T.~W., {Durisen}, R.~H.,  \& {Michael}, S. 2007,
  ApJ, 656, L89

\bibitem[\protect\citeauthoryear{{Boss} \& {Durisen}}{{Boss} \&
  {Durisen}}{2005}]{boss_durisen_2005}
{Boss}, A.~P.,  \& {Durisen}, R.~H. 2005, ApJL, 621, L137

\bibitem[\protect\citeauthoryear{{Bouvier} \& {Wadhwa}}{{Bouvier} \&
  {Wadhwa}}{2010}]{bouvier_wadhwa_2010}
{Bouvier}, A.,  \& {Wadhwa}, M. 2010, Nature Geoscience, 3, 637

\bibitem[\protect\citeauthoryear{{Ciesla} \& {Hood}}{{Ciesla} \&
  {Hood}}{2002}]{ciesla_hood_2002}
{Ciesla}, F.~J.,  \& {Hood}, L.~L. 2002, \icarus, 158, 281

\bibitem[\protect\citeauthoryear{{Ciesla}, {Hood}, \&
  {Weidenschilling}}{{Ciesla} et~al.}{2004}]{ciesla_etal_2004}
{Ciesla}, F.~J., {Hood}, L.~L.,  \& {Weidenschilling}, S.~J. 2004, Meteoritics
  and Planetary Science, 39, 1809

\bibitem[\protect\citeauthoryear{{Connolly}}{{Connolly}}{2012}]{connolly_2012}
{Connolly}, H.~C. 2012, Meteoritics and Planetary Science, 47, 1105

\bibitem[\protect\citeauthoryear{{Connolly} \& {Love}}{{Connolly} \&
  {Love}}{1998}]{connolly_love_1998}
{Connolly}, H.~C., Jr.,  \& {Love}, S.~G. 1998, Science, 280, 62

\bibitem[\protect\citeauthoryear{{Cuzzi} \& {Alexander}}{{Cuzzi} \&
  {Alexander}}{2006}]{cuzzi_alexander_2006}
{Cuzzi}, J.~N.,  \& {Alexander}, C.~M.~O. 2006, Nature, 441, 483

\bibitem[\protect\citeauthoryear{{Cuzzi} \& {Hogan}}{{Cuzzi} \&
  {Hogan}}{2003}]{cuzzi_etal_2003}
{Cuzzi}, J.~N.,  \& {Hogan}, R.~C. 2003, Icarus, 164, 127

\bibitem[\protect\citeauthoryear{{Dauphas} \& {Pourmand}}{{Dauphas} \&
  {Pourmand}}{2011}]{dauphas_pourmand_2011}
{Dauphas}, N.,  \& {Pourmand}, A. 2011, Nature, 473, 489

\bibitem[\protect\citeauthoryear{{Desch} et~al.}{{Desch}
  et~al.}{2005}]{desch_etal_2005}
{Desch}, S.~J., {Ciesla}, F.~J., {Hood}, L.~L.,  \& {Nakamoto}, T. 2005, in
  Astronomical Society of the Pacific Conference Series, Vol. 341, Chondrites
  and the Protoplanetary Disk, ed. A.~N. {Krot}, E.~R.~D. {Scott}, \&
  B.~{Reipurth}, 849

\bibitem[\protect\citeauthoryear{{Desch} \& {Connolly}}{{Desch} \&
  {Connolly}}{2002}]{desch_connolly_2002}
{Desch}, S.~J.,  \& {Connolly}, H.~C., Jr. 2002, Meteoritics and Planetary
  Science, 37, 183

\bibitem[\protect\citeauthoryear{{Desch} et~al.}{{Desch}
  et~al.}{2012}]{desch_etal_mps_2012}
{Desch}, S.~J., {Morris}, M.~A., {Connolly}, H.~C.,  \& {Boss}, A.~P. 2012,
  Meteoritics and Planetary Science, 47, 1139

\bibitem[\protect\citeauthoryear{{Desch} \& {Mouschovias}}{{Desch} \&
  {Mouschovias}}{2001}]{desch_mouschovias_2001}
{Desch}, S.~J.,  \& {Mouschovias}, T.~C. 2001, ApJ, 550, 314

\bibitem[\protect\citeauthoryear{{Galvagni} et~al.}{{Galvagni}
  et~al.}{2012}]{galvagni_etal_2012}
{Galvagni}, M., {Hayfield}, T., {Boley}, A., {Mayer}, L., {Ro{\v s}kar}, R.,
  \& {Saha}, P. 2012, MNRAS, 427, 1725

\bibitem[\protect\citeauthoryear{{Gooding} \& {Keil}}{{Gooding} \&
  {Keil}}{1981}]{gooding_keil_1981}
{Gooding}, J.~L.,  \& {Keil}, K. 1981, Meteoritics, 16, 17

\bibitem[\protect\citeauthoryear{{Grossman}}{{Grossman}}{1988}]{grossman_nature_1988}
{Grossman}, J.~N. 1988, Nature, 334, 14

\bibitem[\protect\citeauthoryear{{Hewins} \& {Radomsky}}{{Hewins} \&
  {Radomsky}}{1990}]{hewins_radomsky_1990}
{Hewins}, R.~H.,  \& {Radomsky}, P.~M. 1990, Meteoritics, 25, 309

\bibitem[\protect\citeauthoryear{{Hood}}{{Hood}}{1998}]{hood_1998}
{Hood}, L.~L. 1998, Meteoritics and Planetary Science, 33, 97

\bibitem[\protect\citeauthoryear{{Hood} et~al.}{{Hood}
  et~al.}{2009}]{hood_etal_2009}
{Hood}, L.~L., {Ciesla}, F.~J., {Artemieva}, N.~A., {Marzari}, F.,  \&
  {Weidenschilling}, S.~J. 2009, Meteoritics and Planetary Science, 44, 327

\bibitem[\protect\citeauthoryear{{Jiang} et~al.}{{Jiang}
  et~al.}{2013}]{jiang_etal_2013}
{Jiang}, Y.-F., {Belyaev}, M., {Goodman}, J.,  \& {Stone}, J.~M. 2013, New
  Astronomy, 19, 48

\bibitem[\protect\citeauthoryear{{Kuebler} et~al.}{{Kuebler}
  et~al.}{1999}]{kuebler_etal_icar_1999}
{Kuebler}, K.~E., {McSween}, H.~Y., {Carlson}, W.~D.,  \& {Hirsch}, D. 1999,
  Icarus, 141, 96

\bibitem[\protect\citeauthoryear{{Kurahashi} et~al.}{{Kurahashi}
  et~al.}{2008}]{kurahashi_etal_2008}
{Kurahashi}, E., {Kita}, N.~T., {Nagahara}, H.,  \& {Morishita}, Y. 2008, GCA,
  72, 3865

\bibitem[\protect\citeauthoryear{{Morris} et~al.}{{Morris}
  et~al.}{2012}]{morris_etal_2012}
{Morris}, M.~A., {Boley}, A.~C., {Desch}, S.~J.,  \& {Athanassiadou}, T. 2012,
  ApJ, 752, 27

\bibitem[\protect\citeauthoryear{{Morris} \& {Desch}}{{Morris} \&
  {Desch}}{2010}]{morris_desch_2010}
{Morris}, M.~A.,  \& {Desch}, S.~J. 2010, ApJ, 722, 1474

\bibitem[\protect\citeauthoryear{{Pollack} et~al.}{{Pollack}
  et~al.}{1994}]{pollack_etal_1994}
{Pollack}, J.~B., {Hollenbach}, D., {Beckwith}, S., {Simonelli}, D.~P.,
  {Roush}, T.,  \& {Fong}, W. 1994, ApJ, 421, 615

\bibitem[\protect\citeauthoryear{{Pollack} et~al.}{{Pollack}
  et~al.}{1996}]{pollack_etal_1996}
{Pollack}, J.~B., {Hubickyj}, O., {Bodenheimer}, P., {Lissauer}, J.~J.,
  {Podolak}, M.,  \& {Greenzweig}, Y. 1996, Icarus, 124, 62

\bibitem[\protect\citeauthoryear{{Scherst{\'e}n} et~al.}{{Scherst{\'e}n}
  et~al.}{2006}]{schersten_etal_2006}
{Scherst{\'e}n}, A., {Elliott}, T., {Hawkesworth}, C., {Russell}, S.,  \&
  {Masarik}, J. 2006, Earth and Planetary Science Letters, 241, 530

\bibitem[\protect\citeauthoryear{Sod}{Sod}{1978}]{sod1978}
Sod, G.~A. 1978, Journal of Computational Physics, 27, 1

\bibitem[\protect\citeauthoryear{{Susa} \& {Nakamoto}}{{Susa} \&
  {Nakamoto}}{2002}]{susa_nakamoto_2002}
{Susa}, H.,  \& {Nakamoto}, T. 2002, ApJL, 564, L57

\bibitem[\protect\citeauthoryear{Toro}{Toro}{2009}]{toro2009riemann}
Toro, E. 2009, Riemann Solvers and Numerical Methods for Fluid Dynamics: A
  Practical Introduction (Springer-Verlag Berlin Heidelberg)

\bibitem[\protect\citeauthoryear{{Villeneuve}, {Chaussidon}, \&
  {Libourel}}{{Villeneuve} et~al.}{2009}]{villeneuve_etal_2009}
{Villeneuve}, J., {Chaussidon}, M.,  \& {Libourel}, G. 2009, Meteoritics and
  Planetary Science Supplement, 72, 5205

\bibitem[\protect\citeauthoryear{{Weidenschilling}, {Marzari}, \&
  {Hood}}{{Weidenschilling} et~al.}{1998}]{weidenschilling_etal_1998}
{Weidenschilling}, S.~J., {Marzari}, F.,  \& {Hood}, L.~L. 1998, Science, 279,
  681

\bibitem[\protect\citeauthoryear{{Weyrauch} \& {Bischoff}}{{Weyrauch} \&
  {Bischoff}}{2012}]{weyrauch_bischoff_2012}
{Weyrauch}, M.,  \& {Bischoff}, A. 2012, Meteoritics and Planetary Science, 47,
  2237

\end{thebibliography}

  \end{document}